\documentclass[twocolumn]{aastex631}
\def\apjl{{ApJ}}                
\def\apjs{{ApJS}}               
\def\ao{{Appl.~Opt.}}           
\def\aap{{A\&A}}                

\usepackage{amsmath}	
\usepackage{amssymb}	

\usepackage{graphicx}
\usepackage{hyperref} 
\usepackage{listings}
\usepackage{color}
\usepackage{graphicx,times}
\usepackage{natbib}
\usepackage{amssymb}
\usepackage[Figureright]{rotating}
\usepackage{newtxtext}
\usepackage[T1]{fontenc}
\usepackage{ae,aecompl}
\usepackage{threeparttable}
\usepackage{multirow}
\usepackage{ulem}
\usepackage{txfonts}

\bibpunct{(}{)}{;}{a}{}{,}
\usepackage{hyperref}
\usepackage{longtable}
\usepackage{booktabs}
\definecolor{dkgreen}{rgb}{0,0.6,0}
\definecolor{gray}{rgb}{0.5,0.5,0.5}
\definecolor{mauve}{rgb}{0.58,0,0.82}
\definecolor{golden}{rgb}{0.86,0.65,0.01}

\shorttitle{Census of distant OC BSSs and maximum $M_{e}$}

\usepackage[Figureright]{rotating}
\usepackage{ulem}
\usepackage{CJK}
\usepackage{amsmath}
\usepackage{natbib}

\usepackage{threeparttable} 
\usepackage{booktabs} 
\begin{document}
\begin{CJK*}{UTF8}{gbsn}

\setlength{\baselineskip}{15pt}
\title{Census of Blue Straggler Stars in Distant Open Clusters and Maximum Fractional Mass Excess of  OC BSS}

\author[0000-0003-0089-2005]{Qian Cui}
\affiliation{School of Physics and Astronomy, China West Normal University, No. 1 Shida Road,    Nanchong 637002, China}

\author[0000-0002-6989-8192]{Zhihong He}
\affiliation{School of Physics and Astronomy, China West Normal University, No. 1 Shida Road,    Nanchong 637002, China}\email{hezh@mail.ustc.edu.cn}

\author{Shunhong Deng}
\affiliation{School of Physics and Astronomy, China West Normal University, No. 1 Shida Road,    Nanchong 637002, China}

\author{Liming Peng}
\affiliation{School of Physics and Astronomy, China West Normal University, No. 1 Shida Road,    Nanchong 637002, China}

\author{Chunyan Li}
\affiliation{School of Physics and Astronomy, China West Normal University, No. 1 Shida Road,    Nanchong 637002, China}

\author{Yangping Luo}
\affiliation{School of Physics and Astronomy, China West Normal University, No. 1 Shida Road,    Nanchong 637002, China}

\author{Kun Wang}
\affiliation{School of Physics and Astronomy, China West Normal University, No. 1 Shida Road,    Nanchong 637002, China}

\begin{abstract}
We identified blue straggler stars (BSSs) in 53 open clusters utilizing data from Gaia DR3. Most of these clusters are situated in the outer regions of the Galactic disc, encompassing structures such as the warp and the Outer arm.
We analyzed their astrometric parameters and determined that 48 of them demonstrate high reliability in radial density profile. Furthermore, through manual isochrone fitting and visual inspection, we confirmed 119 BSS candidates and identified 328 additional possible candidates within these clusters.
Our results contribute to a 46$\%$ increase in the sample size of BSSs in open clusters for regions of the Galactic disc where $R_{gc} > 12 \text{ kpc}$. 
We observed that the new samples are fainter compared to those identified in the past. Additionally, we investigated the maximum fractional mass excess ($M_{e}$) of the BSSs in open clusters, including previously published BSS samples. Our findings indicate a strong correlation between the capability to produce highest-$M_{e}$ BSSs and the mass of their host clusters. This observation appears to reinforce a fundamental principle whereby an increase in the mass of a star cluster correlates with a higher likelihood of stellar mergers. In contrast, we observe minimal correlation between maximum-$M_{e}$ and the cluster age. Among clusters containing BSSs, younger clusters (0.5 to 1~Gyr) display a scarcity of high-$M_{e}$ BSSs. This scarcity may be attributed to the absence of more massive clusters within this age range.
\end{abstract}

 
\section{Introduction} \label{sec:intro}
Blue straggler star is an intriguing class of stellar phenomena that have consistently challenged astronomers and their models of stellar evolution. First identified in globular clusters (GCs) in the 1950s \citep{Sandage1953,BS1958} and later observed in open clusters (OCs) \citep[e.g.,][]{ES1964,Greenstein1964}, 
these stars appear unexpectedly young and massive compared to their counterparts in the same stellar cluster.
Their presence raises substantial questions about the conventional understanding of stellar aging processes because they are located on the blue side of the main sequence within color-magnitude diagrams (CMDs), a region that is typically not associated with older stars. Without additional evidence such as radial velocity, proper motion, or parallax to confirm their membership in a star cluster, this unusual positioning can sometimes lead to their misclassification as field stars~\citep{Carraro2008}. 

The research significance of BSS lies in their potential to offer critical insights into the complex interplay between gravitational dynamics and stellar interactions. Notable examples of BSS formation hypotheses include stellar collisions or mergers~\citep[e.g.][]{Lombardi96,Sills97,Schneider19}; mass transfer in binary systems~\citep[e.g.][]{CH2008,Sen22}; and interactions involving multiple stars~\citep[e.g.][]{Fregeau04,PF2009}.
For the stellar aggregates, GCs and OCs provide distinct environments for studying BSSs. In GCs and the cores of large OCs, the high density of stars increases the likelihood of stellar collisions. In contrast, OCs with lower stellar density are less prone to such collisions, and binary mass transfer processes are more probable mechanisms for the formation of BSSs in these clusters~\citep{Wang24}.
Recent research in isolated environments, such as the halo of the Milky Way \citep{Preston00} and dwarf spheroidal galaxies \citep{Momany07}, reveals intricate formation mechanisms of BSS. These studies have shown that dwarf galaxies exhibit notably higher frequencies of BSS compared to GCs of equivalent luminosity. Additionally, \cite{Clarkson2011} reported the presence of BSS in the Galactic bulge, demonstrating how varied environmental conditions influence BSS genesis. This variation prompted a nuanced analysis of their formation processes.


The ongoing census of BSS within stellar clusters has been significantly advanced by systematic efforts spanning several decades. In GCs, \cite{Piotto2004} compiled a catalog of approximately 3,000 BSSs across 56 such clusters, observing an anticorrelation between BSS frequency and cluster luminosity. 
For the OCs, ~\cite{AL1995,AL2007} made foundational contributions by cataloging 959 and subsequently 1,887 BSSs candidates across 390 and 427 OCs, respectively. Their work demonstrated a correlation between the ratio 
 of the number of stragglers to giants and cluster age and richness, with a consistent ratio of BSS to main sequence stars up to an age of about 0.5~Gyr, after which this ratio begins to increase. 
 
Through the growing catalogs, our understanding of BSS formation across various stellar environments is poised for significant advancement, driving continued research and discovery. The advent of Gaia DR2 \citep{GAIADR2} and DR3 \citep{GAIADR3} has significantly refined BSS research in Galactic clusters, enhancing both precision and scope in astrometric surveys. Gaia DR2 facilitated \cite{Rain21} in cataloging 897 BSSs across 111 OCs, uncovering correlations between BSS occurrence and factors such as cluster age and mass. Concurrently, \cite{Jadhav21} identified 868 BSSs in 228 clusters, noting a power-law relationship between cluster mass and BSS quantity. Expanding upon these advancements, \cite{Licy23} utilized Gaia DR3 to uncover an additional 138 previously undocumented BSSs in 50 newly identified clusters, reflecting a 10\% increase in known BSSs and a 17\% enhancement in the number of clusters containing BSS. The enriched data set enables more statistical analyses and theoretical models, fostering a deeper understanding of BSS formation processes and elucidating stellar population dynamics within diverse Galactic environments.

In the process of searching for star clusters and verifying their CMDs, the presence of BSSs can confound automated algorithms tasked with isochrone fitting. This interference often necessitates a visual inspection to correctly identify and mitigate their impact \citep[e.g.,][]{Castro20,He23b}. In our recent study, we address this issue by examining 53 OCs that show the presence of prominent BSSs~\citep[][H23]{He23b}, most of which are situated in clusters located over 4~kpc away from the sun. Our study aims to catalogue these BSSs, evaluating their probabilities and physical properties. Different from the previous censuses \citep[e.g.,][]{Rain21,Licy23}, these clusters primarily occupy the outer regions of the Galactic disk, thus providing a valuable extension to existing BSS samples in OCs, potentially enriching our understanding of their evolutionary pathways and distribution across the Milky Way.

The structure of this article is organized as follows: In Section \ref{sec:Data Source}, we introduce our star cluster samples along with the BSS samples selected for comparison. Section \ref{sec:method} describes the methodologies employed in our analysis. In Section \ref{sec:results}, we present the cataloging results, along with a comprehensive analysis and statistical evaluation of the identified BSSs. Finally, Section \ref{sec:summary} offers a summary of our findings.

\section{OC Sample} \label{sec:Data Source}

Historically, OC catalogs have identified between 2,000 and 3,000 samples, as detailed by \cite{Dias02} and \cite{Kharchenko13}. However, the limited precision of ground-based telescopic measurements has occasionally led to the misclassification of some objects as asterisms~\citep{CA20}. Enhancing the situation, the launch of the Gaia spacecraft in 2013 introduced an significant improved precision of 24 microarcseconds~\citep{Gaia16}. Utilizing the high-precision data from Gaia, subsequent studies have successfully re-identified approximately half of the previously cataloged clusters~\citep{CG18}. Additionally, the advancements from the Gaia mission have significantly improved the accuracy of new cluster identifications, facilitating the discovery of many new stellar aggregates. With important discoveries reported each year in recent researches \citep[e.g.,][]{Sim19,LP19,Castro20,castro22,hao22,Ferreira20,Kounkel20,Qin21,Qin23,he21a,He22b,He22a,He23a,He23b,Hunt21,HR23}. As a result, the current catalog now includes over 7,000 identified OCs and candidates, marking a considerable expansion in our understanding of these astronomical objects.

\begin{figure*}
    \centering
    \includegraphics[width=0.91\linewidth]{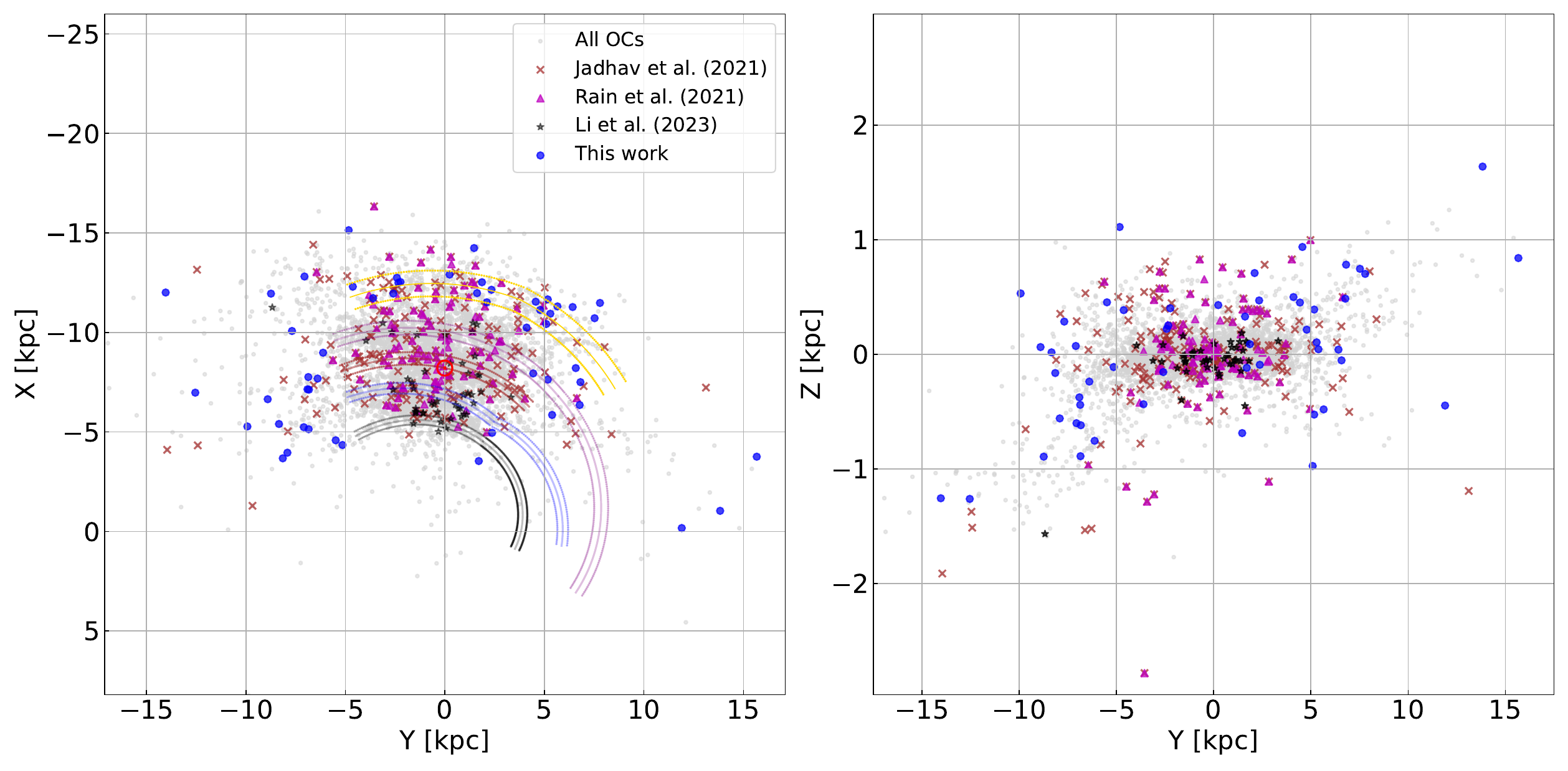}
  \caption{The spatial distribution of OCs containing BSSs in the Milky Way. Left panel: A face-on view of the Galactic plane, showing all OCs~\citep{He23_warp} as gray points and OCs with BSSs as colored symbols, relative to the spiral arms. This is based on the spiral arm model from~\cite{Reid2019}, which includes the Scutum-Centaurus (black), Sagittarius (blue), Local (brown), Perseus (magenta), and Outer (yellow) arms. The sun's position is marked at (-8.15, 0)~kpc. The newly identified OCs with BSSs studied in this work are primarily located in the outer regions of the Galaxy. Right panel: An edge-on view displays the vertical distribution of all OCs and OCs with BSSs along the Z-axis. The distribution indicates that most OC BSSs are confined to the Galactic disk with minimal vertical displacement, although some are associated with Galactic warp structures.}
    \label{fig:bss in mw}
\end{figure*}

In our previous work (H23), we presented the "Two Gaussian Fitting for Isolated Groups" (TGFIG) method.
To mitigate the influence of field stars and other cluster members near the clusters, H23 removed known clusters from the search area prior to performing the clustering analysis. Additionally, recognizing that distant clusters are susceptible to foreground star contamination, H23 conducted a second data clip using parallax measurements. Specifically, stars with parallax values outside of 5 sigma (the standard deviation of the parallax distribution) were excluded from the dataset before a new clustering was performed to identify cluster members. These steps ensured that cluster members were maximized while field stars were minimized, thereby enabling the identification of more distant clusters and their members~\citep[see details in H23 and][]{Negueruela25}.
By employing TGFIG, we conducted a thorough search for stars in the Galactic plane ($\left| b \right| < 10^\circ$) utilizing data from Gaia DR3. This analysis led to the identification of 1,488 new OCs, of which 53 clusters are potential hosts for BSSs.
Figure~\ref{fig:bss in mw} illustrates the spatial distribution of these 53 OCs within the Milky Way, along with previously known OCs that contain BSSs. The distribution reveals a notable concentration of our identified clusters in the outer spiral arms, with some samples located at higher vertical scale height, suggesting their presence within the context of the Galactic warp structure.

\begin{figure*}
    \centering
    \includegraphics[width=1\linewidth]{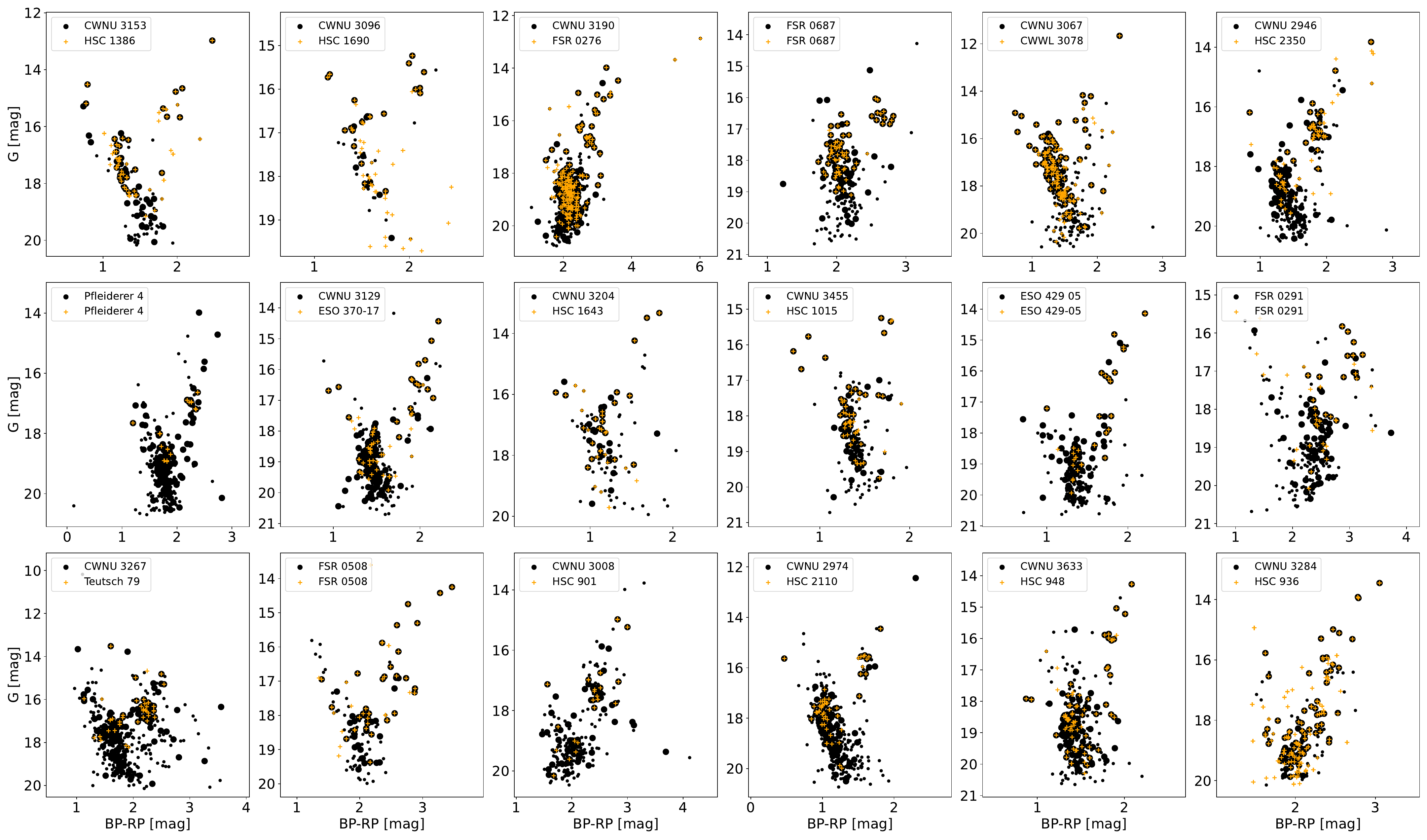}
    \caption{The cross-match results between H23 and HR23 and the corresponding distribution of member stars in the H-R diagram. Star cluster members from H23 are depicted as black dots, with larger dots representing core members and smaller dots indicating border members. The orange pluses denote member stars from the clusters in HR23. Three clusters are identified as reliable in HR23: CWNU~3190 (FSR~0276), CWNU~2974 (HSC~2110), and FSR~0687.}
    \label{fig:xmatch}
\end{figure*}

Simultaneously, \citet[][HR23]{HR23} compiled a catalog containing 7,167 clusters, with 4,105 classified as high-confidence clusters. This classification signifies that these clusters exhibit well-defined Hertzsprung-Russell (H-R) diagrams. They feature a distinct main sequence and display consistent radial velocities among their members. In our study, we cross-matched 53 clusters with the HR23 catalog and found 18 common clusters (Figure~\ref{fig:xmatch}). 
Among these common clusters, three are also identified as reliable true clusters by HR23. However, the other 15 are not labeled as reliable OCs. Despite this, all 53 clusters in our study are marked as high-reliability clusters in H23. To further evaluate the characteristics of these clusters, we conduct additional investigations using proper motion probabilities and density profile analysis (see Section~\ref{sec:method}).
In addition, the CMDs in two studies reveal a limited number of outliers, such as red stragglers and sub-subgiant star in CWNU~3153; similar instances are observed in other H23 clusters~\citep{Negueruela25}. These outliers may be attributed to differential extinction within the clusters and/or field star contamination.

Additionally, H23 employed the DBSCAN algorithm to evaluate the membership of the newly identified objects based on Gaia DR3 data. Members were categorized into core and border groups. Core members are defined as stars that have a minimum number of neighboring stars within a defined range, utilizing five astrometric parameters: [$l$, $b$, $\varpi$, $\mu_\alpha$, $\mu_\delta$]. In contrast, border members are the neighboring stars associated with core members. As shown in Figure~\ref{fig:xmatch}, differences in selection criteria and clustering methodologies applied to the Gaia DR3 data led to variations in the number of identified cluster members. We found that the counts of brighter member stars were quite similar between the two catalogs. However, the H23 catalog indicated a larger number of fainter member stars. This discrepancy may be attributed to the increased uncertainties associated with fainter stars, along with differences in the membership determination techniques used in both studies.
To facilitate isochrone fitting, we selected members from the H23 membership catalog as our research samples. Importantly, most BSSs are situated in the brighter region of the CMDs of the star clusters. This suggests that the differences in member counts are unlikely to significantly impact our findings.


\section{Method} \label{sec:method}
To assess the concentration of star clusters and confirm their authenticity as real clusters, we investigated the density profile of above 53 OCs. To mitigate the impact of field stars, we began by calculating the probability of stars within a specified range of each cluster, utilizing the proper motions associated with the clusters. This probability calculation informed our selection of stars for density classification. We based our analysis on data from Gaia DR3 and followed a series of preprocessing steps.
Initially, we extracted stars from a 30 $\times$ 30 $arcmin^2$ region centered on the position of each cluster. Subsequently, we implemented several selection criteria to ensure the reliability of the astrometric solutions. Specifically, we: i) excluded sources whose parallax values were outside the 3$\sigma$ range of the cluster's central parallax, with $\sigma$ denoting the measurement uncertainty of Gaia DR3 sources; and ii) eliminated sources with proper motion uncertainties greater than 1~mas~yr$^{-1}$.
\subsection{Memebership Probabilities in Proper Motions}
For the sources selected based on the above criteria, we used the approach inspired by \cite{Dinescu1996} and \cite{Balaguer1998}  to calculate member star frequency $\phi_c$ and field star frequency $\phi_f$:

\begin{align}
\phi_c &= \frac{N_{c} }{2\pi \sqrt{(\sigma_c^2 + \epsilon_{xi}^2)(\sigma_c^2 + \epsilon_{yi}^2)}}  \nonumber \\
&\quad \times \exp \left( -\frac{1}{2} \left[ \frac{(\mu_{xi} - \mu_{xc})^2}{\sigma_c^2 + \epsilon_{xi}^2} + \frac{(\mu_{yi} - \mu_{yc})^2}{\sigma_c^2 + \epsilon_{yi}^2} \right] \right)
\end{align} 

\begin{align}
\phi_f &= \frac{N_{f}}{2\pi \sqrt{1 - \gamma^2} \sqrt{(\sigma_{xf}^2 + \epsilon_{xi}^2)(\sigma_{yf}^2 + \epsilon_{yi}^2)}} \nonumber \\
&\quad \times \exp \left( -\frac{1}{2(1 - \gamma^2)} \left[ \frac{(\mu_{xi} - \mu_{xf})^2}{\sigma_{xf}^2 + \epsilon_{xi}^2} \right. \right. + \frac{(\mu_{yi} - \mu_{yf})^2}{\sigma_{yf}^2 + \epsilon_{yi}^2} 
 \left. \left. - 2\gamma^2 \right] \right)
\end{align}
where ($\mu_{xi}$, $\mu_{yi}$) are the proper motion values of the $i_{th}$ star, while ($\epsilon_{xi}$, $\epsilon_{yi}$) represent their uncertainties. The proper motion center for the cluster is represented by ($\mu_{xc}$, $\mu_{yc}$), whereas ($\mu_{xf}$, $\mu_{yf}$) are the proper motion values for the field stars. The  proper motion dispersion for cluster stars is denoted by $\sigma_c$, while $\sigma_{xf}$ and $\sigma_{yf}$ correspond to the field stars' dispersions. We employed the correlation of proper motion parameters ($pmra\_pmdec\_corr$) to calculate the correlation coefficient, denoted as $\gamma$. $N_c$ and $N_f$ denote the normalized counts of cluster and field stars, respectively, with $N_c + N_f = 1$. Lastly, the membership probability for the $i_{th}$ star is determined by:
\begin{equation} 
P_{\mu}(i) = \frac{\phi_c(i)}{\phi_c(i) + \phi_f(i) } 
\end{equation}

Based on the formula above, we calculate the membership probability for each star in the region of each cluster.
Our findings indicate that most of the member stars identified through the clustering algorithm (H23) exhibit high proper motion membership probabilities (>80$\%$), which correspond to stars that closely align with the median proper motion of the cluster. It is noteworthy that BSSs tend to be brighter within clusters, resulting in smaller astrometric uncertainties for these stars. Therefore, their astrometric parameters are generally situated nearer to the central values of the cluster. In contrast, fainter member stars may display greater uncertainties in proper motion, which could lead to their underrepresentation in previous analyses utilizing the DBSCAN algorithm and potentially result in lower probability values. Consequently, we excluded stars with a probability below 50$\%$ from our subsequent analysis, which focused on calculating the radial density profile. Although the 50$\%$ cutoff may exclude some potential fainter members, it has a negligible impact on the identification of BSSs.

\subsection{Density Assessment}

To determine the radial density profile of the cluster, we analyzed its radial density distribution utilizing stellar membership probabilities ($P_{\mu}(i)$) derived from the proper motion data in Gaia DR3. We selected stars with $P_{\mu}(i) \geq 50\%$ to minimize contamination from field stars. The initial center of the cluster was defined as the median Galactic longitude and latitude of the core member stars identified in H23.
For each star, we calculated the distance $r$ from the cluster center by using their coordinates along with the cluster's angular distance. To construct the radial density profile, we organized the stars into concentric rings centered on the cluster. The radii of these rings increased in increments corresponding to a linear distance of 0.25 or 0.5~pc, depends on the background density of the field. For each ring, we determined its area $S$ and counted the number of stars $N$ contained within it. The surface density $\rho$ was then computed as $\rho = N / S$.

After constructing the surface density $\rho$, we fitted the data using the model proposed by \cite{King1962} and iterated the process multiple times until the center of the cluster no longer showed significant variation:
\begin{equation}\label{eq_density}
\rho(r)=k \Bigg(\frac{1}{\sqrt{1 + (r/r_c)^2}}-\frac{1}{\sqrt{1 + (r_t/r_c)^2}}\Bigg)^2+ \rho_{\text{bg}}
\end{equation}
where \( r_c \) represents the core radius and \( r_t \) is the tidal radius where the cluster density blends into the background, $k$ is the scaling constant, and \( \rho_{\text{bg}} \) denotes the background density. To ensure the robustness of the fit, we tested a range of bin widths (0.25 to 1.5 pc in steps of 0.25 pc). The best-fit parameters ($r_c$, $r_t$, $k$, and $\rho_{\text{bg}}$) were determined by minimizing the residual sum of squares. The uncertainties were then quantified using the median absolute deviation~\footnote{$\sigma_x = 1.4826 \cdot \text{median}(|x_i - \text{median}(x)|)$ , where $x_i$ represents the parameter values (either $r_c$ or $r_t$) obtained for each bin width.} of the fitted parameters across different bins.

In the fitting process, we determined the average background surface density (\(\rho_{\text{bg}}\)) for radii greater than 20~pc. However, for some clusters, the fitted tidal radius \(r_t\) could not be constrained to a common value, resulting in excessively large values that approached infinity. This condition implies that the second term in Equation~\ref{eq_density} becomes zero, making it challenging to derive an accurate \(r_t\). To address this issue, we applied a correction by identifying the radius where the absolute gradient of the fitting curve stabilizes. This corresponding radius is designated as the corrected value (\(r_t'\)).
Furthermore, to account for the discrepancy between the actual surface density and the theoretical King model, we considered the standard deviation of the fluctuations, denoted as \(\sigma_{\rho}\). We introduced a new parameter, the boundary radius \(r_b\), defined as the radius at which the difference between the fitted surface density and the actual surface density equals the \(\sigma_{\rho}\). This can be expressed as \(\rho(r_b) = \rho_{\text{bg}} + \sigma_{\rho}\).

\begin{figure*}
\begin{center}
\includegraphics[width=0.462\linewidth]{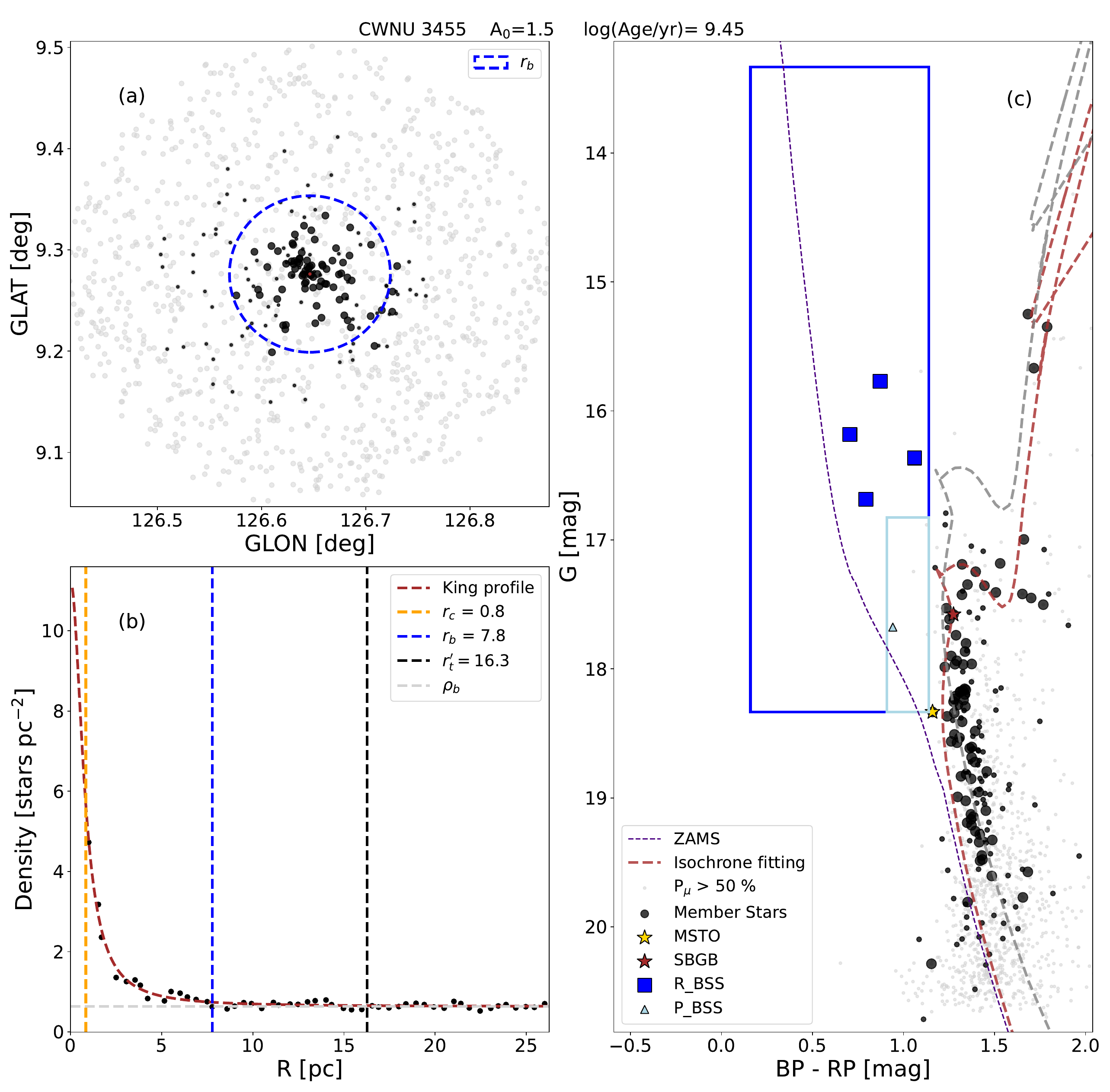}
\includegraphics[width=0.462\linewidth]{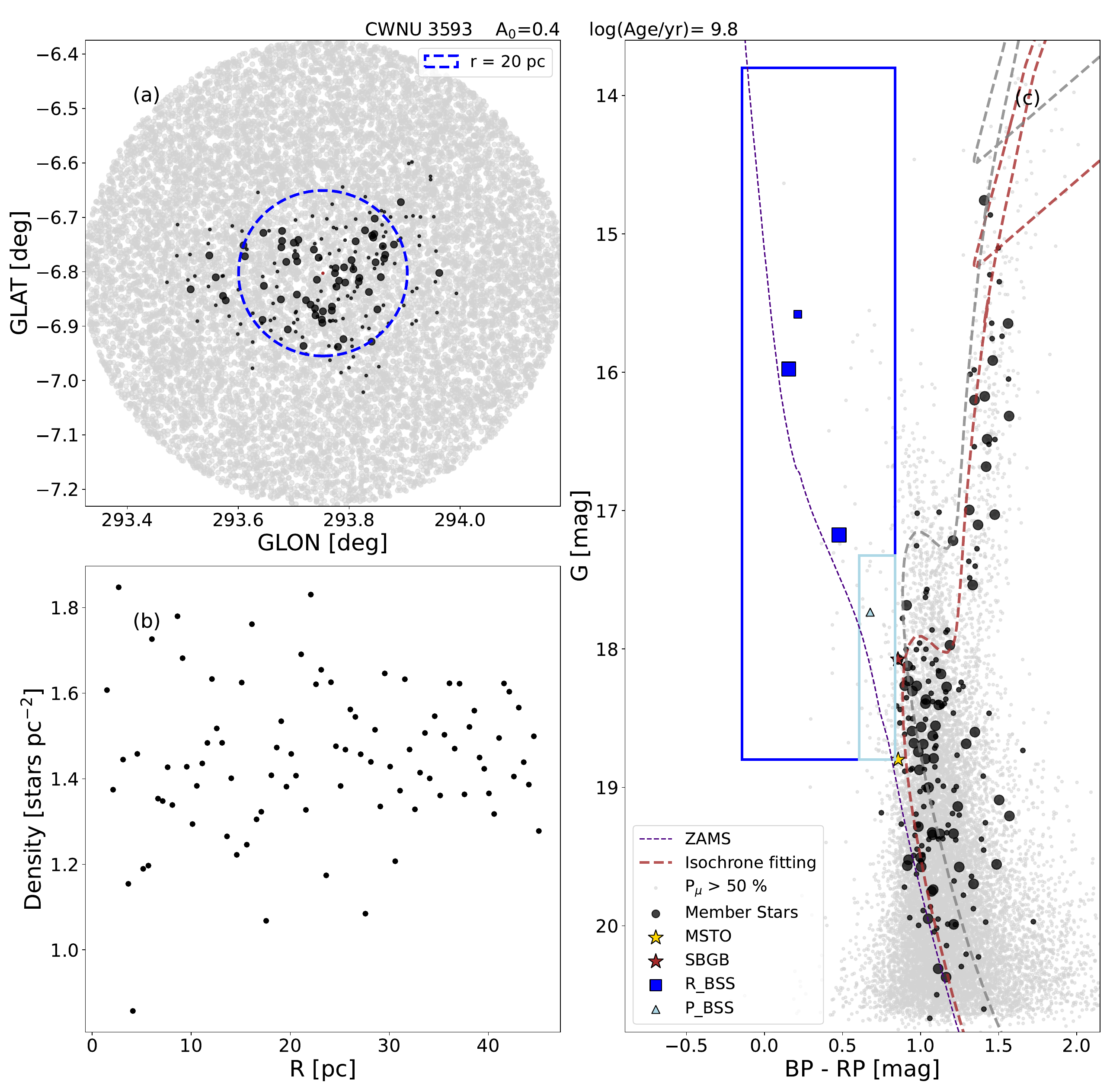}
\caption{Examples of Class A (left panels) and Class B (right panels) OCs. The subplots display various aspects of the member stars. Panels (a) shows the member stars in Galactic coordinates. Panels (b) presents the density profile along with the corresponding King model fitting, highlighting the boundary radius \(r_b\). Specifically, CWNU~3455 is a newly identified OC from H23, situated within the Galactic northern warp. This cluster exhibits a high concentration of member stars. In contrast, CWNU~3593 displays poor concentration, resulting in the King model's failure to accurately fit its density profile. Panels (c) depicts the various star types on the H-R diagram. Gray dots represent the stars with a membership probability greater than 50$\%$. Black dots indicate core (larger) and border (smaller) member stars. Large and small blue squares denote BSS candidates among core and border members, respectively. The purple dashed line marks the ZAMS, while the brown and gray dashed line illustrates the best-fit isochrone line and its binary sequence, respectively. The full figure sets for other 51 OCs can be seen in Appendix~\ref{appdendixa}.}
\label{fig_profile_isochrone}
\end{center}
\end{figure*}

As illustrated in Figure~\ref{fig_profile_isochrone}, we obtained the radial density profile and the fitted King model based on the optimal parameters for the cluster. We categorized the OCs according to their density profiles. Class~A comprises clusters whose profiles align with the King model. In contrast, Class~B includes clusters that display density profiles not well characterized by the King model. This discrepancy may be due to either a dense background of field stars or possible misclassification of these objects as genuine clusters.
For the Class~B clusters, which account for 10$\%$ of the total sample, require further validation to confirm their status as authentic OC.
For the Class~A OCs, we found that the tidal radius generally ranges from 10 to 30~pc, while the core radius is typically less than 5~pc. These findings are consistent with the established understanding of OC radii in prior studies~\citep[e.g.][]{Kharchenko13,Tarricq22,HR23}. Furthermore, the boundary radius is substantially larger than the core radius for many clusters, indicating that the density of these star clusters is significantly higher than that of the surrounding field stars.

\subsection{Isochrone Fitting}
The isochrone fitting was performed by aligning key features of the cluster's CMD, specifically the main sequence turn off (MSTO) point and the onset of the sub-giant branch (SBGB), with the corresponding theoretical isochrones. Since the previous automatic fitting process may affected by the binary or straggler stars, to derive precise cluster parameters, we employed a manual fitting approach. We used the Gaia EDR3 passbands and theoretical isochrone models from \cite{Bressan2012}, applied to the H-R diagram of each cluster.  Isochrone data with an age range of [6.0, 10.1] dex in steps of 0.05~dex, were used for this process. Cluster metallicities were selected based on the Galactic metallicity distribution relationship established by \cite{Hayden2015}. In this step, in order to reduce possible field star contamination in the border members, we only used the core members for the fitting. Besides, due to the overestimation of BP-band photometry for fainter stars~\citep{Riello21}, a shift in the CMD towards the bluer side may occur. Consequently, we opted to exclude members with G-band exceeds 19 to 19.5~mag when performing isochrone fits.

As described in our previous works~\citep{He23a,He23b}, For each OC, the extinction coefficient was derived from the polynomial function:
\begin{equation}\label{extinction_coefficient}
\begin{split}
c = c_1 + c_2 \ast bp\_rp_0 + c_3 \ast bp\_rp_0^2 + c_4 \ast bp\_rp_0^3 \\+ c_5 \ast A_0
+ c6 \ast A_0^2 + c_7  \ast A_0^3 + c_8  \ast bp\_rp_0  \ast A_0 \\+ c_9  \ast A_0  \ast bp\_rp_0^2 + c_{10}  \ast bp\_rp_0  \ast A_0^2
\end{split}
\end{equation}
where $c_1$ to $c_{10}$ values( shown in Table~\ref{extinction coefficient}) were adopted from the public auxiliary data provided by ESA/Gaia/DPAC/CU5 and prepared by Carine Babusiaux.


\section{Results and Analysis} \label{sec:results}

\subsection{Identification of BSSs}
After performing isochrone fitting to the H-R diagram of the cluster members, we identified the MSTO and the starting point of the SBGB within the diagram (Figure~\ref{fig_profile_isochrone} (c)). We utilized the Zero-Age Main Sequence (ZAMS) as a reference for classifying BSS. Stars situated to the bluer side of the ZAMS are considered reliable candidates for BSS. Ultimately, we identified BSS by their positions in the H-R diagram. Specifically, BSS are found in the upper left region of the MSTO, appearing both brighter and bluer than the MSTO.
In accordance with the classification established by~\citet{Jadhav21}, we categorize the BSS into two distinct groups. The first group, denoted as \textit{P\_BSS}, includes stars that are situated within 0.25 magnitudes bluer than the MSTO and no more than 0.752 magnitudes fainter than SBGB. This definition aims to reduce contamination from the binary sequence. In contrast, \textit{R\_BSS} comprises stars that exceed these brightness and color thresholds, specifically remaining no more than 1 magnitude bluer than the MSTO and no more than 5 magnitudes brighter than it (Colored boxes in Figure~\ref{fig_profile_isochrone} (c)). Further classification was performed based on follow criteria:
\begin{itemize}
    \item Type I: Stars classified as \( \textit{R\_BSS} \) and core member.
    
    \item Type II: All other \( \textit{R\_BSS} \) or \( \textit{P\_BSS} \) stars not satisfying the Type I criteria.
\end{itemize}
This dual classification framework combines photometric and astrometric to improve the reliability of BSS identification. Additionally, considering the overestimation of BP-band photometry~\citep{Riello21} at the faint end of the CMD, particularly for G < 19 - 19.5~mag, the color [BP-RP] may shift toward the blue. This shift can lead to unreliable parameters. Therefore, we have excluded stars fainter than 19~mag from the BSS candidates (all are Type II) in eight OCs analyzed in this study.

%

Based on the list of 53 potential clusters provided in H23, we conducted further analysis and identified 447 BSSs, including 119 Type I BSSs among these clusters.  The relevant properties of the clusters are listed in Table~\ref{tab:all_clu}, and parameters of Type I and Type II stragglers are listed in Table ~\ref{tab:bss_pro}. The density profile and positions of various star types on the H-R diagram for each OC can be seen in Appendix~\ref{appdendixa}.
Figure~\ref{fig:age_gamg} illustrates that most clusters containing BSSs exhibit a logarithmic age distribution between 8.7~(0.5~Gyr) and 9.7~(50.0~Gyr). While the plot of cluster age against the number of BSSs suggests a positive correlation, a precise quantitative relationship remains elusive. Additionally, this study refines the limiting magnitude for identifying BSSs towards fainter stars.

\begin{deluxetable*}{cccccccccccccc}
\tablecaption{List of all clusters with newly identified BSSs. \label{tab:all_clu}}
\tabletypesize{\scriptsize}
\tablehead{
\colhead{\textbf{Cluster}} & \colhead{\textbf{GLON}} & \colhead{\textbf{GLAT}} & \colhead{\textbf{Parallax}} & \colhead{\textbf{pmRA}} & \colhead{\textbf{pmDE}} & \colhead{\textbf{log(Age/yr)}} & \colhead{\textbf{\boldmath{$A_0$}}} & \colhead{\textbf{OC\_Type}} & \colhead{\textbf{\boldmath{$r_c$}}} & \colhead{\textbf{\boldmath{$r_b$}}} & \colhead{\textbf{\boldmath{$r_t$} ($r_t'$)}} & \colhead{\textbf{N\_TypeI}} & \colhead{\textbf{N\_TypeII}} \\
\colhead{} & \colhead{[deg]} & \colhead{[deg]} & \colhead{[mas]} & \colhead{[mas yr$^{-1}$]} & \colhead{[mas yr$^{-1}$]} & \colhead{} & \colhead{[mag]} & \colhead{} & \colhead{[pc]} & \colhead{[pc]} & \colhead{[pc]} & \colhead{} & \colhead{}
}

\startdata
Berkeley 26 & 207.699 & 2.354 & 0.18 & 0.06 & 0.42 & 9.65 & 2.00 & A & \(1.04 \pm 0.36\) & \(5.78 \pm 2.49\) & \(9.82 \pm 0.10\) & 4 & 20 \\
        CWNU 2946 & 286.158 & 2.929 & 0.08 & -4.24 & 2.02 & 9.35 & 1.55 & A & \(6.84 \pm 0.87\) & \(9.03 \pm 0.27\) & \(21.36 \pm 0.33\) & 2 & 2 \\
        CWNU 2961 & 273.269 & -3.714 & 0.14 & -3.52 & 3.42 & 9.40 & 1.20 & A & \(1.52 \pm 0.06\) & \(10.24 \pm 0.14\) & \(20.20 \pm 0.16\) & 2 & 7 \\
        CWNU 2974 & 262.278 & -7.022 & 0.13 & -2.55 & 4.40 & 9.50 & 0.80 & A & \(1.49 \pm 0.16\) & \(6.24 \pm 0.19\) & \(11.99 \pm 0.56\) & 1 & 3 \\
        CWNU 2987 & 228.100 & 3.474 & 0.16 & -0.67 & 1.44 & 9.30 & 0.66 & A & \(0.84 \pm 0.05\) & \(6.87 \pm 0.29\) & \(14.62 \pm 0.09\) & 2 & 2 \\
        CWNU 3008 & 113.145 & 4.680 & 0.10 & -1.22 & -0.25 & 9.30 & 3.50 & A & \(7.58 \pm 0.72\) & \(9.53 \pm 0.21\) & \(22.05 \pm 0.44\) & 2 & 8 \\
        CWNU 3009 & 124.156 & -4.891 & 0.14 & -1.23 & -0.05 & 9.30 & 1.30 & A & \(0.81 \pm 1.13\) & \(3.29 \pm 0.25\) & \(13.57 \pm 1.06\) & 1 & 1 \\
        CWNU 3048 & 157.106 & 4.428 & 0.23 & 0.44 & -1.64 & 9.35 & 2.94 & A & \(0.54 \pm 0.57\) & \(6.97 \pm 0.48\) & \(11.51 \pm 1.55\) & 1 & 4 \\
        CWNU 3064 & 149.648 & 5.699 & 0.21 & 0.05 & -0.02 & 9.45 & 1.32 & A & \(0.69 \pm 0.16\) & \(4.44 \pm 0.30\) & \(15.00 \pm 2.23\) & 5 & 16 \\
        CWNU 3067 & 147.947 & 10.025 & 0.23 & 0.56 & -0.11 & 9.25 & 1.76 & A & \(3.57 \pm 0.02\) & \(9.94 \pm 4.53\) & \(21.46 \pm 0.21\) & 3 & 1 \\
        CWNU 3085 & 292.325 & 0.517 & 0.11 & -6.50 & 1.92 & 8.95 & 2.85 & A & \(1.03 \pm 0.16\) & \(5.90 \pm 0.42\) & \(20.81 \pm 1.56\) & 1 & 7 \\
        CWNU 3096 & 214.251 & -2.026 & 0.22 & -0.31 & -0.31 & 9.15 & 2.47 & A & \(1.33 \pm 0.24\) & \(7.98 \pm 0.26\) & \(16.78 \pm 0.50\) & 2 & 0 \\
        CWNU 3102 & 114.051 & -9.984 & 0.20 & -2.17 & -0.86 & 9.45 & 0.75 & A & \(0.98 \pm 0.69\) & \(3.89 \pm 0.21\) & \(13.55 \pm 1.95\) & 2 & 2 \\
        CWNU 3109 & 225.277 & -4.989 & 0.17 & -0.09 & 1.11 & 9.55 & 1.40 & A & \(1.83 \pm 0.42\) & \(6.61 \pm 0.12\) & \(12.44 \pm 0.53\) & 1 & 0 \\
        CWNU 3114 & 66.921 & 0.393 & 0.15 & -3.38 & -5.98 & 8.75 & 2.80 & A & \(0.85 \pm 0.06\) & \(8.24 \pm 0.94\) & \(17.69 \pm 0.40\) & 1 & 8 \\
        CWNU 3129 & 255.945 & 2.017 & 0.11 & -2.06 & 1.84 & 9.25 & 2.10 & A & \(2.07 \pm 0.10\) & \(12.95 \pm 2.87\) & \(16.49 \pm 0.46\) & 2 & 4 \\
        CWNU 3142 & 117.775 & 0.928 & 0.16 & -0.74 & -0.78 & 9.45 & 2.32 & A & \(1.88 \pm 0.19\) & \(4.93 \pm 0.08\) & \(11.13 \pm 0.12\) & 2 & 6 \\
        CWNU 3153 & 177.157 & 5.046 & 0.19 & 0.31 & -0.76 & 9.20 & 1.89 & A & \(2.15 \pm 0.84\) & \(5.60 \pm 0.25\) & \(18.23 \pm 2.25\) & 4 & 2 \\
        CWNU 3154 & 20.255 & -1.398 & 0.18 & -0.20 & -1.74 & 9.15 & 4.10 & A & \(1.39 \pm 0.02\) & \(10.33 \pm 0.25\) & \(20.12 \pm 0.10\) & 1 & 2 \\
        CWNU 3160 & 62.781 & 6.010 & 0.05 & -2.59 & -4.85 & 9.30 & 1.10 & A & \(0.52 \pm 0.12\) & \(10.47 \pm 4.13\) & \(15.37 \pm 0.18\) & 3 & 3 \\
        CWNU 3184 & 74.334 & 2.945 & 0.06 & -2.46 & -4.12 & 8.65 & 3.68 & A & \(4.12 \pm 1.16\) & \(12.25 \pm 2.68\) & \(22.49 \pm 0.90\) & 0 & 1 \\
        CWNU 3190 & 87.309 & 5.739 & 0.20 & -3.44 & -4.98 & 9.25 & 3.95 & A & \(2.43 \pm 0.02\) & \(8.02 \pm 0.21\) & \(14.82 \pm 0.07\) & 2 & 2 \\
        CWNU 3200 & 288.257 & 0.087 & 0.10 & -4.66 & 2.02 & 8.80 & 2.65 & A & \(7.22 \pm 0.22\) & \(38.13 \pm 0.19\) & \(45.79 \pm 0.38\) & 2 & 9 \\
        CWNU 3204 & 208.239 & 2.814 & 0.17 & -0.01 & -0.81 & 9.50 & 0.87 & A & \(1.73 \pm 0.13\) & \(5.68 \pm 0.43\) & \(16.26 \pm 0.39\) & 3 & 5 \\
        CWNU 3252 & 206.987 & 4.580 & 0.16 & 0.42 & -0.24 & 9.50 & 0.48 & A & \(1.13 \pm 0.06\) & \(3.71 \pm 0.10\) & \(12.25 \pm 0.27\) & 3 & 6 \\
        CWNU 3267 & 306.493 & -1.024 & 0.14 & -6.34 & -1.02 & 8.90 & 2.71 & A & \(1.16 \pm 0.28\) & \(4.16 \pm 0.01\) & \(17.29 \pm 2.62\) & 3 & 17 \\
        CWNU 3282 & 236.558 & -4.124 & 0.12 & -0.36 & 1.19 & 9.45 & 1.90 & A & \(1.97 \pm 0.13\) & \(10.50 \pm 0.07\) & \(18.91 \pm 1.15\) & 3 & 6 \\
        CWNU 3284 & 117.024 & 6.109 & 0.22 & -2.51 & -0.07 & 9.70 & 2.90 & A & \(1.49 \pm 0.35\) & \(7.06 \pm 0.35\) & \(21.14 \pm 42.55\) & 2 & 13 \\
        CWNU 3296 & 278.536 & -5.160 & 0.12 & -2.60 & 2.96 & 9.25 & 1.50 & A & \(0.73 \pm 0.42\) & \(5.67 \pm 0.06\) & \(15.54 \pm 0.52\) & 1 & 1 \\
        CWNU 3455 & 126.646 & 9.276 & 0.15 & -0.63 & 0.16 & 9.45 & 1.50 & A & \(0.84 \pm 0.41\) & \(7.77 \pm 1.54\) & \(16.26 \pm 0.27\) & 4 & 1 \\
        CWNU 3510 & 279.598 & 0.374 & 0.10 & -4.50 & 2.58 & 9.35 & 2.65 & A & \(1.55 \pm 1.07\) & \(6.19 \pm 0.53\) & \(11.86 \pm 0.26\) & 3 & 6 \\
        CWNU 3523 & 275.363 & -5.741 & 0.05 & -2.72 & 3.00 & 9.15 & 0.81 & A & \(2.81 \pm 0.68\) & \(7.94 \pm 0.32\) & \(21.83 \pm 1.44\) & 2 & 7 \\
        CWNU 3555 & 108.834 & 5.314 & 0.11 & -1.88 & -0.93 & 9.10 & 2.90 & A & \(1.01 \pm 0.58\) & \(7.57 \pm 0.26\) & \(18.61 \pm 0.09\) & 1 & 5 \\
        CWNU 3556 & 254.641 & -4.966 & 0.08 & -1.10 & 1.53 & 9.15 & 2.25 & A & \(2.60 \pm 0.09\) & \(11.56 \pm 0.15\) & \(19.82 \pm 0.03\) & 2 & 3 \\
        CWNU 3629 & 84.265 & 4.242 & 0.16 & -3.08 & -4.22 & 9.20 & 3.55 & A & \(1.68 \pm 0.24\) & \(6.79 \pm 0.21\) & \(20.06 \pm 0.87\) & 1 & 7 \\
        CWNU 3633 & 119.170 & -4.312 & 0.15 & -2.78 & -0.38 & 9.45 & 1.60 & A & \(1.54 \pm 0.78\) & \(8.89 \pm 0.54\) & \(20.75 \pm 0.30\) & 2 & 8 \\
        CWNU 3825 & 56.158 & -1.787 & 0.07 & -3.07 & -5.71 & 9.00 & 3.95 & A & \(4.79 \pm 0.17\) & \(10.51 \pm 2.40\) & \(23.91 \pm 0.10\) & 2 & 2 \\
        CWNU 3863 & 84.565 & 6.479 & 0.13 & -3.95 & -4.67 & 9.25 & 2.80 & A & \(1.45 \pm 0.50\) & \(5.18 \pm 0.21\) & \(11.48 \pm 4.39\) & 2 & 4 \\
        CWNU 4001 & 297.927 & -3.598 & 0.10 & -5.51 & 0.56 & 9.50 & 2.14 & A & \(1.15 \pm 0.41\) & \(6.14 \pm 0.27\) & \(17.06 \pm 1.08\) & 4 & 6 \\
        CWNU 4203 & 36.610 & -1.363 & 0.25 & -1.81 & -4.09 & 9.25 & 5.67 & A & \(1.05 \pm 0.08\) & \(3.96 \pm 5.39\) & \(20.95 \pm 0.59\) & 2 & 10 \\
        ESO 429 05 & 246.506 & -5.403 & 0.09 & -1.51 & 1.67 & 9.40 & 1.60 & A & \(1.96 \pm 0.25\) & \(11.01 \pm 0.65\) & \(20.91 \pm 0.66\) & 2 & 9 \\
        FSR 0291 & 90.508 & -0.507 & 0.14 & -3.22 & -4.24 & 8.80 & 4.80 & A & \(1.48 \pm 0.07\) & \(8.28 \pm 0.17\) & \(21.27 \pm 0.18\) & 2 & 20 \\
        FSR 0508 & 121.922 & 2.111 & 0.16 & -2.32 & -0.64 & 9.25 & 3.38 & A & \(2.12 \pm 0.64\) & \(9.11 \pm 0.45\) & \(19.78 \pm 25.89\) & 1 & 15 \\
        FSR 0687 & 156.921 & 0.968 & 0.19 & 0.67 & 0.27 & 9.05 & 3.95 & A & \(3.67 \pm 0.47\) & \(10.16 \pm 0.99\) & \(22.20 \pm 2.74\) & 1 & 4 \\
        Pfleiderer 4 & 115.964 & 0.269 & 0.12 & -1.30& -0.50 & 9.45 & 2.68 & A & \(1.35 \pm 0.15\) & \(12.52 \pm 1.76\) & \(21.53 \pm 1.53\) & 6 & 15 \\
        Saurer 1 & 214.686 & 7.386 & 0.07 & -0.26 & -0.33 & 9.80 & 0.43 & A & \(2.03 \pm 0.18\) & \(8.09 \pm 0.88\) & \(13.67 \pm 0.64\) & 3 & 1 \\
        Saurer 4 & 298.787 & -1.021 & 0.08 & -6.89 & 0.80 & 8.80 & 4.10 & A & \(1.64 \pm 0.23\) & \(7.38 \pm 0.95\) & \(35.26 \pm 1.56\) & 1 & 5 \\
        Teutsch 48 & 274.191 & -2.176 & 0.14 & -4.14 & 3.74 & 9.10 & 2.81 & A & \(0.86 \pm 0.16\) & \(9.57 \pm 4.62\) & \(16.24 \pm 3.58\) & 4 & 8 \\
        CWNU 3593 & 293.753 & -6.803 & 0.10 & -5.20 & 1.98 & 9.80 & 0.40 & B & ~ & ~ & ~ & 2 & 2 \\
        CWNU 3911 & 303.050 & 3.939 & 0.15 & -6.26 & -0.31 & 9.55 & 1.42 & B & ~ & ~ & ~ & 2 & 3 \\
        CWNU 3989 & 278.297 & -3.120 & 0.13 & -3.92 & 3.17 & 9.35 & 2.04 & B & ~ & ~ & ~ & 1 & 15 \\
        CWNU 4150 & 75.136 & 3.931 & 0.13 & -3.15 & -5.30 & 9.35 & 1.40 & B & ~ & ~ & ~ & 1 & 3 \\
        CWNU 4172 & 166.457 & -6.350 & 0.14 & 0.45 & -0.56 & 9.35 & 1.74 & B & ~ & ~ & ~ & 7 & 11 \\
          \hline \hline
\enddata

\end{deluxetable*}

\
\begin{deluxetable*}{cccccccccccc}
\tablecaption{Example of BSSs in an OC. \label{tab:bss_pro}}
\tablehead{
\colhead{\textbf{Cluster}} & \colhead{\textbf{Gaia DR3 ID}} & \colhead{\textbf{GLON}} & \colhead{\textbf{GLAT}} & \colhead{\textbf{Parallax}} & \colhead{\textbf{pmRA}} & \colhead{\textbf{pmDE}} & \colhead{\textbf{Gmag}} & \colhead{\textbf{BP-RP}} &\colhead{\textbf{RUWE}}& \colhead{\textbf{\(M_e\)}} & \colhead{\textbf{Type}} \\
\colhead{} & \colhead{} & \colhead{[deg]} & \colhead{[deg]} & \colhead{[mas]} & \colhead{[mas yr$^{-1}$]} & \colhead{[mas yr$^{-1}$]} & \colhead{[mag]} & \colhead{[mag]} & \colhead{} & \colhead{}
}
\startdata
\multirow{5}{*}{CWNU 3455} & 534151377367490816 & 126.598 & 9.255 & 0.16 & -0.63 & 0.15 & 16.69 & 0.79 &1.047 & 0.71 & I \\
                            & 534522462542061184 & 126.725 & 9.239 & 0.19 & -0.73 & 0.16 & 16.18 & 0.71 &1.044& 1.10 & I \\
                            & 534525898515891072 & 126.671 & 9.302 & 0.14 & -0.62 & 0.17 & 15.77 & 0.87 &0.998 &1.49 & I \\
                            & 534526551351938432 & 126.645 & 9.282 & 0.12 & -0.58 & 0.10 & 16.36 & 1.06 & 0.967&0.95 & I \\
                            \cline{2-12}
                            & 534522462539369344 & 126.729 & 9.237 & 0.21 & -0.68 & 0.62 & 17.68 & 0.94 & 1.026&0.18 & II \\
                            \hline \hline
\enddata
\end{deluxetable*}


\begin{figure*}
    \centering
    \includegraphics[width=0.985\linewidth]{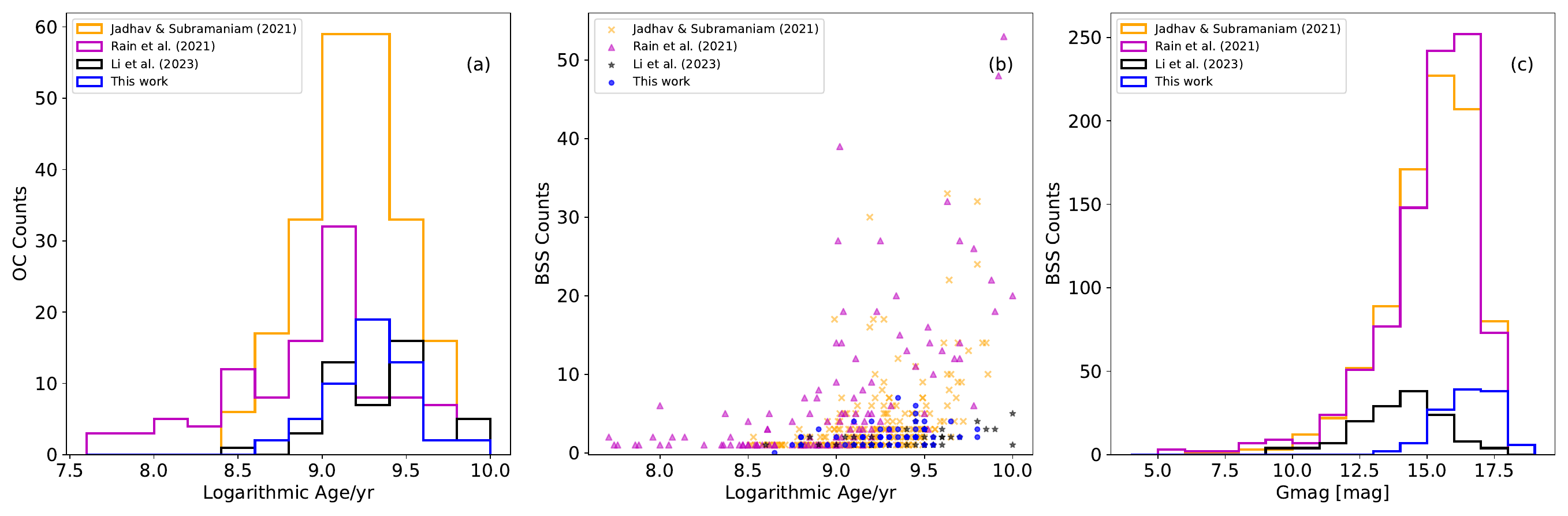}
\caption{Statistics of BSSs in OCs from various studies. Panel (a) shows the age distribution of OCs containing BSSs. Panel (b) illustrates the relationship between cluster age and the number of BSSs. Panel (c) displays the apparent magnitudes of BSSs across different studies. The colors orange, magenta, black, and blue correspond to categories from ~\cite{Jadhav21}, ~\cite{Rain21}, ~\cite{Licy23}, and Type~I BSSs derived in this work, respectively.
}
    \label{fig:age_gamg}
    
\end{figure*}

\subsection{Maximum Fractional Mass Excess}
\begin{figure*} 
    \centering
    \includegraphics[width=0.985\linewidth]{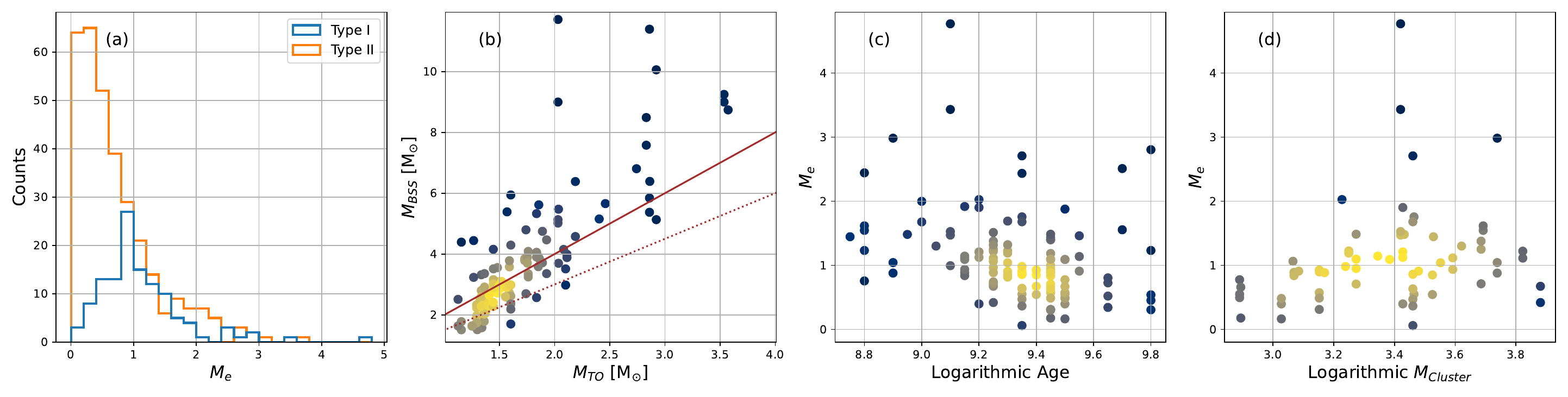}
\caption{(a) Distribution of Type~I and Type~II BSSs relative to $M_{e}$. (b) Mass of Type~I BSSs ($M_{BSS}$) versus the TO mass of the cluster ($M_{TO}$). The brown solid and dotted lines signify the fractional mass excess $M_{e} = 1$ and $M_{e} = 0.5$, respectively. (c) $M_{e}$ of Type~I BSSs plotted against cluster age. (d) $M_{e}$ of Type~I BSSs plotted against cluster mass (logarithmic mass/M$_\odot$). The color of the scatter points reflects the degree of spatial density in the distribution. Warmer colors signify a greater degree of crowding among the data points.}
    \label{fig_me_plot1}
\end{figure*}

Previously, ~\cite{Jadhav21} introduced a new parameter known as "fractional mass excess" ($M_{e}$), which normalizes the mass of BSS with the mass at the MSTO of a cluster. This approach provides a valuable framework for investigating the formation mechanisms of BSSs. They classified BSSs into three categories based on their $M_{e}$ values and proposed corresponding formation mechanisms: BSSs with low $M_{e}$ values are likely formed through mass transfer, while those with high $M_{e}$ values may result from binary mergers. Additionally, BSSs with extreme $M_{e}$ values could originate from multiple mergers or mass transfer events. The formation of BSSs is characterized by a process of mass accretion during the main-sequence phase. Building upon the work of ~\cite{Jadhav21}, we utilize the fractional mass excess $M_{e}$ to explore the evolutionary pathways of BSSs:
\begin{equation}\label{ME_equa}
M_{e} =\frac{M_{BSS} - M _{TO}}{M _{TO}} 
\end{equation}
where $M_{BSS}$ and $M_{TO}$ represent the masses of the BSSs and the turn-off star, respectively. The values for these masses are obtained by assuming the target star is a single star and identifying the corresponding mass from the ZAMS by matching its observed magnitude.
Using Equation~\ref{ME_equa}, we calculated $M_{e}$ for all BSSs.  For Type~I stars, there are ~55 stars with $M_{e}$ > 1, ~47 stars with 0.5 < $M_{e}$ < 1, and 17 stars with $M_{e}$ < 0.5.  As illustrate in Figure~\ref{fig_me_plot1}~(a)-(c), Most clusters exhibit $M_{e}$ values less than 1, consistent with the findings of ~\citet{Jadhav21}. Notably, we observe that Type~II BSSs are more prevalent in OCs with smaller $M_{e}$ values (Figure~\ref{fig_me_plot1} (a)). However, the presence of field star contamination may contribute to the increased number of Type~II BSSs observed.  Consequently, our analysis will focus exclusively on Type~I BSSs to better understand their characteristics.
Figure \ref{fig_me_plot1}~(d) illustrates the relationship between cluster mass~\footnote{We adopted a methodology similar to that of ~\citet{Jadhav21} for calculating the masses of OCs. This approach involved comparing the apparent luminosity function (LF) of the clusters with theoretical models, which enabled us to derive their masses (for details, see Section A3 in ~\citet{Jadhav21}). The LF range we considered spans from G = 19~mag to the MSTO, with a minimum cutoff of 1~mag. It is noteworthy that many of the clusters in our study are distant and older. This presented challenges in determining the masses of those clusters where the MSTO is either very close to or exceeds 19 mag.} and the fractional mass excess. It appears that $M_{e}$ generally increases with cluster mass, the conclusion is similar to that of ~\citet{Jadhav21}. 

Subsequently, we then focus on the maximum fractional mass excess of BSSs, which indicates the mass limitation for BSS formation within a star cluster. It is essential to recognize that the currently observed BSSs in an individual cluster may not fully represent its potential for BSS formation given the existing conditions. However, statistical analysis of a large sample can provide valuable qualitative insights.
Given the limited number of OCs available in our study, we employed the uniform search results for BSSs from the all-sky census conducted by ~\citet{Jadhav21}. Their work provided statistical data on the fractional mass excess of all BSSs. 
They noted that the frequency of extreme-$M_{e}$ events ($M_{e} > 1$) increases consistently with cluster mass.
Furthermore, their findings revealed that the fraction of extreme-$M_{e}$ BSSs decreases as the cluster exceeds 1~Gyr in age, with the peak fraction occurring at approximately 1~Gyr.

\begin{figure*} 
    \centering
    \includegraphics[width=0.618\linewidth]{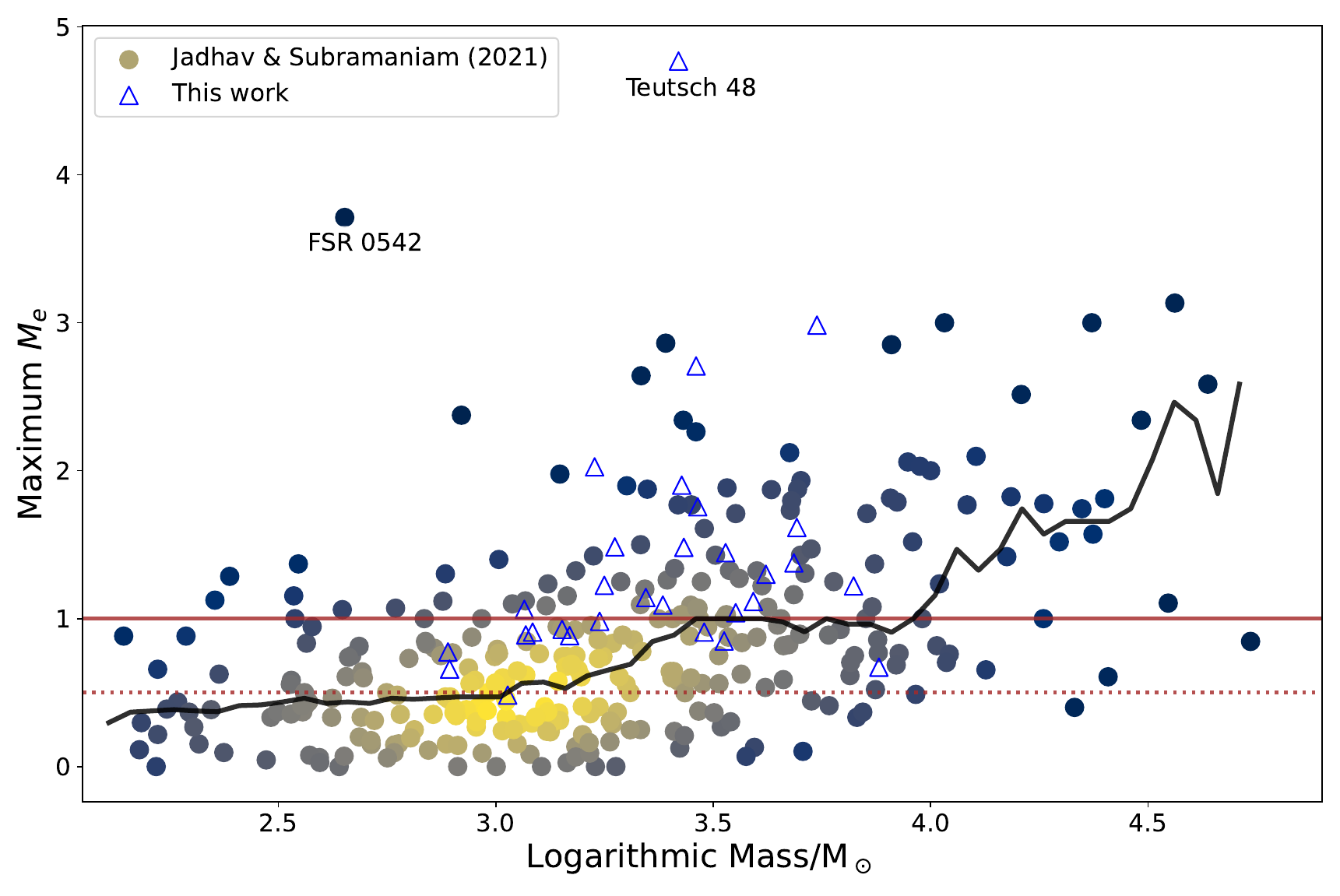}
\caption{The maximum $M_{e}$ of BSSs in all-aged OCs as a function of cluster mass. The black line represents the running median values of maximum $M_{e}$ relative to cluster mass. The triangles denote OCs containing Type~I BSSs identified in this study, while solid circles represent BSSs from ~\citet{Jadhav21}. The brown solid and dotted lines indicate $M_{e} = 1$ and $M_{e} = 0.5$, respectively. The colors indicate scatter point density, with cooler colors representing sparse areas and warmer colors indicating crowded areas. Two significant outliers, FSR~0542 and Teutsch~48, are also marked in the figure.}
    \label{fig_me_plot2}
\end{figure*}

However, our main objective is to investigate the distribution characteristics of the maximum mass of BSSs in OCs that harbor them. Figure~\ref{fig_me_plot2} illustrates that the samples we examined show significant $M_{e}$ values across different cluster masses. Furthermore, the maximum $M_{e}$ generally increases with the mass of the clusters. The overall sample of OCs hosting BSSs also reveals a clear positive correlation between the maximum $M_{e}$ and the cluster mass.
We also observed a small number of outliers significantly deviating from the median $M_{e}$ values, likely due to field star contamination in older, distant, and heavily extinguished star clusters (Figure~\ref{fig_delta_me}). If relatively bright field stars are misidentified as member stars while the turn-off point is faint, it can lead to inflated $M_{e}$ values. 
Consequently, these BSSs with extremely high $M_{e}$ require more precise astrometric parameters to ascertain whether they are true member stars. If they are confirmed as true members, their exceptionally high $M_{e}$ would render them intriguing subjects for research, potentially linked to the process of extensive stellar mergers and mass transfers.

\begin{figure*} 
    \centering
    \includegraphics[width=0.985\linewidth]{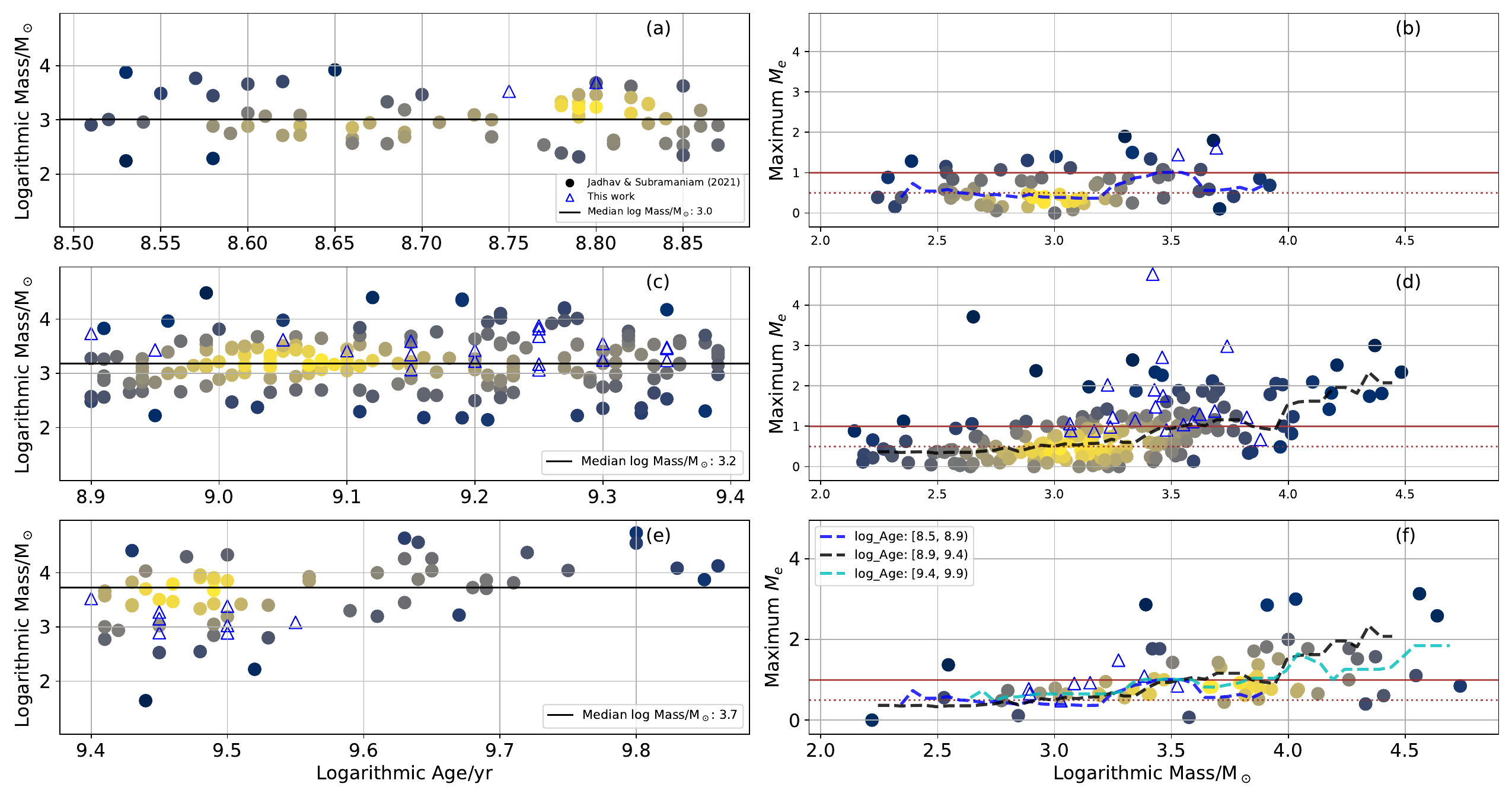}
\caption{Left panels: The mass of OCs plotted against various age ranges. The black line represents the median logarithmic mass /M$_\odot$ within each age range. Right panels: The maximum fractional mass excess of OC BSSs in different age ranges presented as a function of cluster mass. The dashed lines illustrate the running median values of maximum $M_{e}$ in relation to cluster mass. The triangles indicate OCs containing Type~1 BSSs identified in this study, while solid circles denote BSSs cataloged in ~\citet{Jadhav21}. The brown solid and dotted lines correspond to $M_{e} = 1$ and $M_{e} = 0.5$, respectively. The colors indicate scatter point density, with cooler colors representing sparse areas and warmer colors indicating crowded areas.}
    \label{fig_me_plot3}
\end{figure*}

\begin{figure*} 
    \centering
    \includegraphics[width=0.985\linewidth]{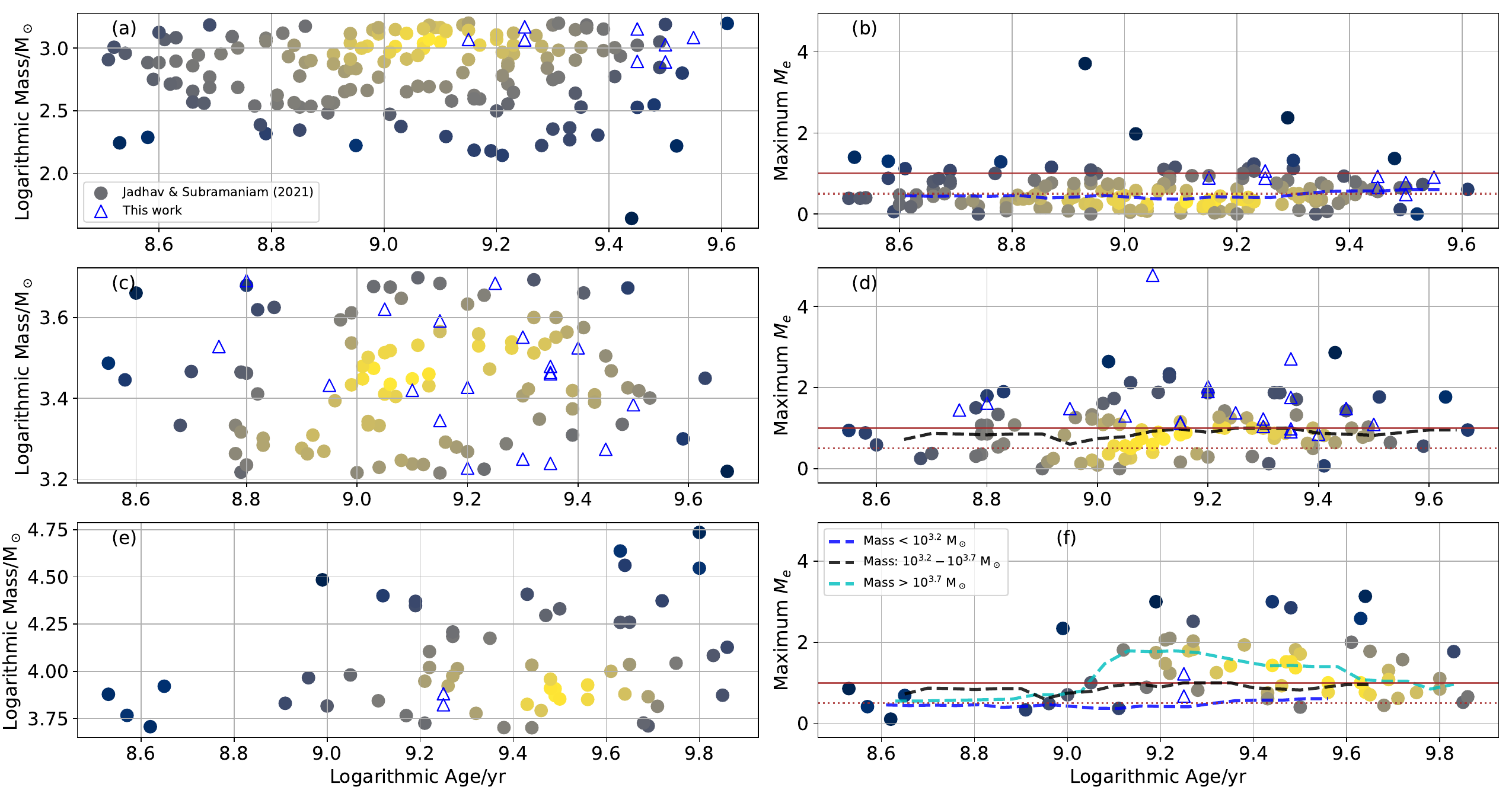}
\caption{Left panels: The OCs plotted within various mass ranges. Right panels: The maximum fractional mass excess of OC BSSs in different mass ranges presented as a function of cluster age. The dashed lines illustrate the running median values of maximum $M_{e}$ in relation to cluster age. The triangles indicate OCs containing Type~1 BSSs identified in this study, while solid circles denote BSSs cataloged in ~\citet{Jadhav21}. The brown solid and dotted lines correspond to $M_{e} = 1$ and $M_{e} = 0.5$, respectively. The colors indicate scatter point density, with cooler colors representing sparse areas and warmer colors indicating crowded areas.}
    \label{fig_me_plot4}
\end{figure*}

Meanwhile, We observe an increase in the mass of OCs containing BSSs with age (Figure~\ref{fig_me_plot3}, left panels). This observation aligns with expectations, given that OCs frequently traverse the plane of the Milky Way and its spiral arms. Such movements expose them to external gravitational influences, which can lead to the evaporation of member stars. As a result, it is primarily the more massive OCs that tend to survive these processes over time and remain detectable.
To address potential age selection bias, we categorize the clusters into three relative age intervals: the lower interval (logarithmic age: 8.5~(0.3~Gyr) - 8.9~(0.8~Gyr)); the central interval (logarithmic age: 8.9~(0.8~Gyr) - 9.4~(2.5~Gyr)); and the oldest interval (logarithmic age: 9.4~(2.5~Gyr) - 9.9~(7.9~Gyr)). Figure~\ref{fig_me_plot3}(b) illustrates that in relatively young OCs, the maximum $M_{e}$ is not significantly affected by variations in cluster mass. This lack of sensitivity might be due to a scarcity of young, massive star clusters that contain BSSs. This also could suggests that the formation of BSSs in younger clusters may have less dependence on mass.
Although it is noteworthy that within the younger cluster interval, there is a distinct increase in $M_{e}$ around 10$^{3.2}$~M$_\odot$. However, this observation should be interpreted with caution, as the limited sample size in this region necessitates further validation. In contrast, in the other two intervals, a clear trend is evident: maximum $M_{e}$ increases with cluster mass. When comparing the $M_{e}$ trends across clusters of different age ranges, a consistent pattern becomes apparent. Conversely, we find no discernible correlation between maximum $M_{e}$ of BSSs and the ages of their corresponding OCs (Figure~\ref{fig_me_plot4}). This finding indicates that the formation of BSS with highest $M_{e}$ is primarily driven by the mass of the clusters.

\section{Summary} \label{sec:summary}
In this study, we have expanded the catalog of BSSs within OCs located in the Galactic outer disk. By utilizing photometric data from Gaia DR3, we fitted isochrones to optical CMDs to accurately select BSS candidates based on their distinct positions in these diagrams. Our sample includes 53 newly identified clusters, alongside recently discovered OCs, resulting in the cataloging of 119 Type~I and 328 Type~II BSS candidates.
As part of our analysis, we derived the radial density profiles for all studied clusters and applied the King model. Through this modeling, we identified 48 OCs as highly concentrated clusters. However, some clusters displayed significant deviations from the model, which may be attributed to non-spherical mass distributions or tidal disturbances affecting their structures.

In accordance with previous research, we identified correlations between the frequency of new BSSs and established findings~\citep{Jadhav21,Rain21,Licy23}. The new BSSs we identified in this study exhibit fainter luminosities. However, they are consistent with other properties, such as the age and mass of the associated OCs. Additionally, inspired by ~\citet{Jadhav21}, we investigated the maximum fractional mass excess in the BSS samples. Notably, we found that the ability of a cluster to produce BSSs with relatively high mass primarily depends on its overall cluster mass.

We anticipate that this study will provide valuable samples for future investigations into BSSs within OCs. Additionally, the continuous release of updated Gaia data will offer more accurate astrometric parameters, assisting in the identification of potential BSS candidates. 
Furthermore, more precise astrometric measurements that can also help exclude some field stars among the Type~II BSS candidates.
And the discovery of more OCs will also enrich the available BSS samples. 

\section{Acknowledgements}
We sincerely thank the referee for the insightful suggestions. This work was supported by National Natural Science Foundation of China through grants 12303024, the Natural Science Foundation of Sichuan Province (2024NSFSC0453), and the "Young Data Scientists" project of the National Astronomical Data Center (NADC2023YDS-07); Y. Luo is supported by the NSFC under grant 12173028, the CSST project: CMS-CSST-2021-A10; K. Wang is supported by the NSFC 12373035.
This work has made use of data from the European Space Agency (ESA) mission GAIA (\url{https://www.cosmos.esa.int/gaia}), processed by the GAIA Data Processing and Analysis Consortium (DPAC,\url{https://www.cosmos.esa.int/web/gaia/dpac/consortium}). Funding for the DPAC has been provided by national institutions, in particular the institutions participating in the GAIA Multilateral Agreement. This work has made use of: \texttt{TOPCAT} \citep{Taylor2005_TOPCAT}, \texttt{Astropy} \citep{AstropyCollaboration2018}, \texttt{Matplotlib} \citep{Hunter2007_Matplotlib}, \texttt{numpy} \citep{Harris2020} and \texttt{pandas}\citep{mckinney2011pandas}.


\appendix
\setcounter{figure}{0}
\renewcommand{\thefigure}{A\arabic{figure}}

\setcounter{table}{0}
\renewcommand{\thetable}{A\arabic{table}}


\section{Figure set of studied clusters}\label{appdendixa}
Similar to Figures~\ref{fig_profile_isochrone}, Figures~\ref{figa1} to \ref{figa3} illustrate the member distribution of 51 OCs in Galactic coordinates. These figures also present the King model used to fit the density profiles of the clusters, as well as CMDs that indicate the positions of different types of stars.

\begin{figure*}
\begin{center}
\includegraphics[width=0.235\linewidth]{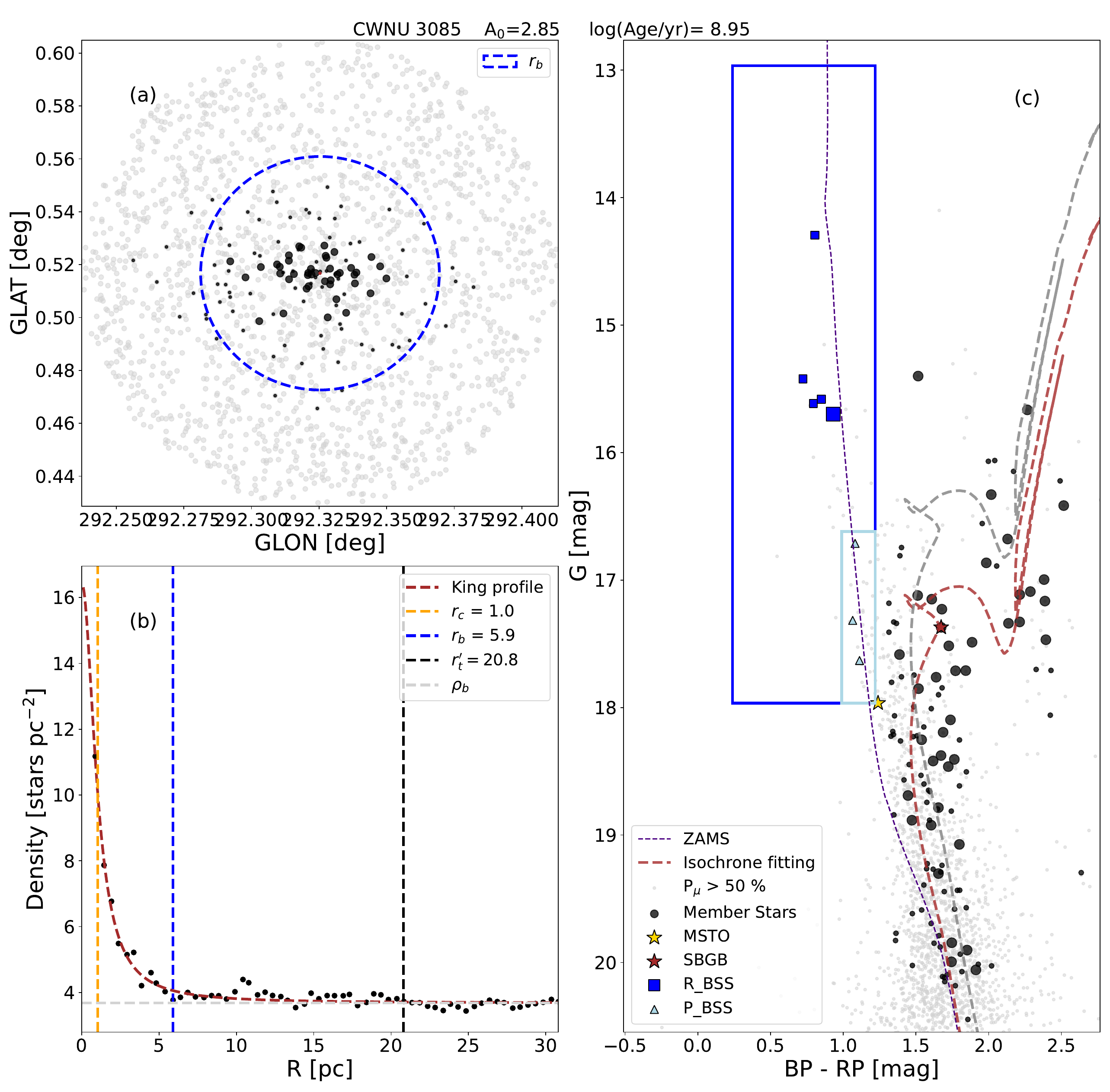}
\includegraphics[width=0.235\linewidth]{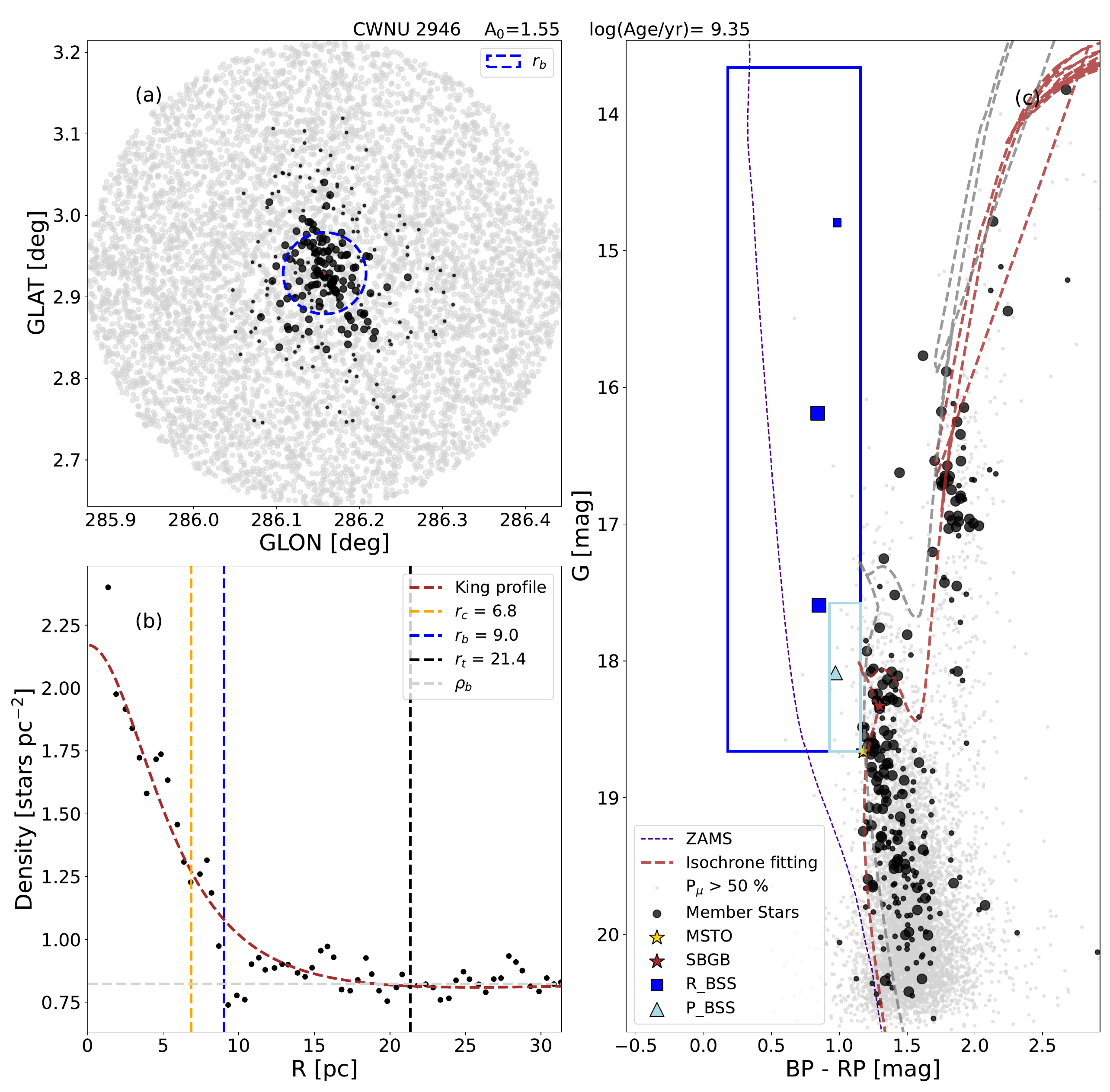}
\includegraphics[width=0.235\linewidth]{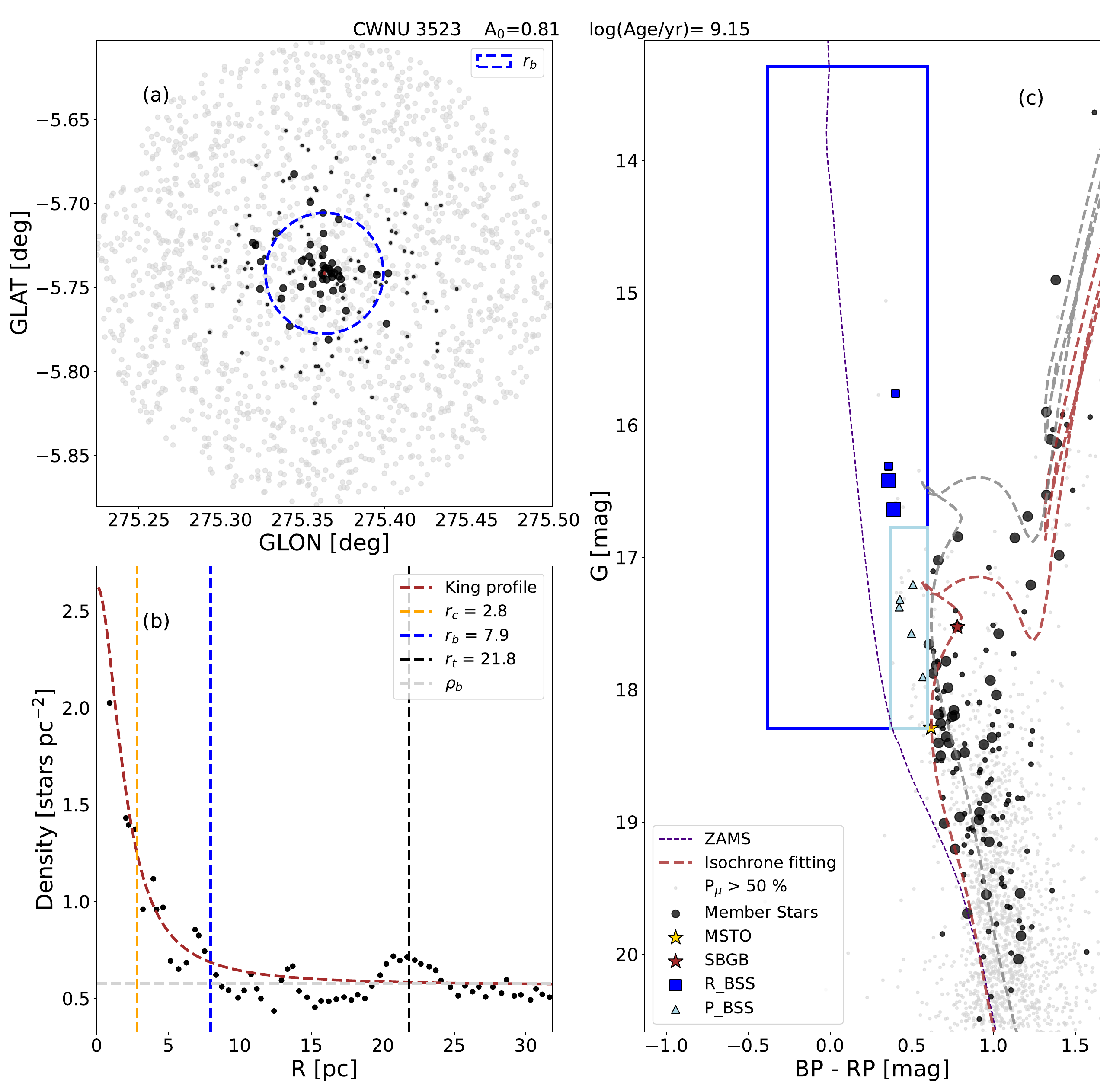}
\includegraphics[width=0.235\linewidth]{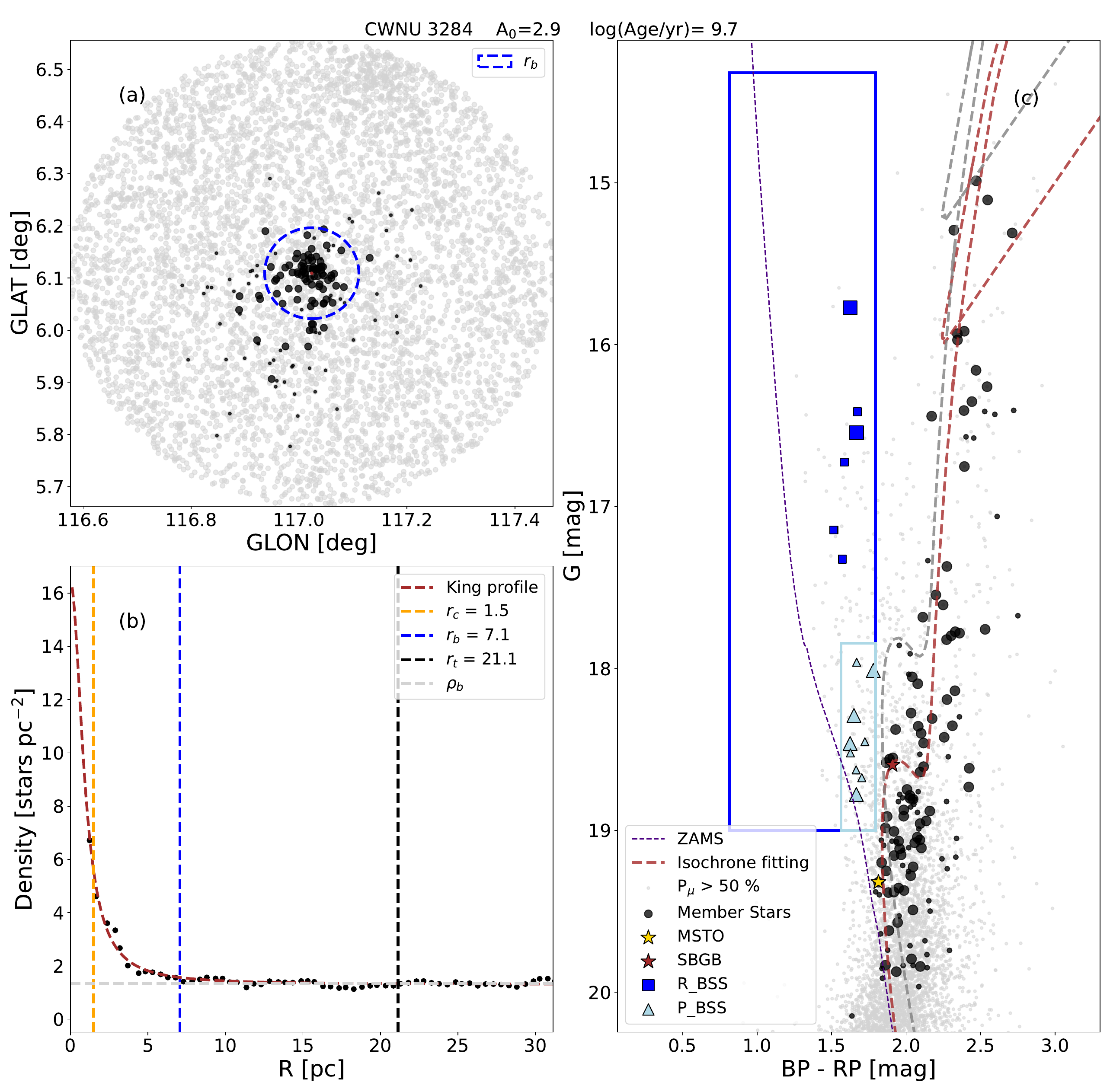}
\includegraphics[width=0.235\linewidth]{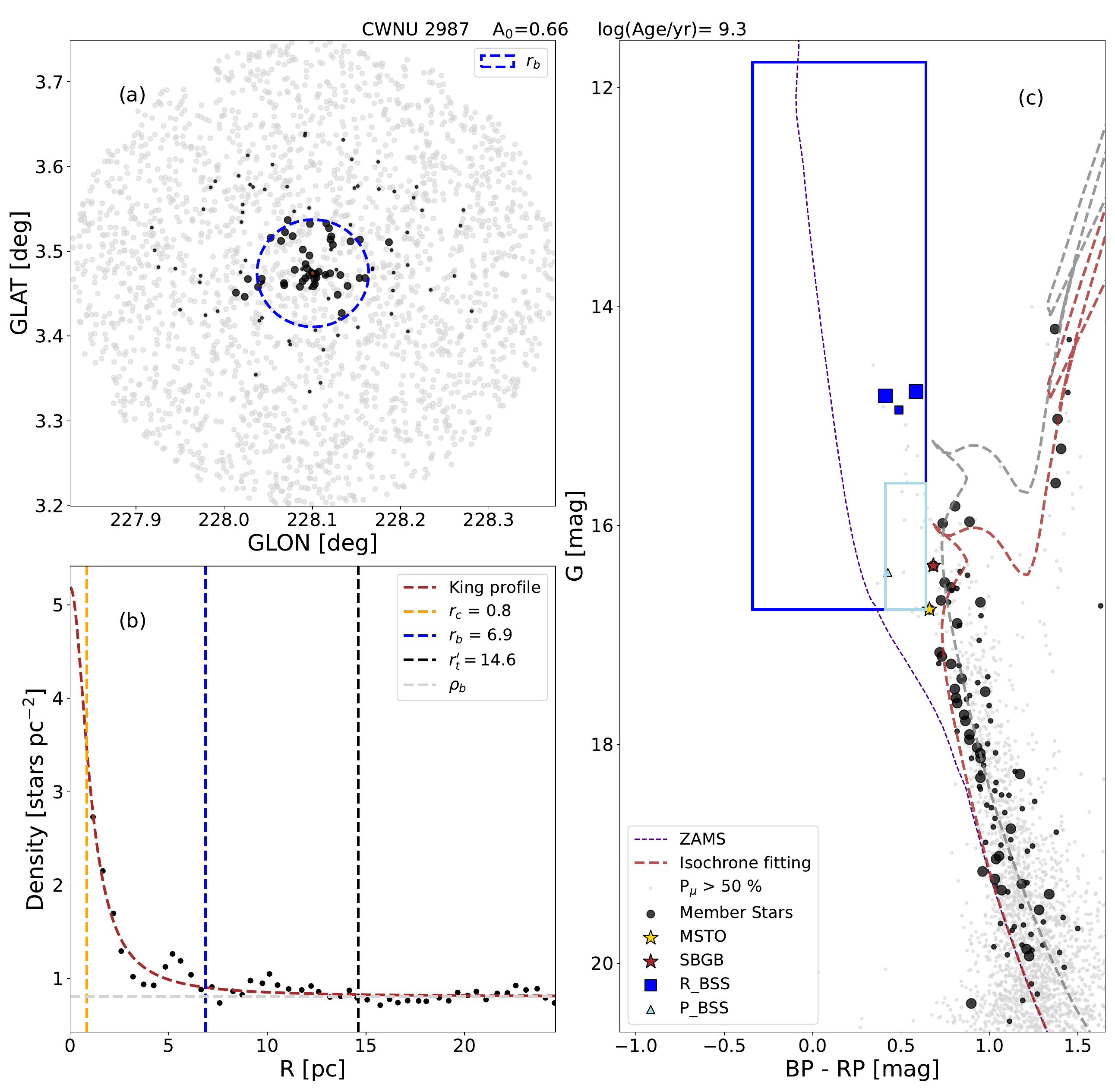}
\includegraphics[width=0.235\linewidth]{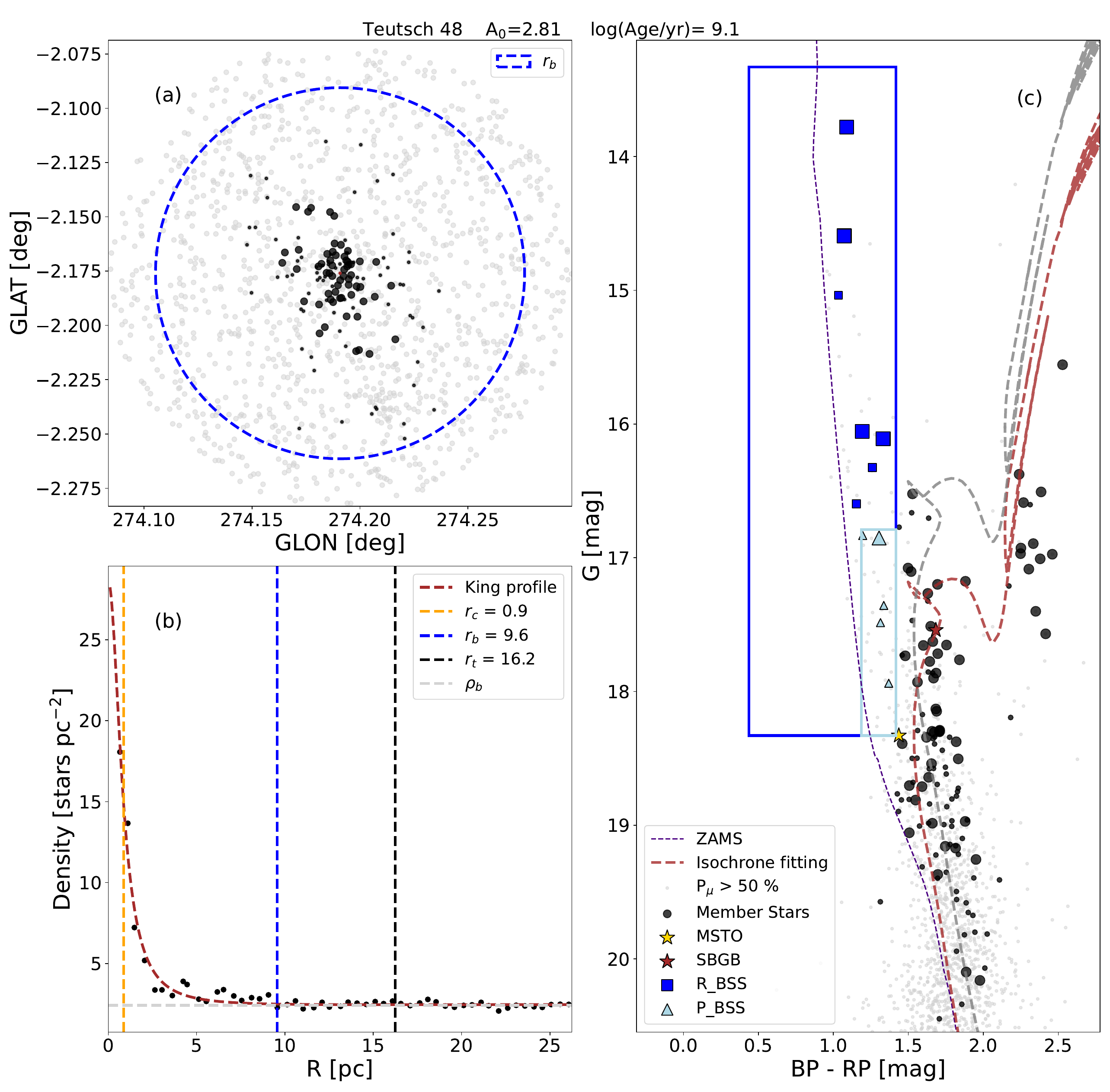}
\includegraphics[width=0.235\linewidth]{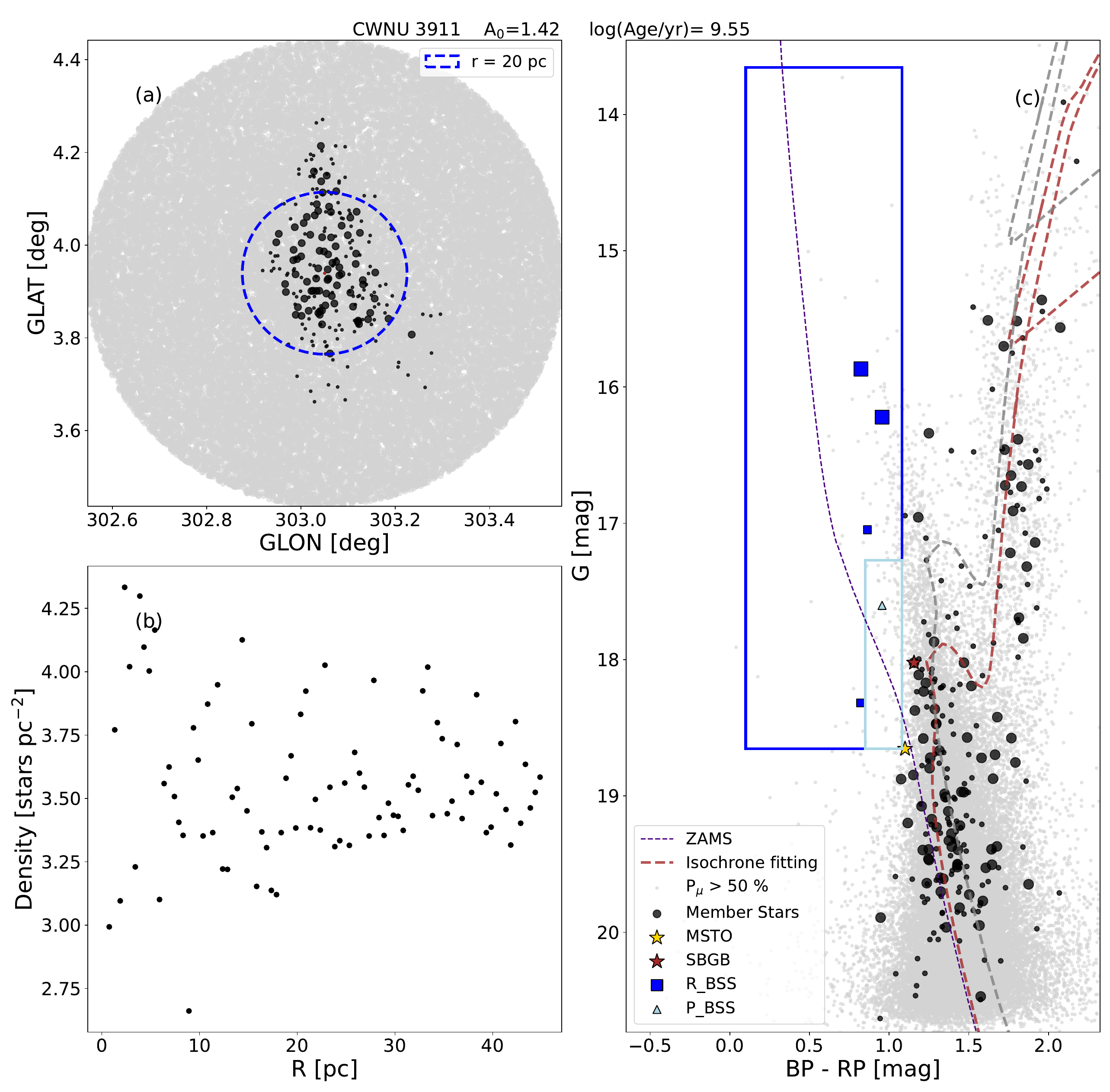}
\includegraphics[width=0.235\linewidth]{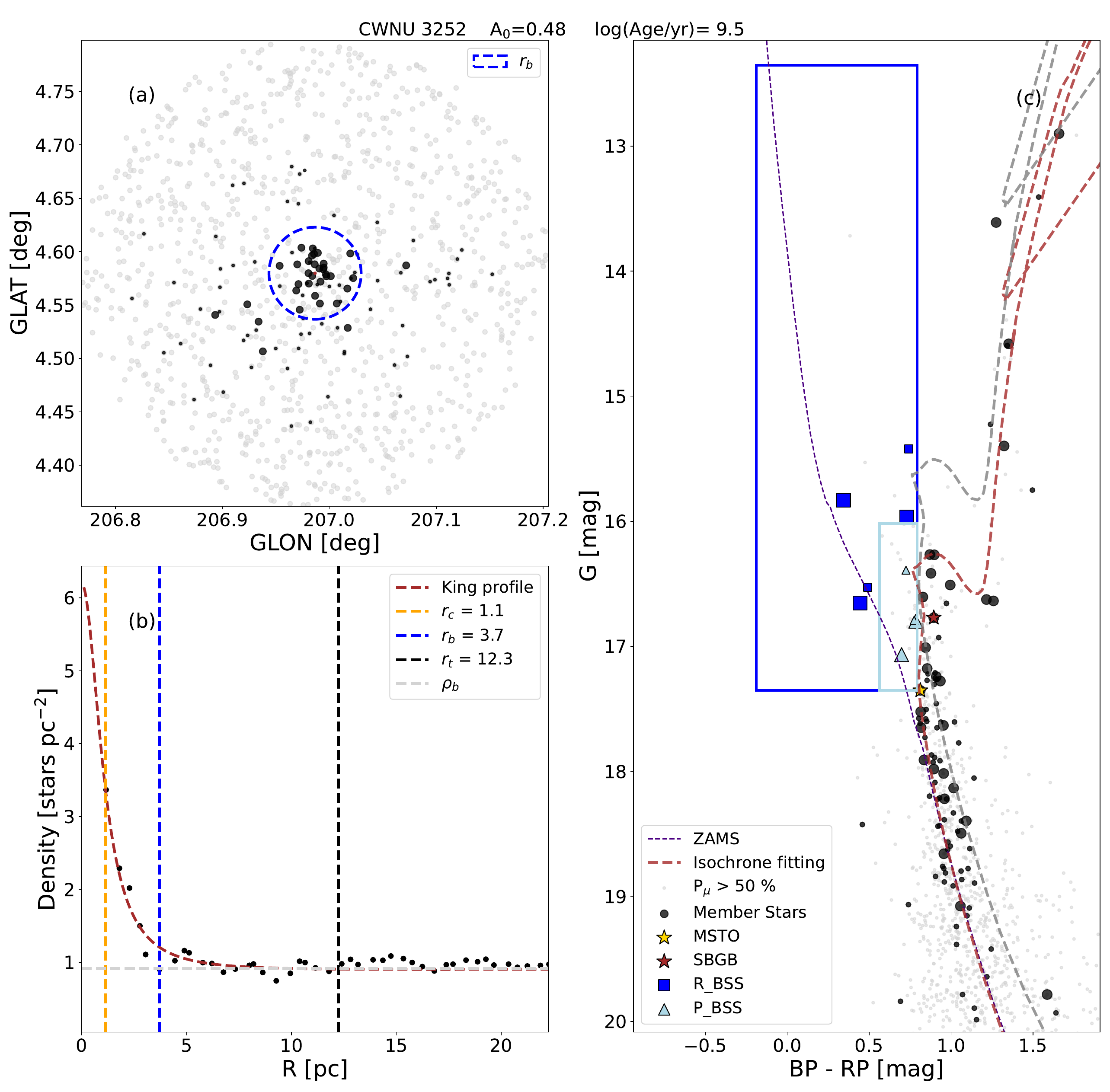}
\includegraphics[width=0.235\linewidth]{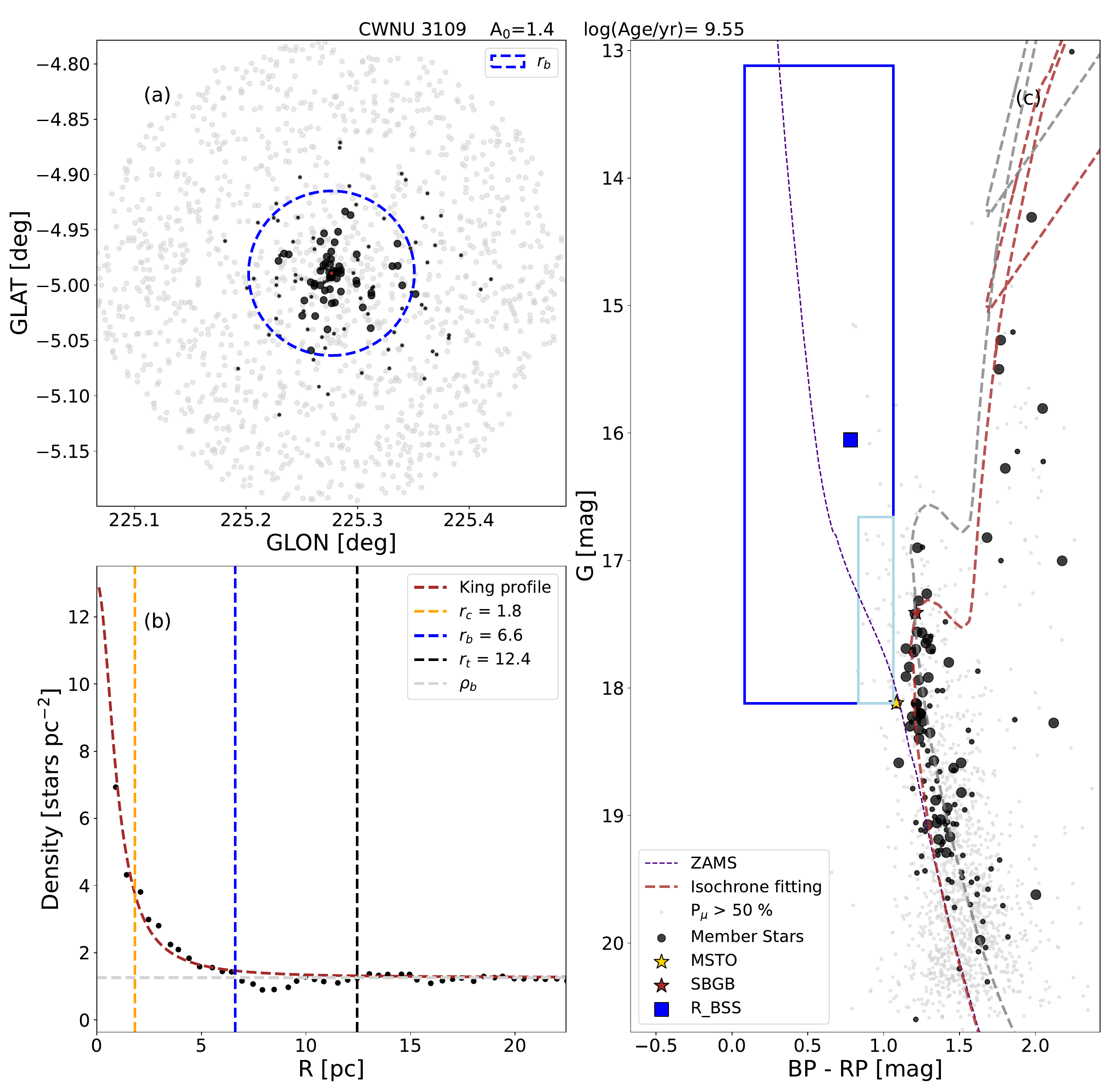}
\includegraphics[width=0.235\linewidth]{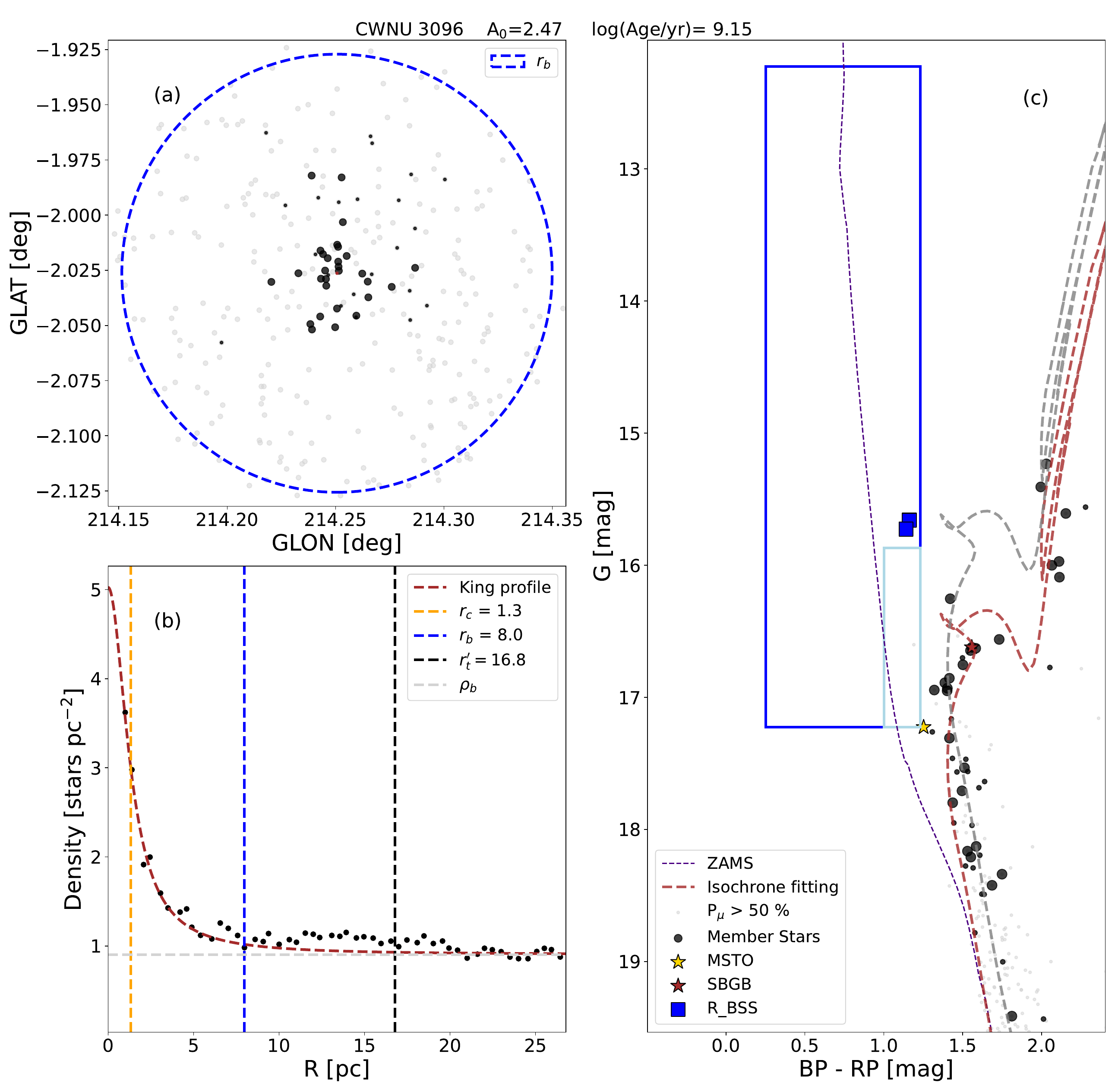}
\includegraphics[width=0.235\linewidth]{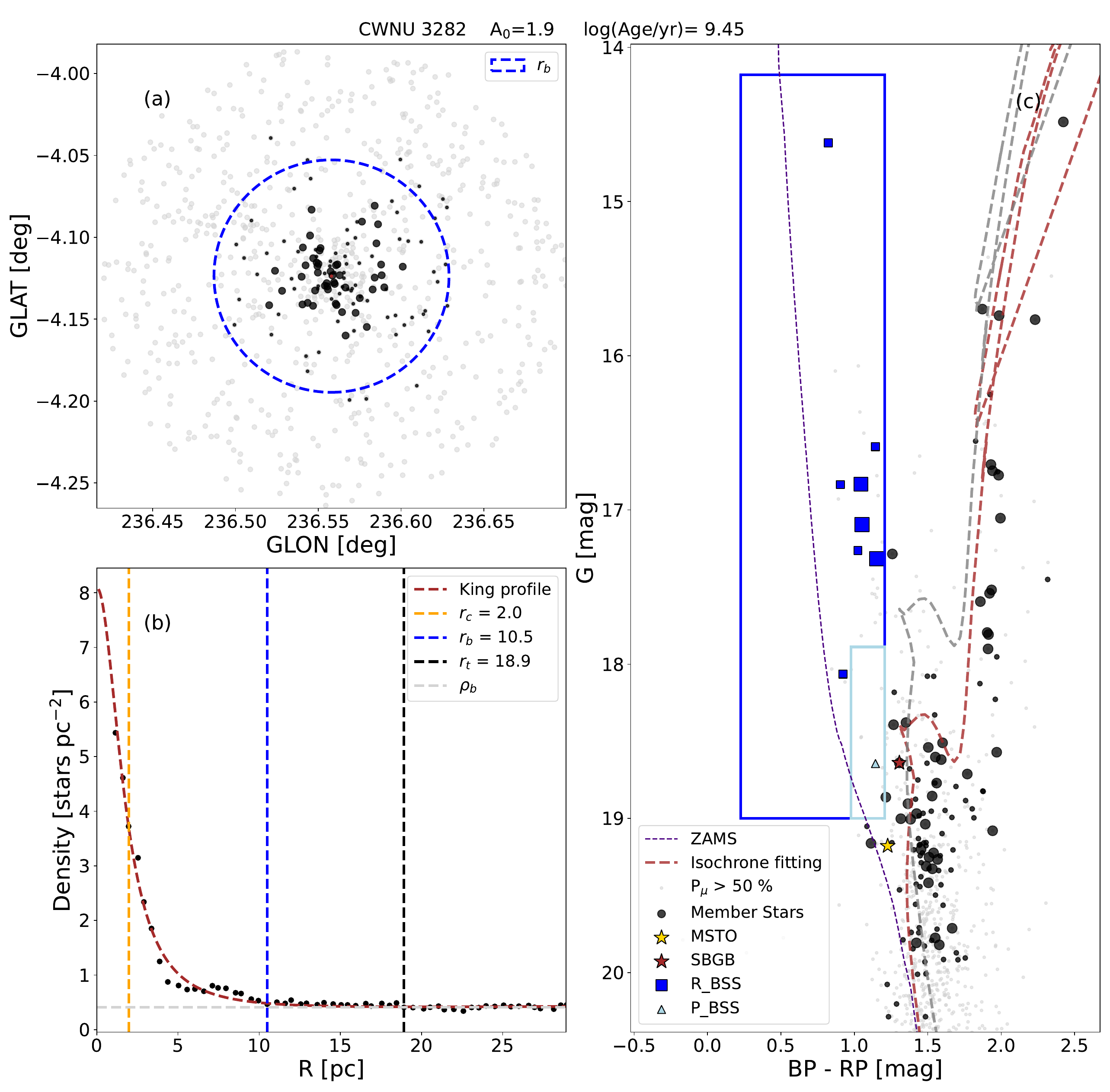}
\includegraphics[width=0.235\linewidth]{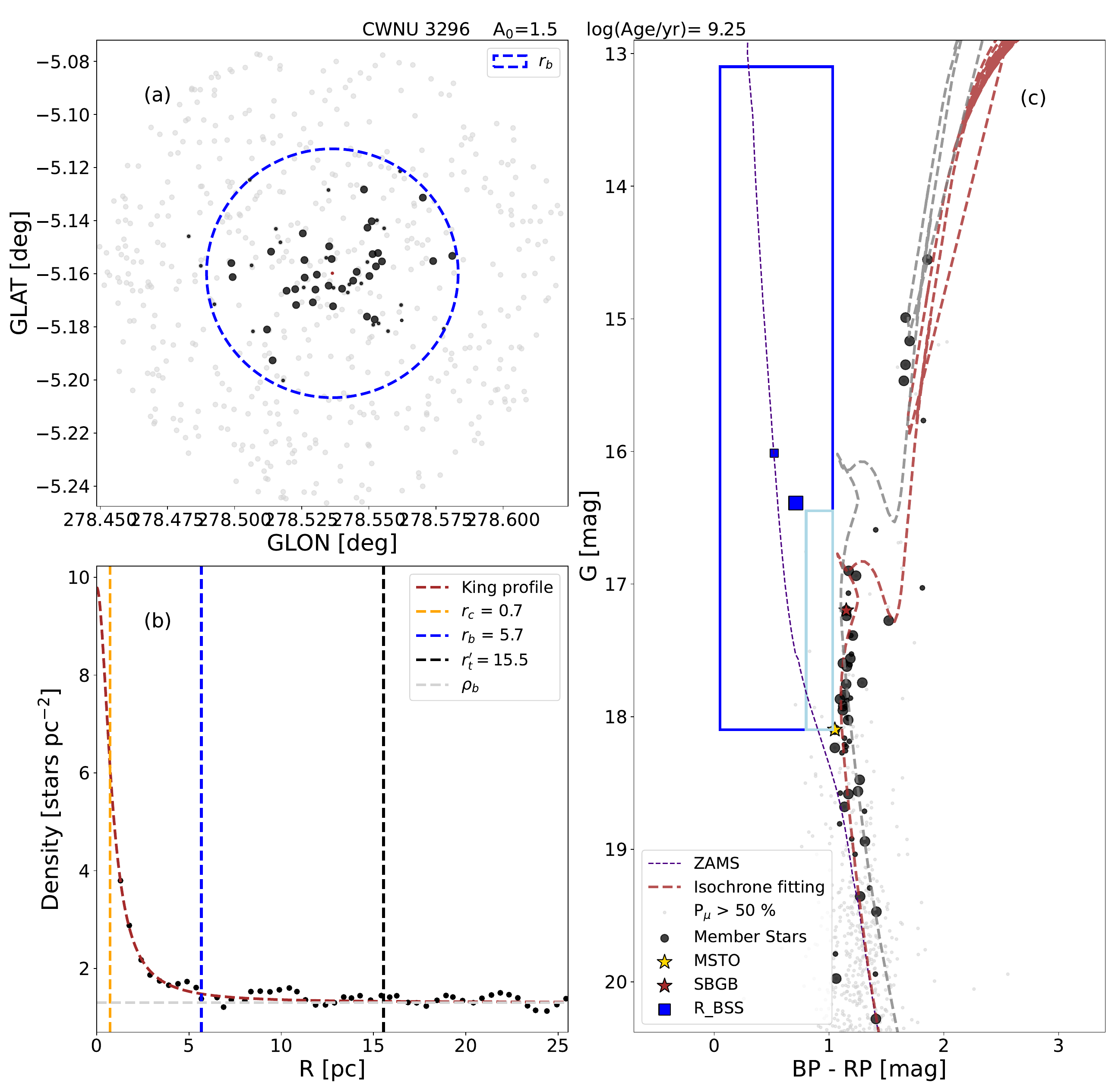}
\includegraphics[width=0.235\linewidth]{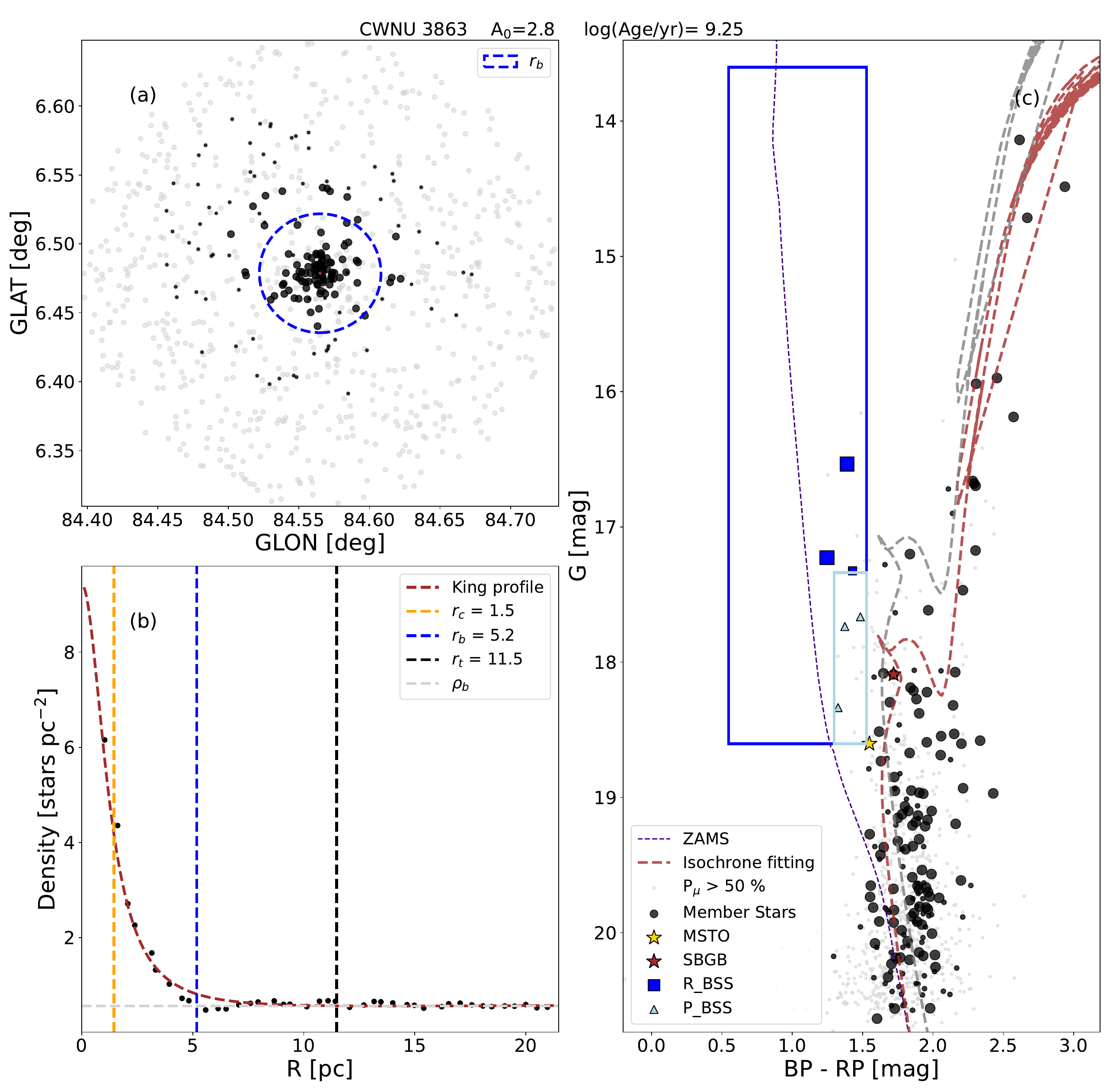}
\includegraphics[width=0.235\linewidth]{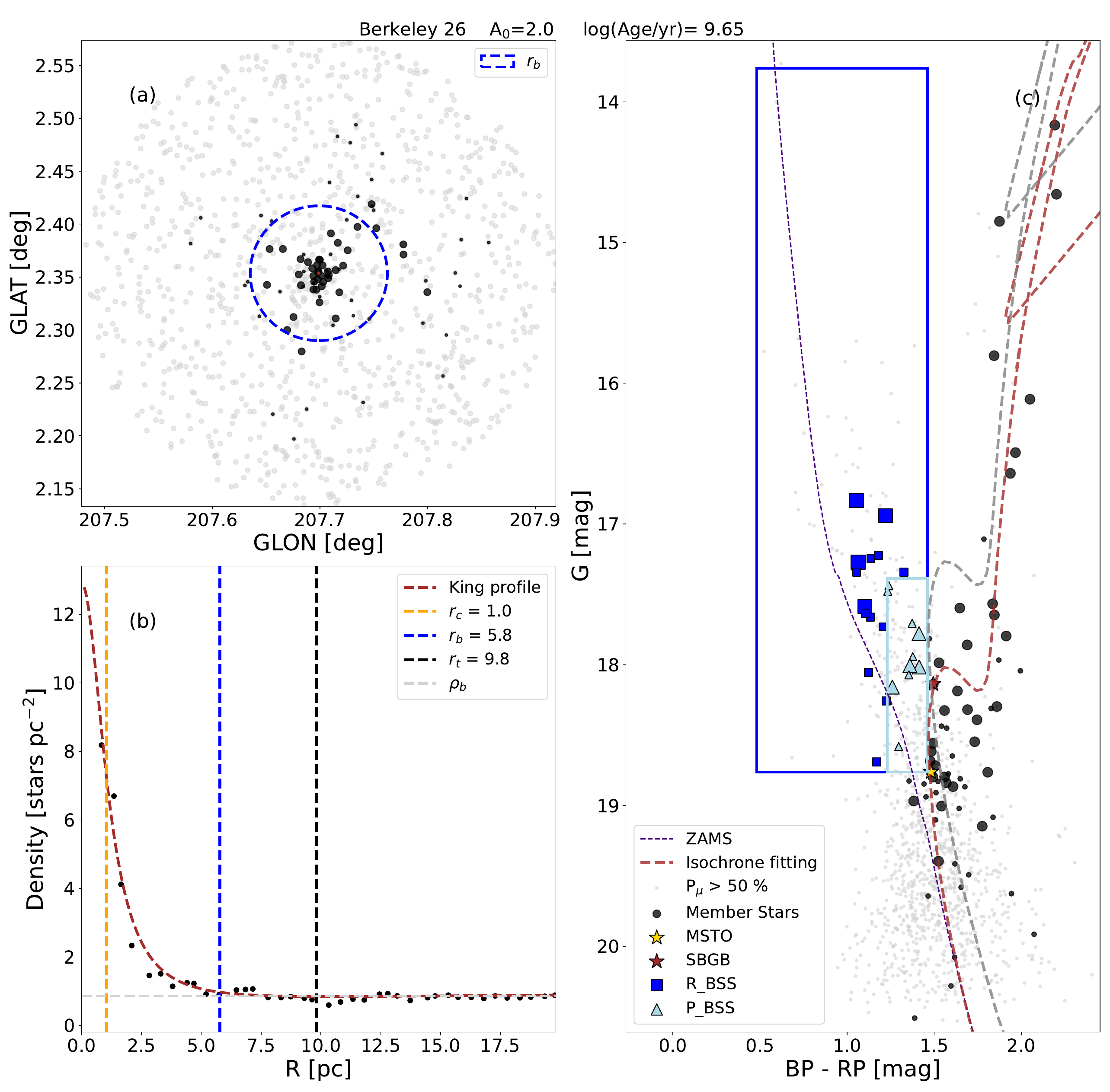}
\includegraphics[width=0.235\linewidth]{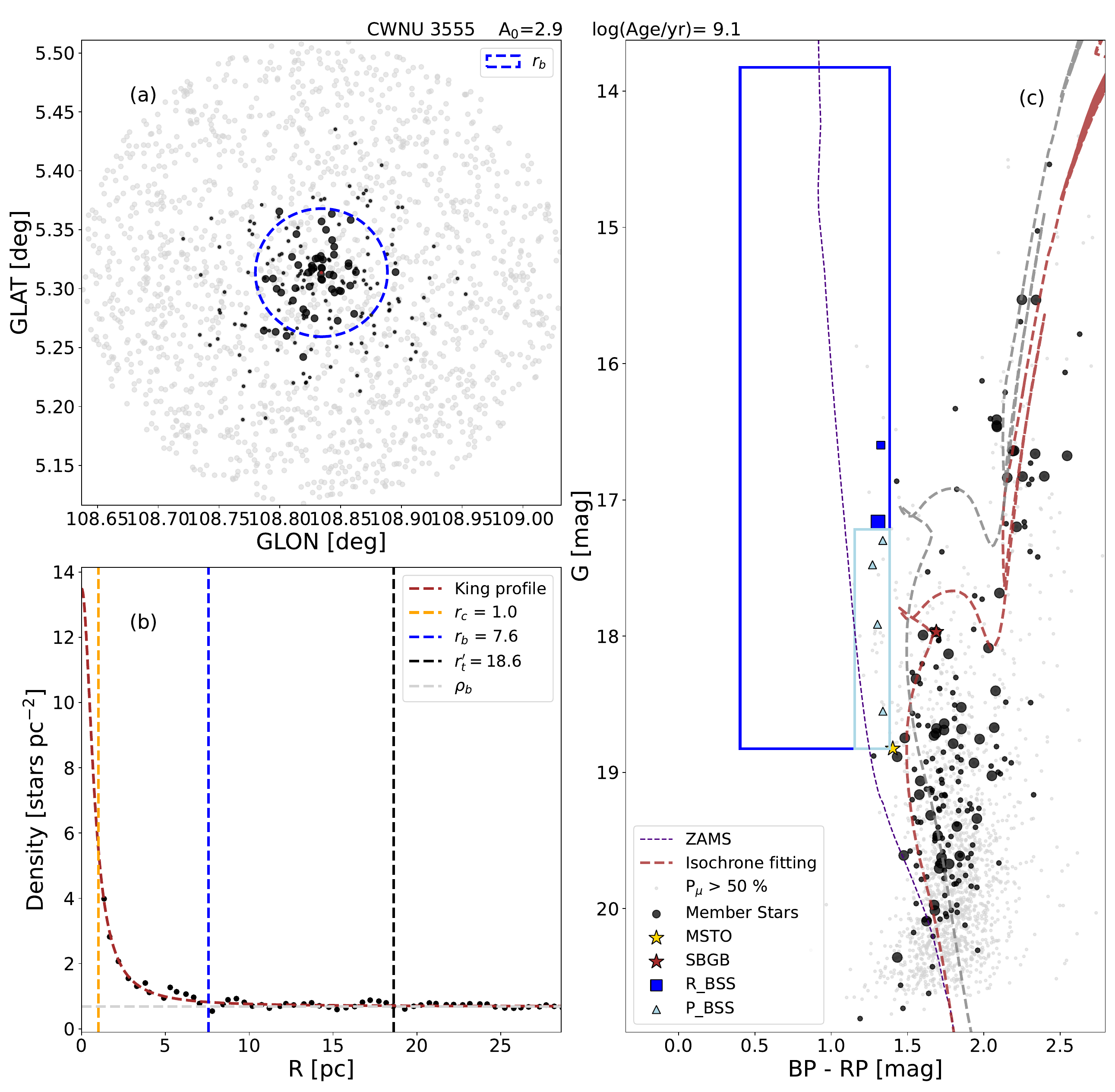}
\includegraphics[width=0.235\linewidth]{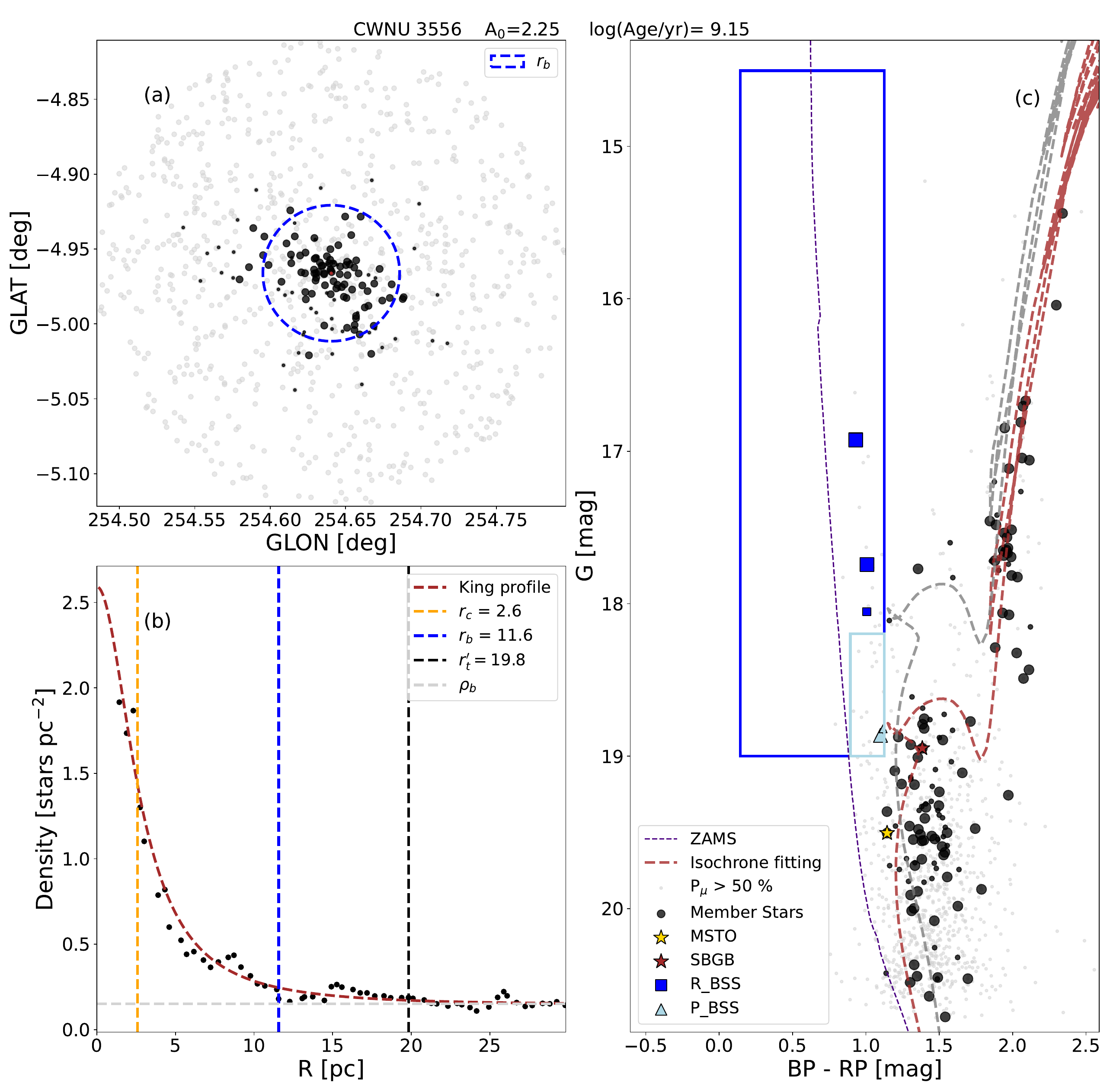}
\includegraphics[width=0.235\linewidth]{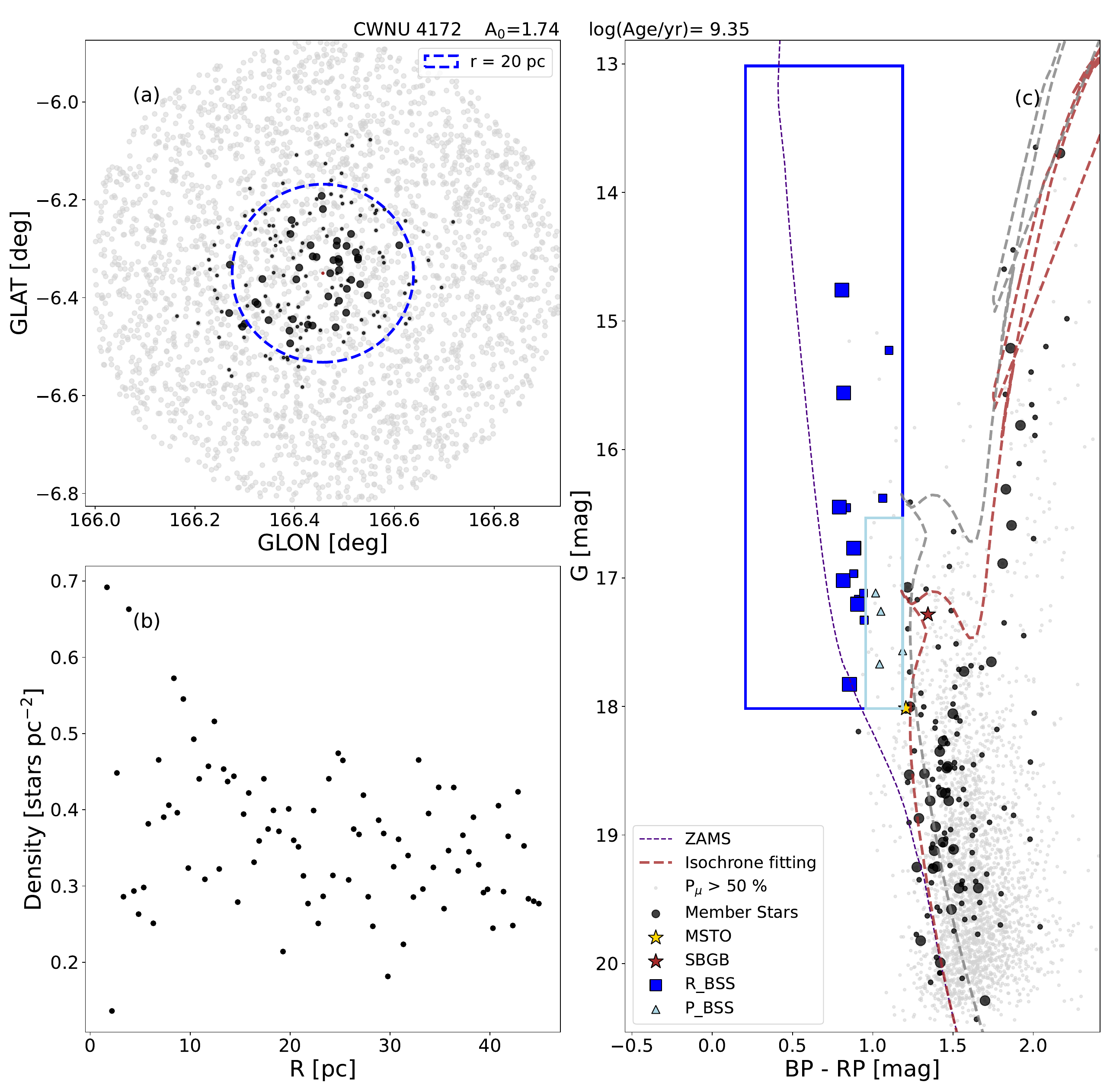}
\includegraphics[width=0.235\linewidth]{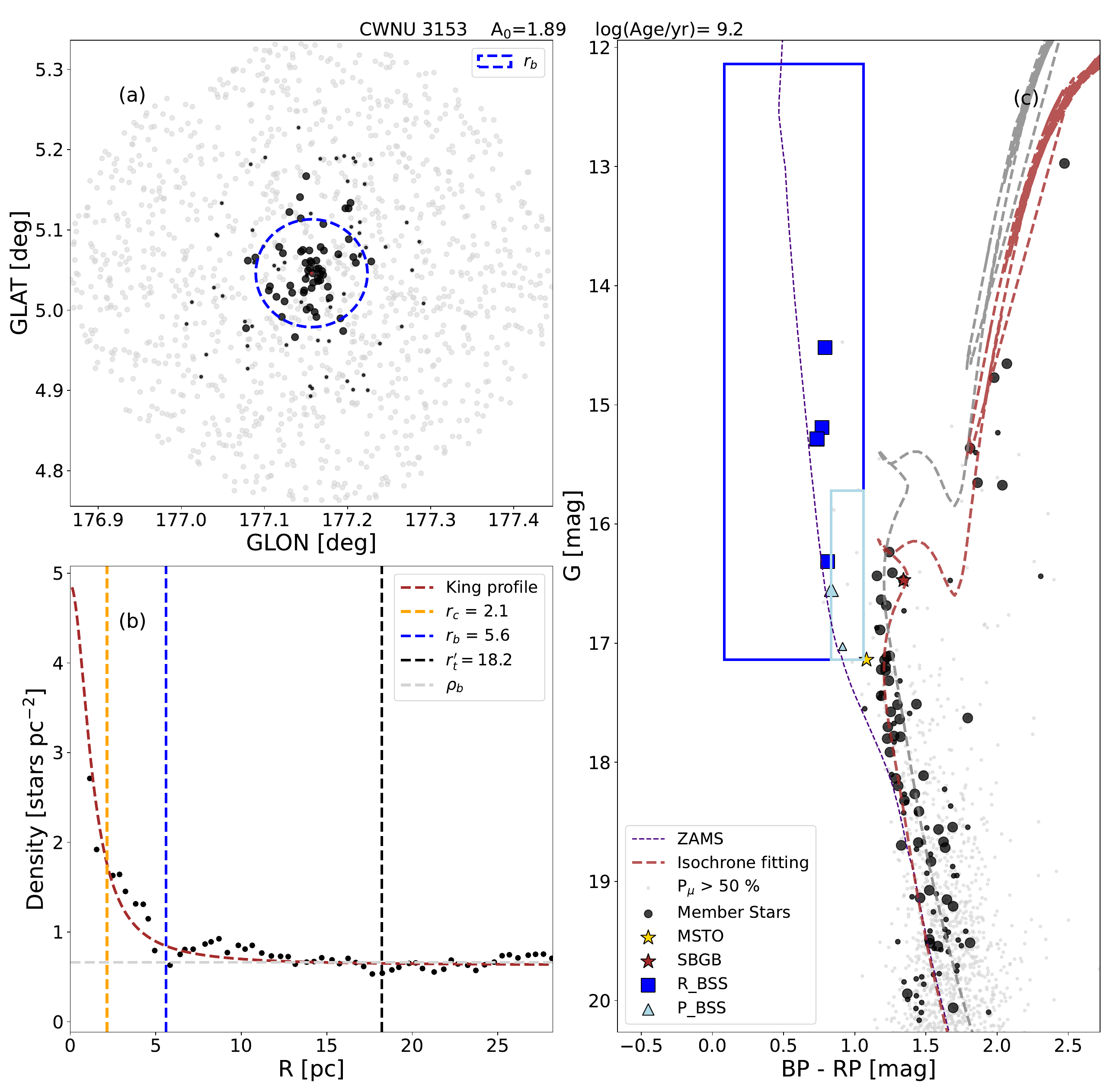}
\includegraphics[width=0.235\linewidth]{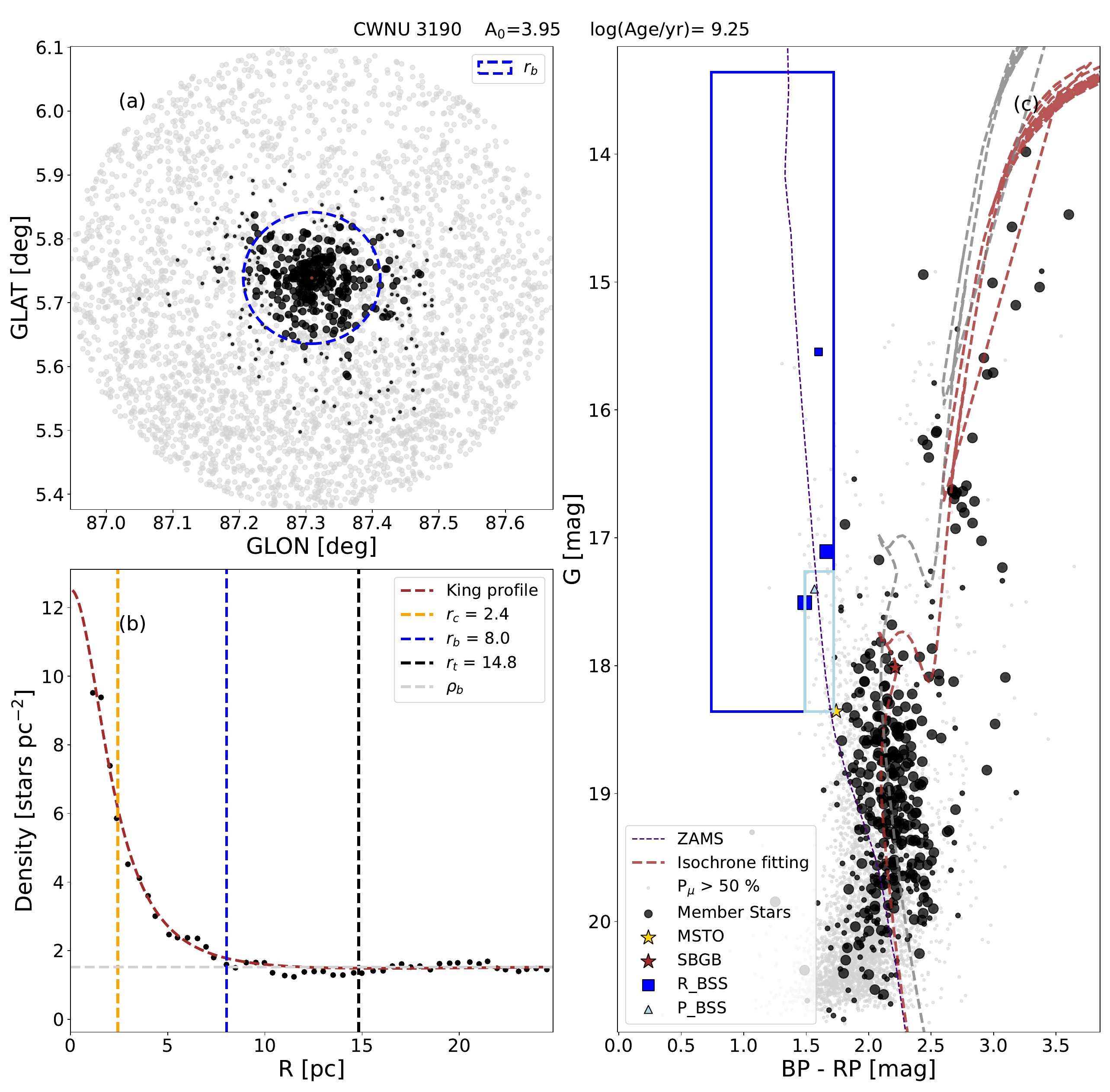}
\includegraphics[width=0.235\linewidth]{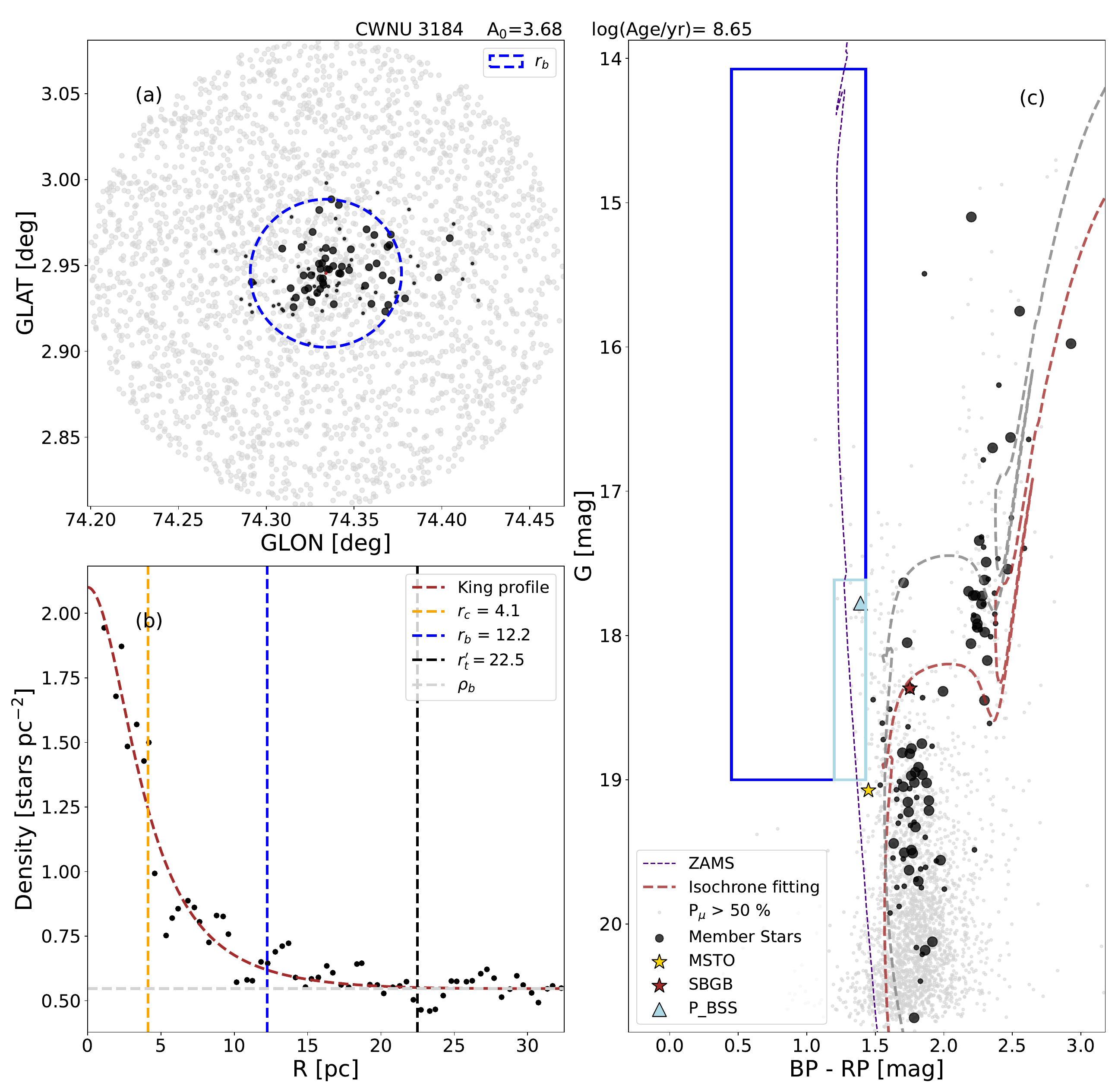}
\caption{Panel (a) shows the core member stars in Galactic coordinates. Panel (b) presents the density profile along with the corresponding King model fitting, highlighting the boundary radius $R_b$. Panel (c) depicts the H-R diagram and includes the best isochrone fitting. As described in Section \ref{sec:results}, the boxes represent the candidates for BSS: the blue boxes indicate \textit{R\_BSS}, while the light blue boxes indicate \textit{P\_BSS}; along with the positions of various star types on the H-R diagram. Gray dots indicate stars with a membership probability greater than 50$\%$.
}
\label{figa1}
\end{center}
\end{figure*}

\begin{figure*}
\begin{center}
\includegraphics[width=0.235\linewidth]{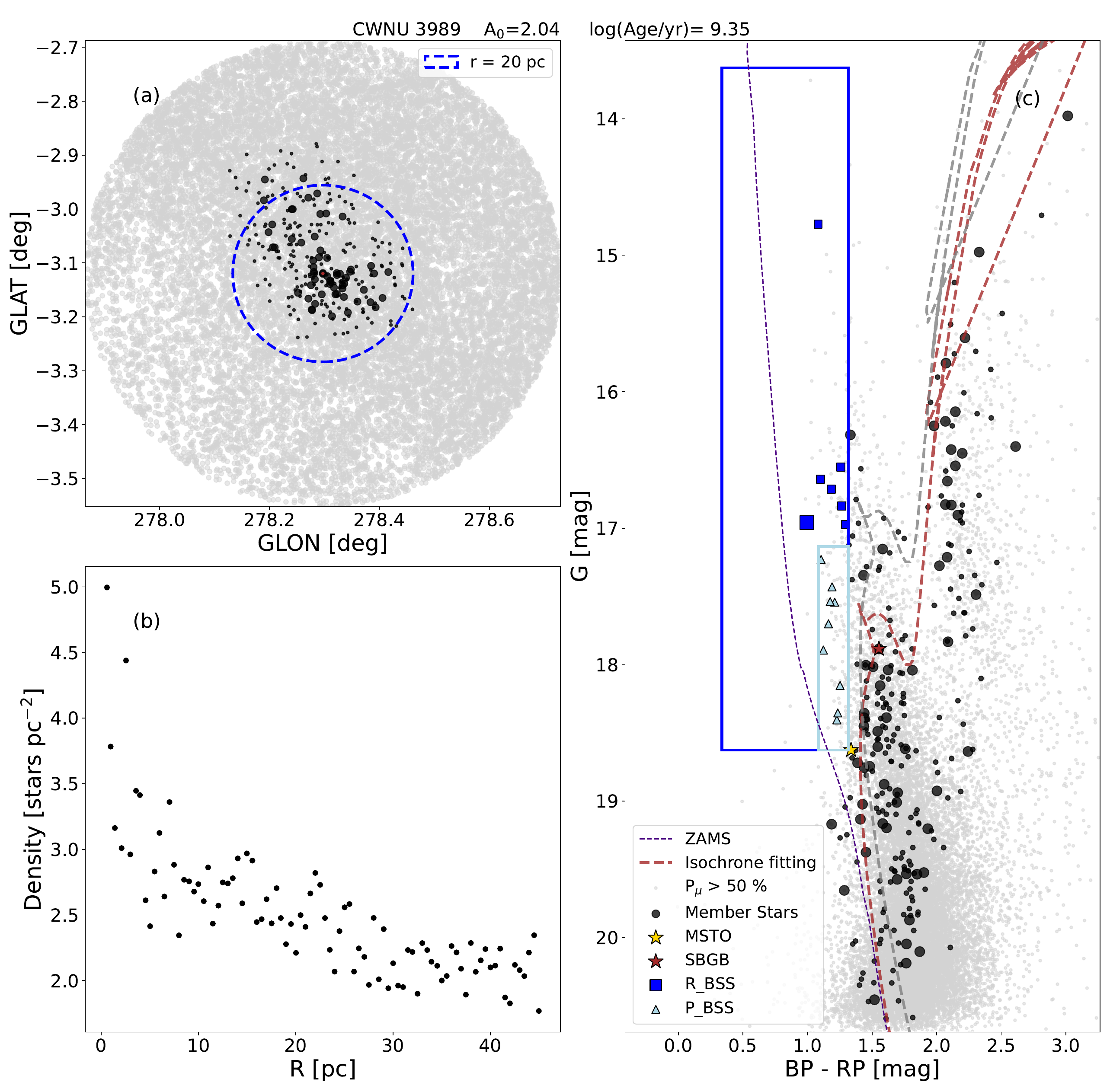}
\includegraphics[width=0.235\linewidth]{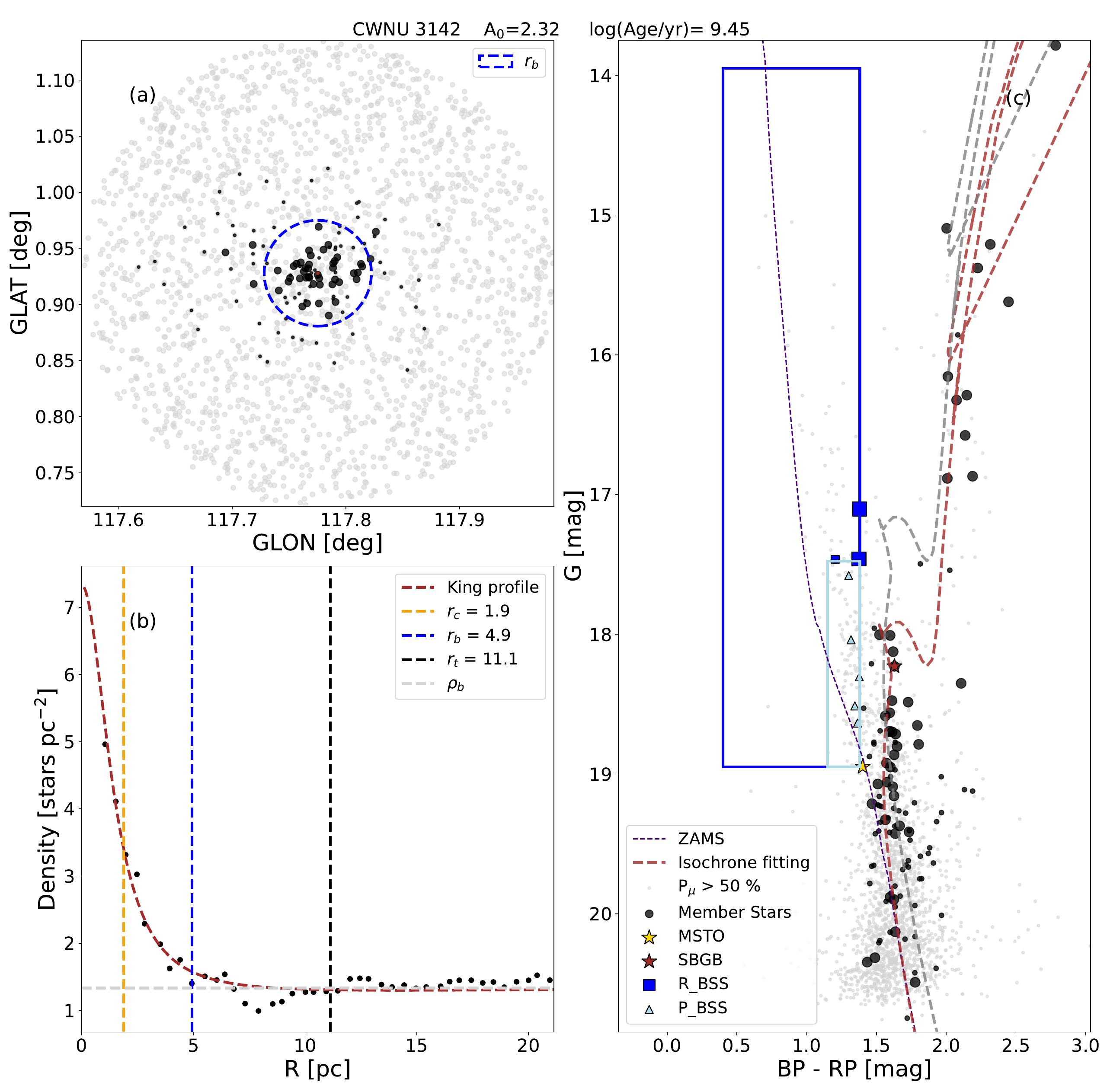}
\includegraphics[width=0.235\linewidth]{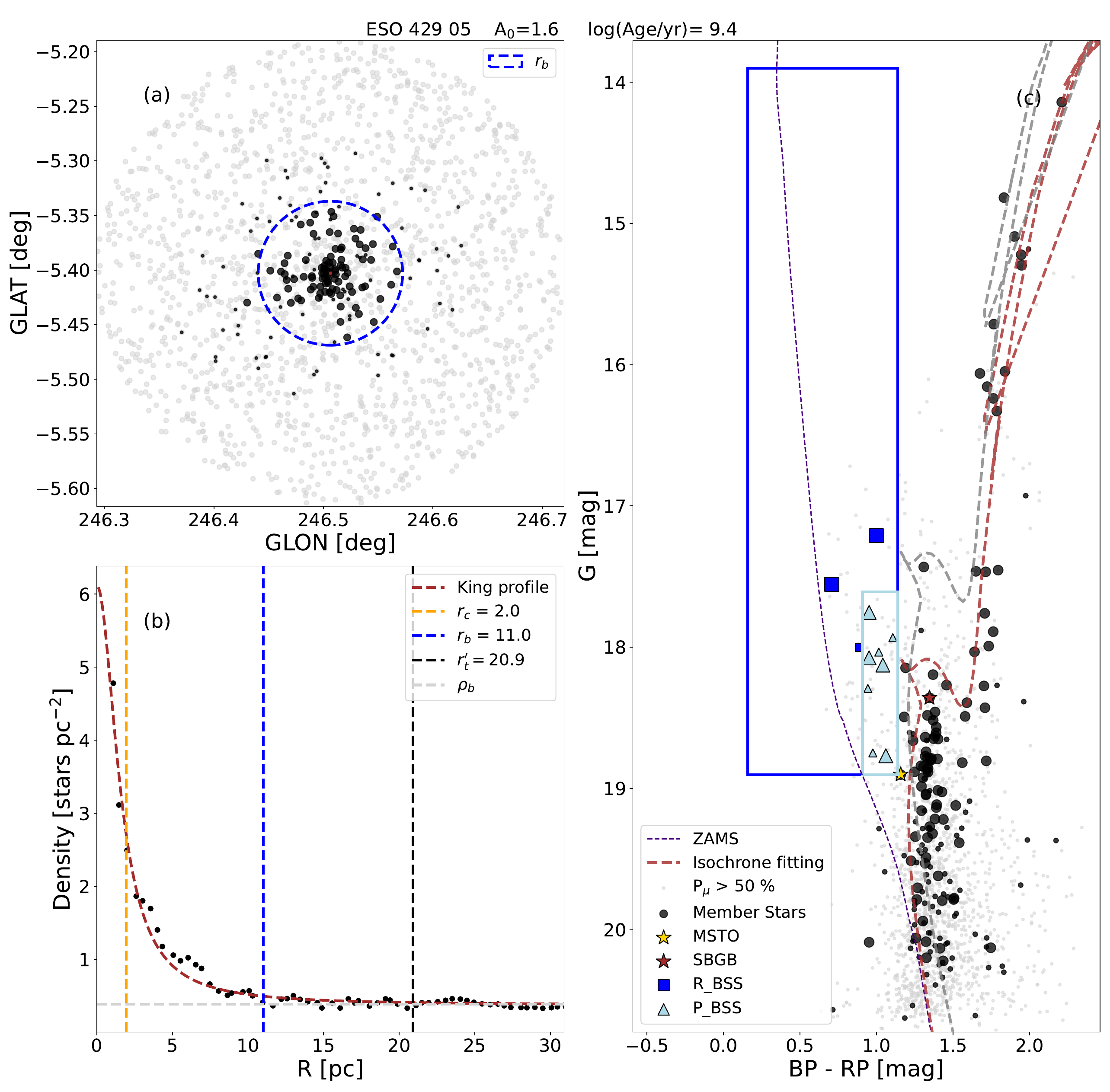}
\includegraphics[width=0.235\linewidth]{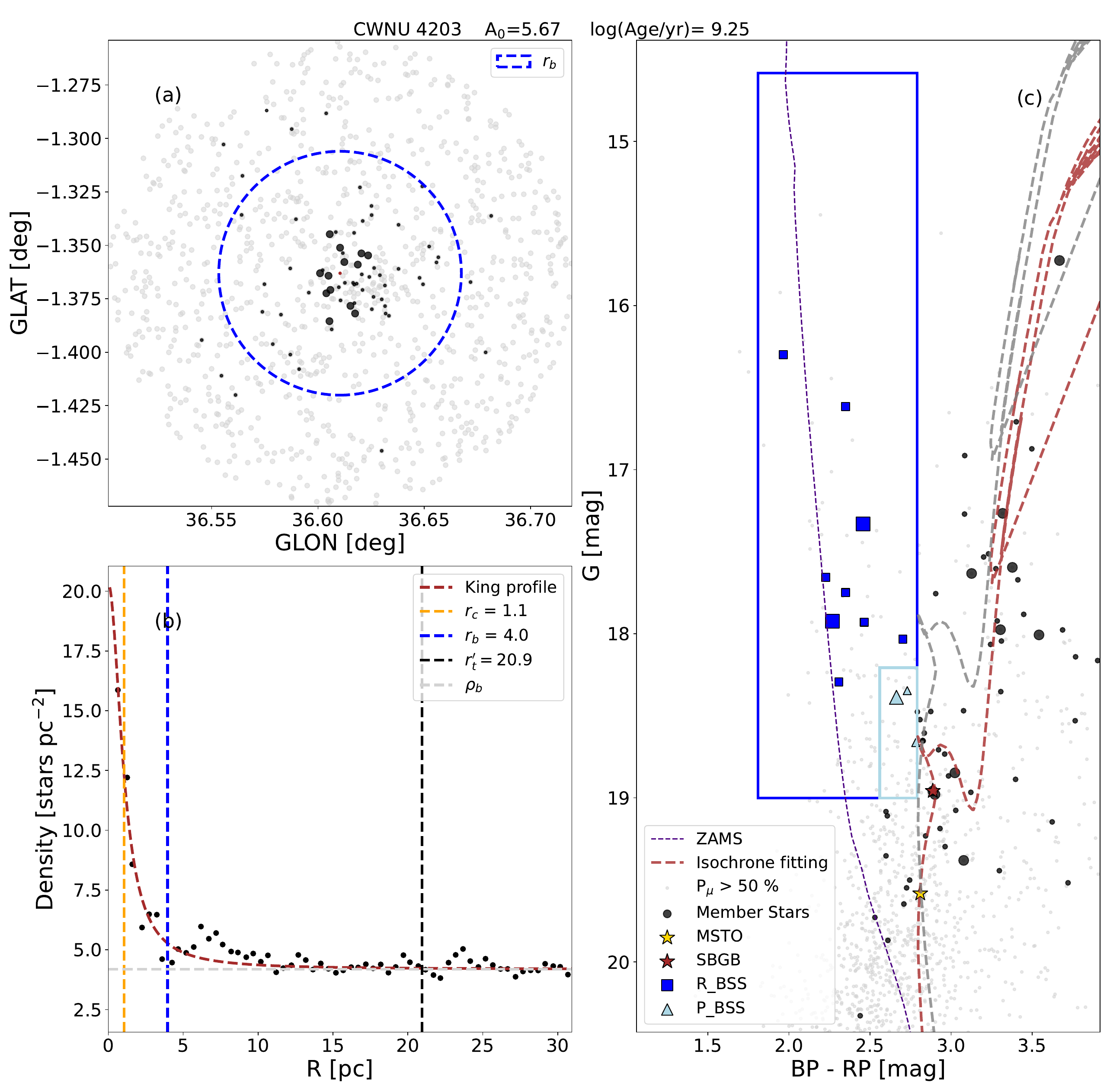}
\includegraphics[width=0.235\linewidth]{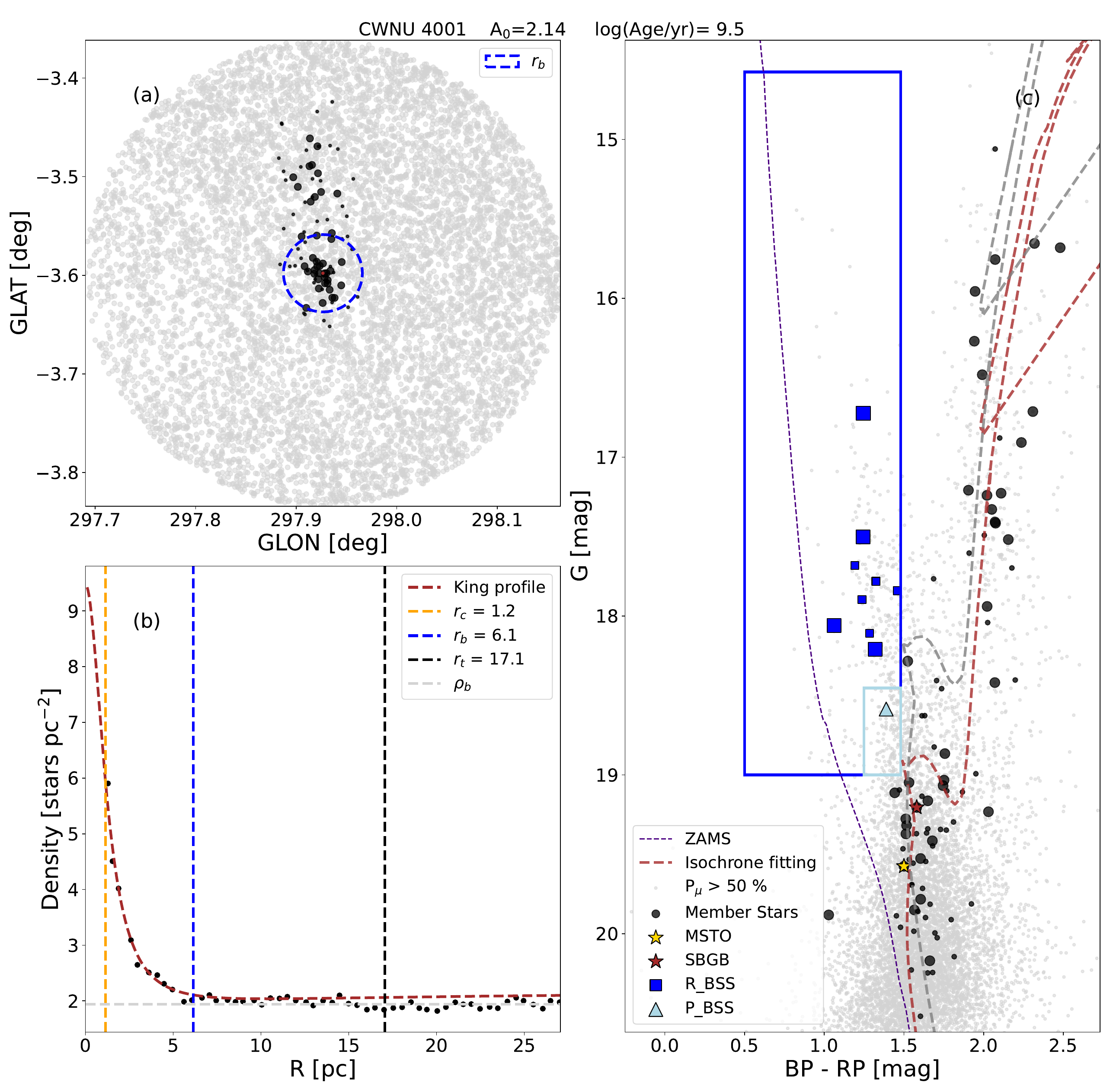}
\includegraphics[width=0.235\linewidth]{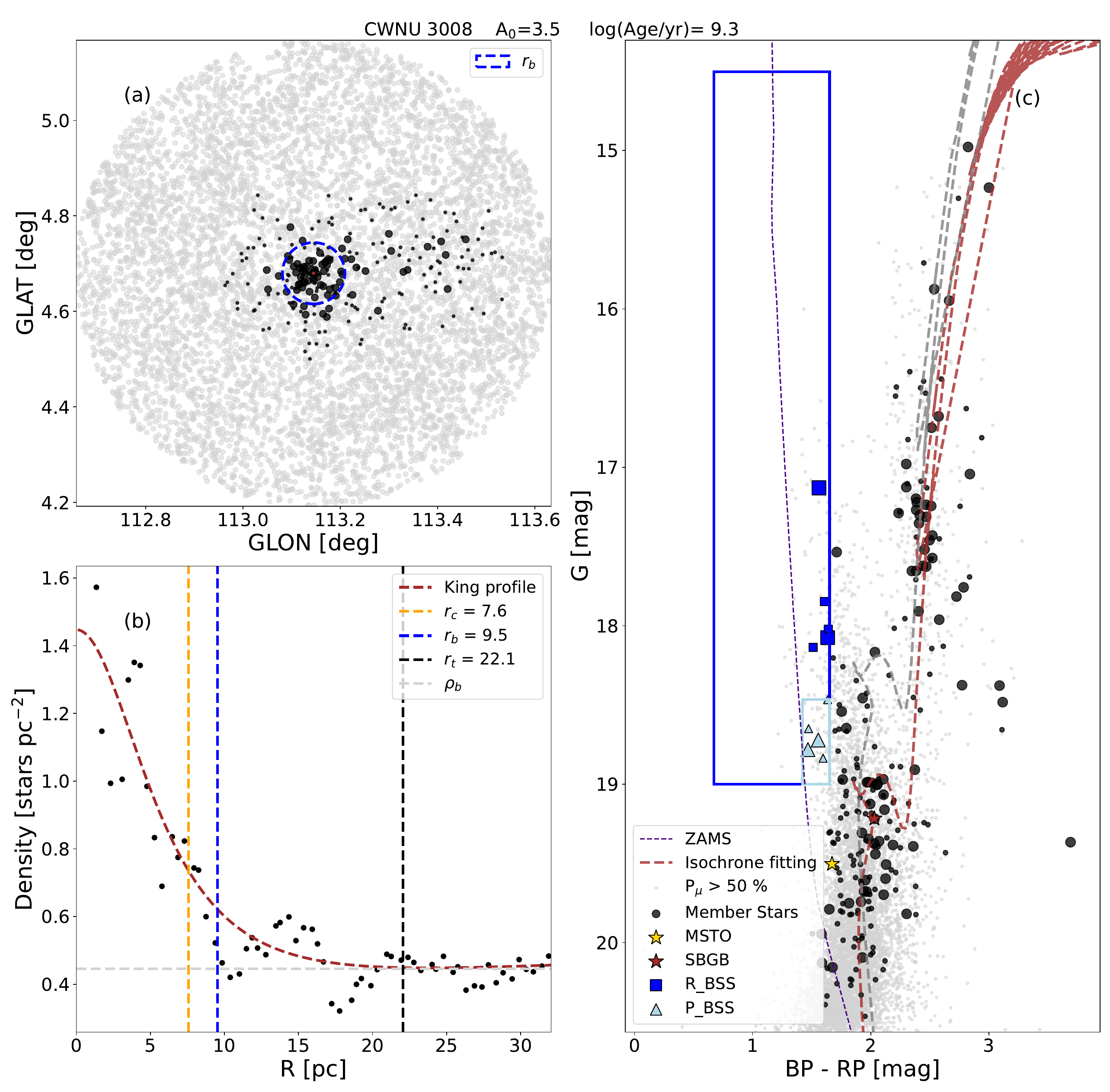}
\includegraphics[width=0.235\linewidth]{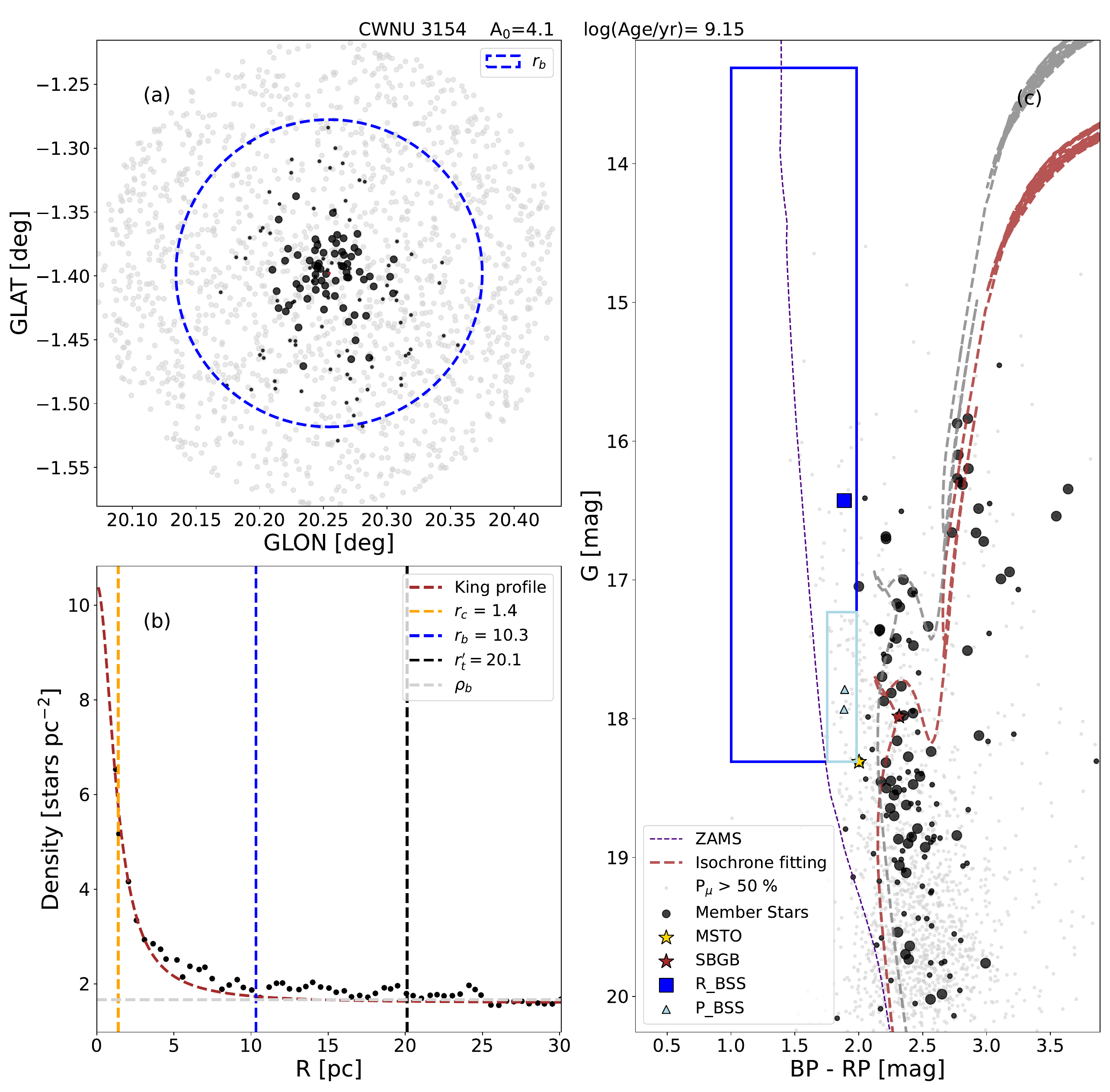}
\includegraphics[width=0.235\linewidth]{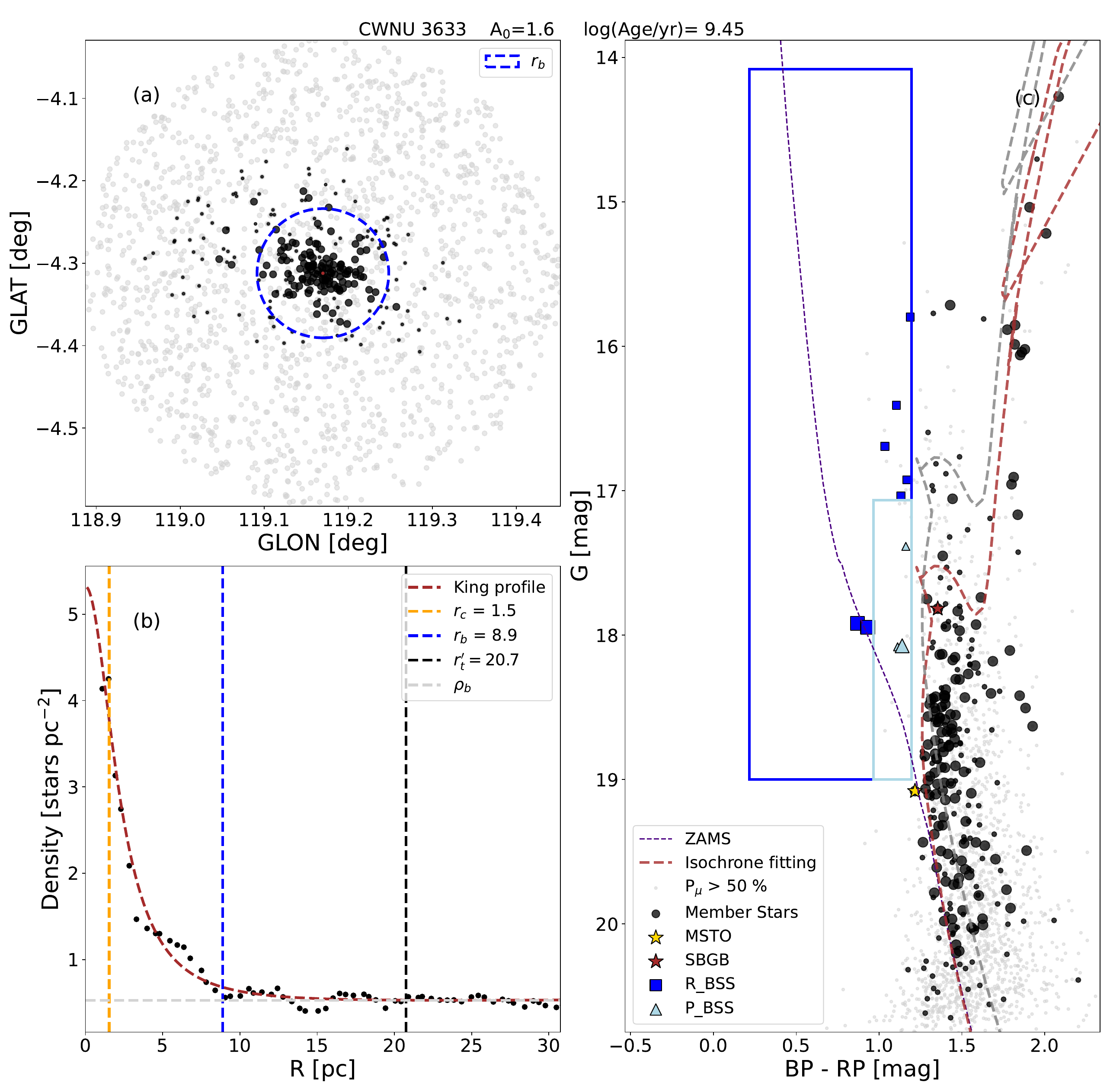}
\includegraphics[width=0.235\linewidth]{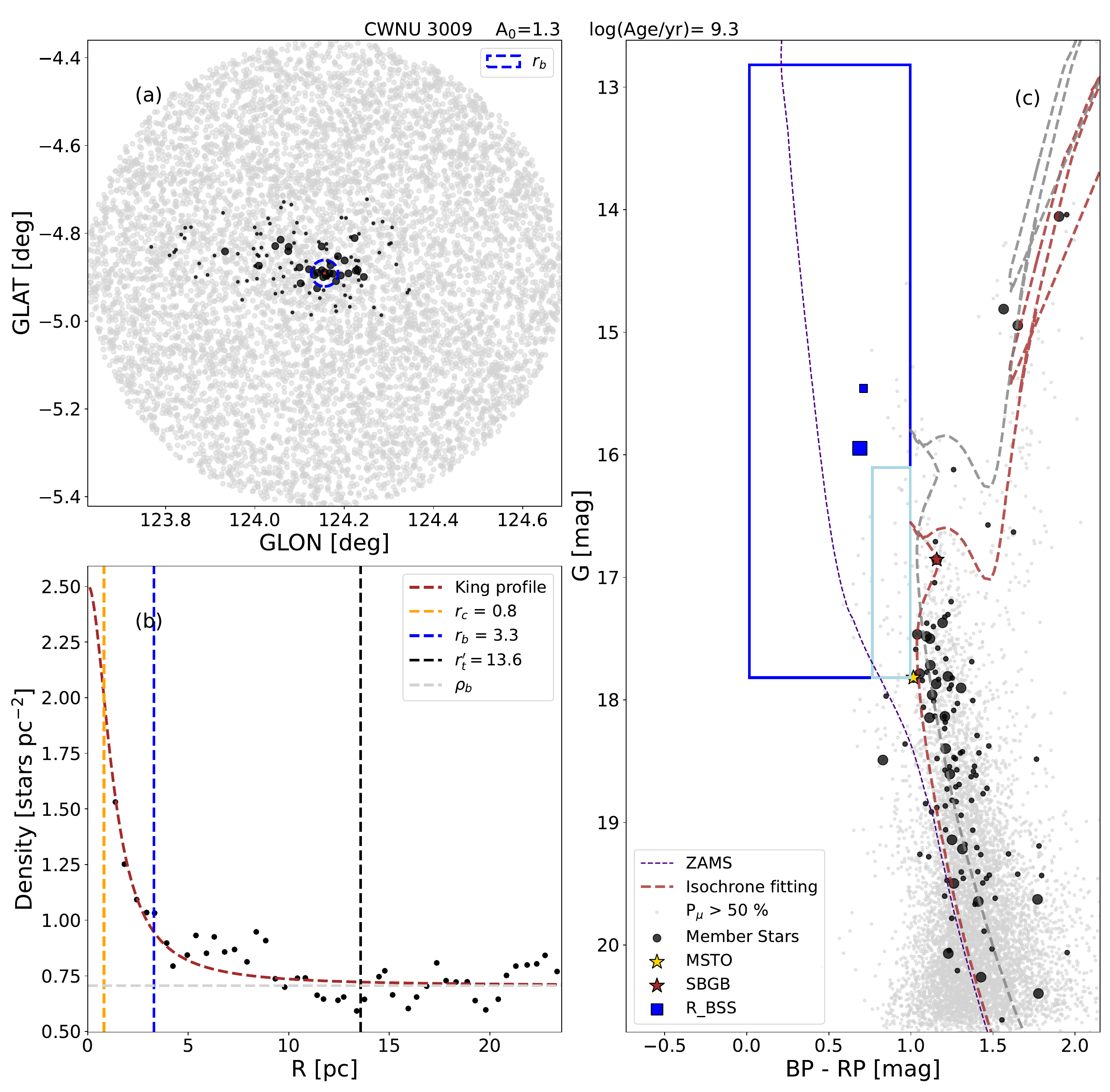}
\includegraphics[width=0.235\linewidth]{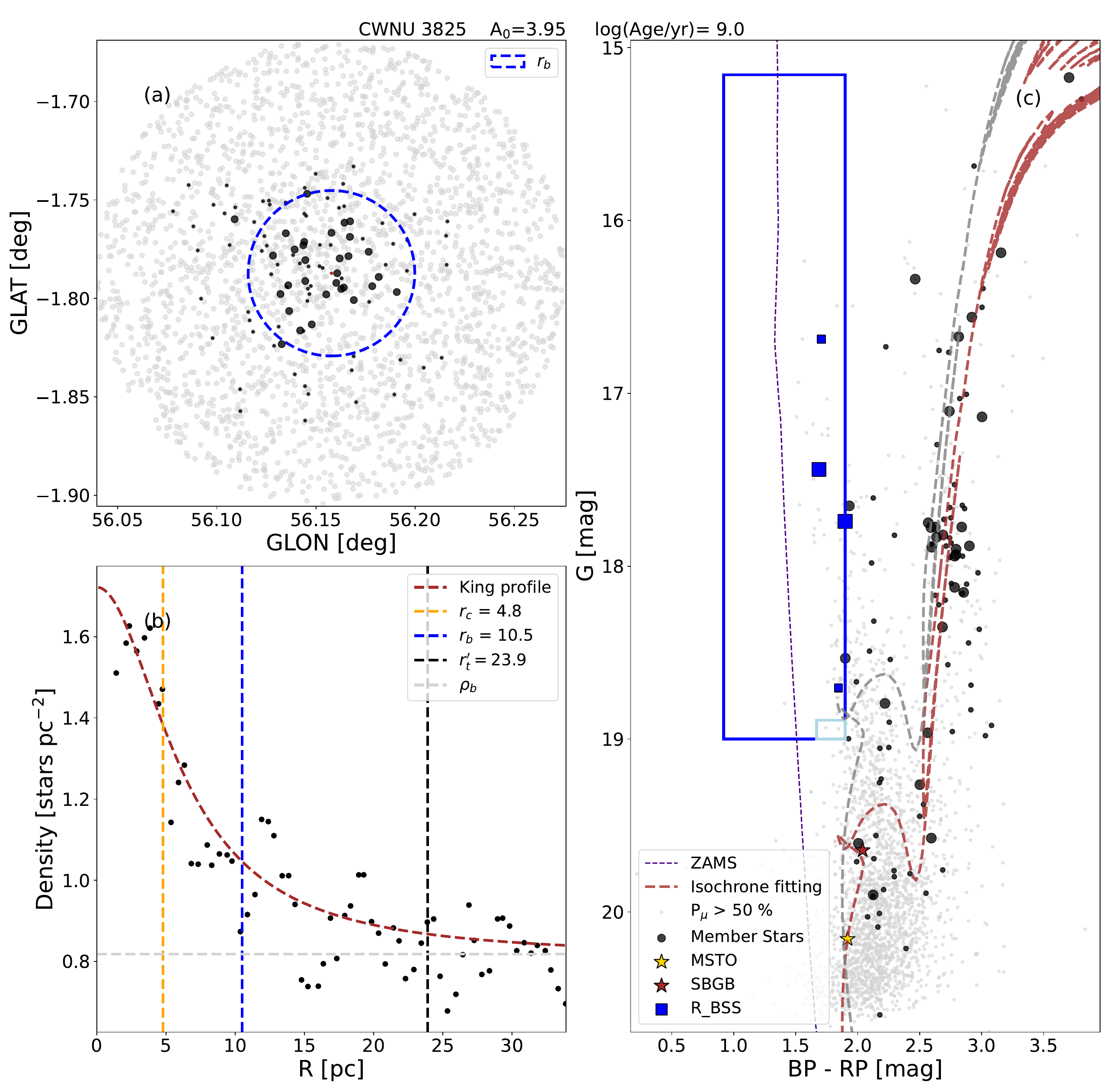}
\includegraphics[width=0.235\linewidth]{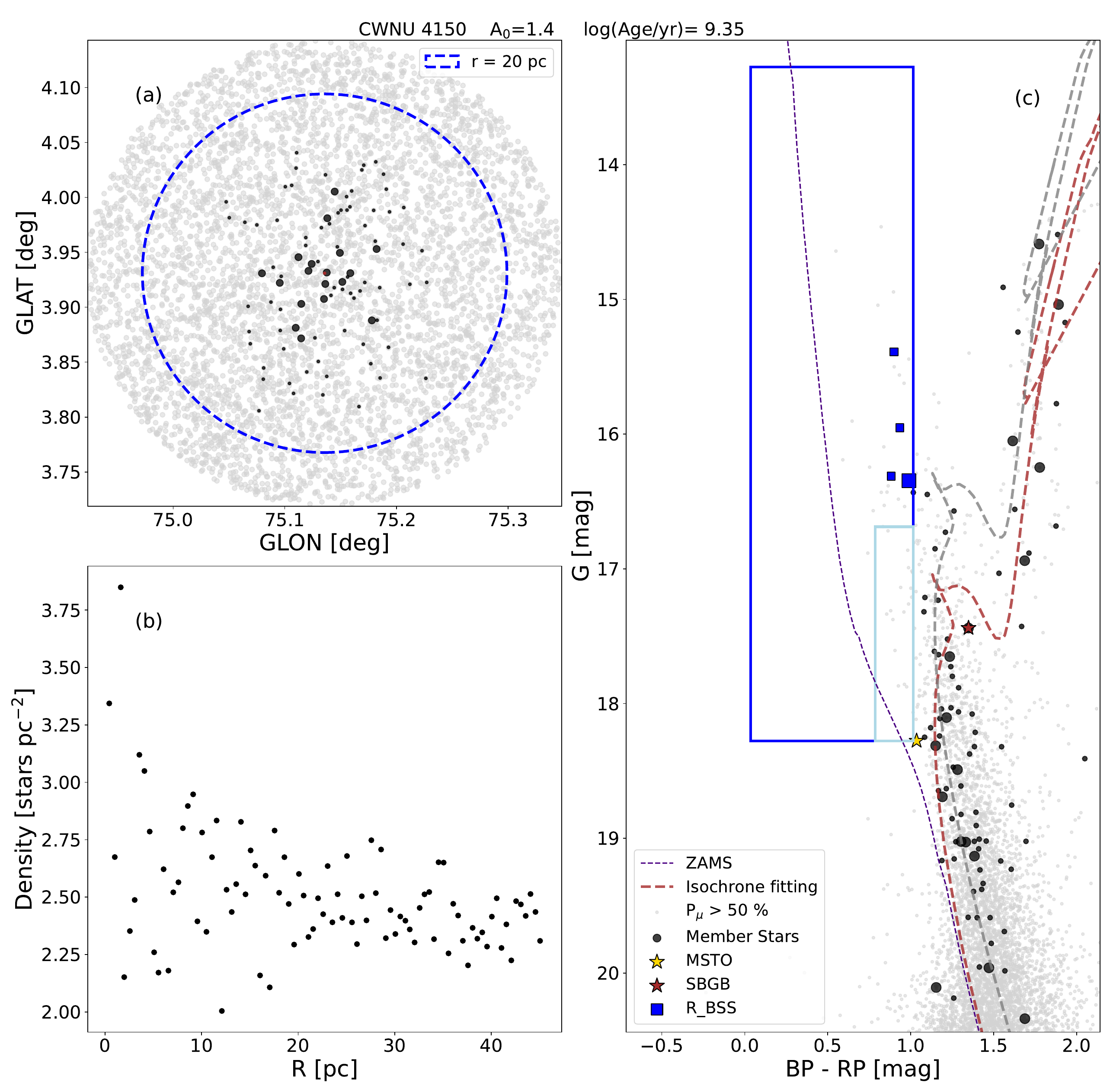}
\includegraphics[width=0.235\linewidth]{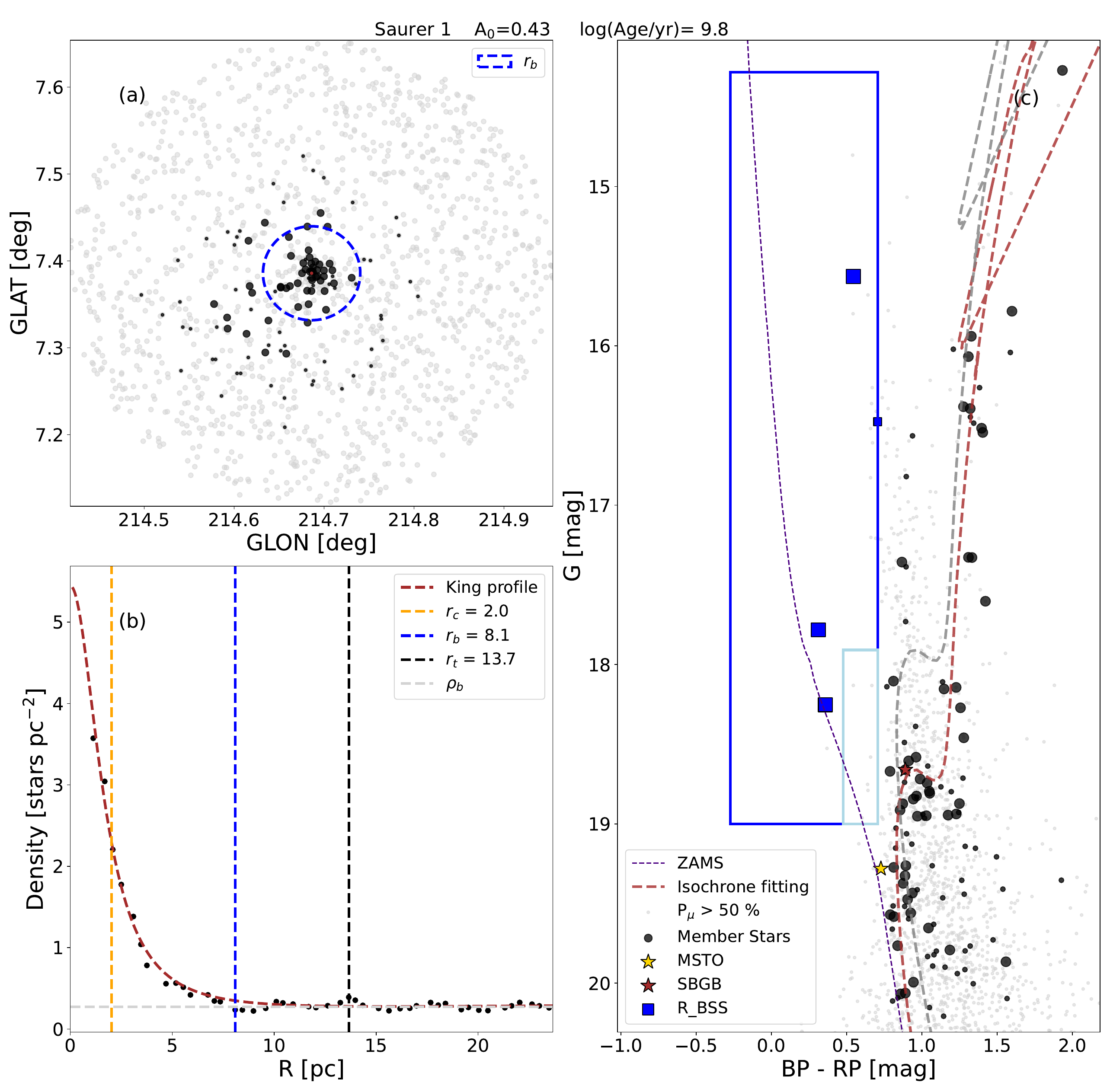}
\includegraphics[width=0.235\linewidth]{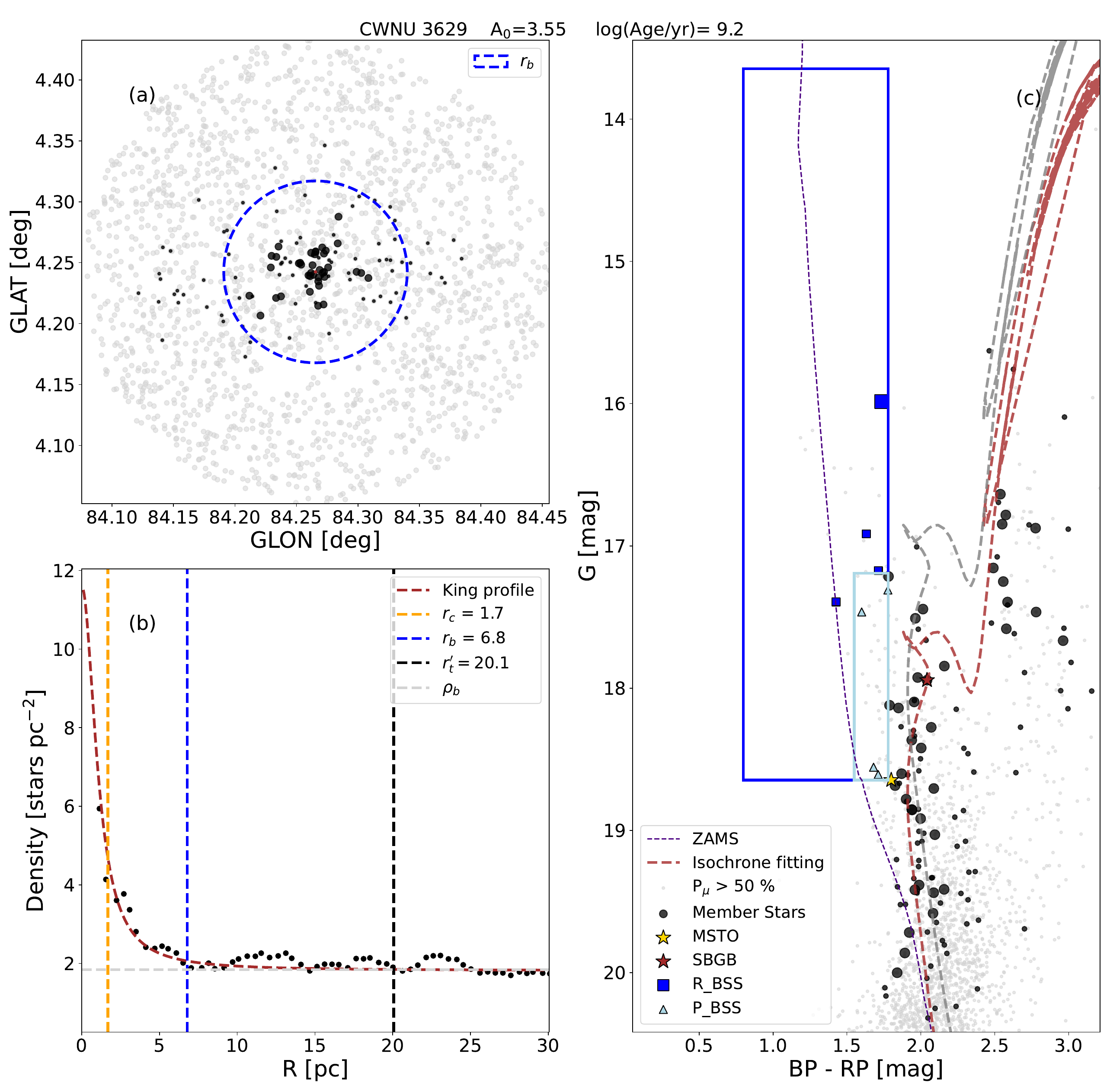}
\includegraphics[width=0.235\linewidth]{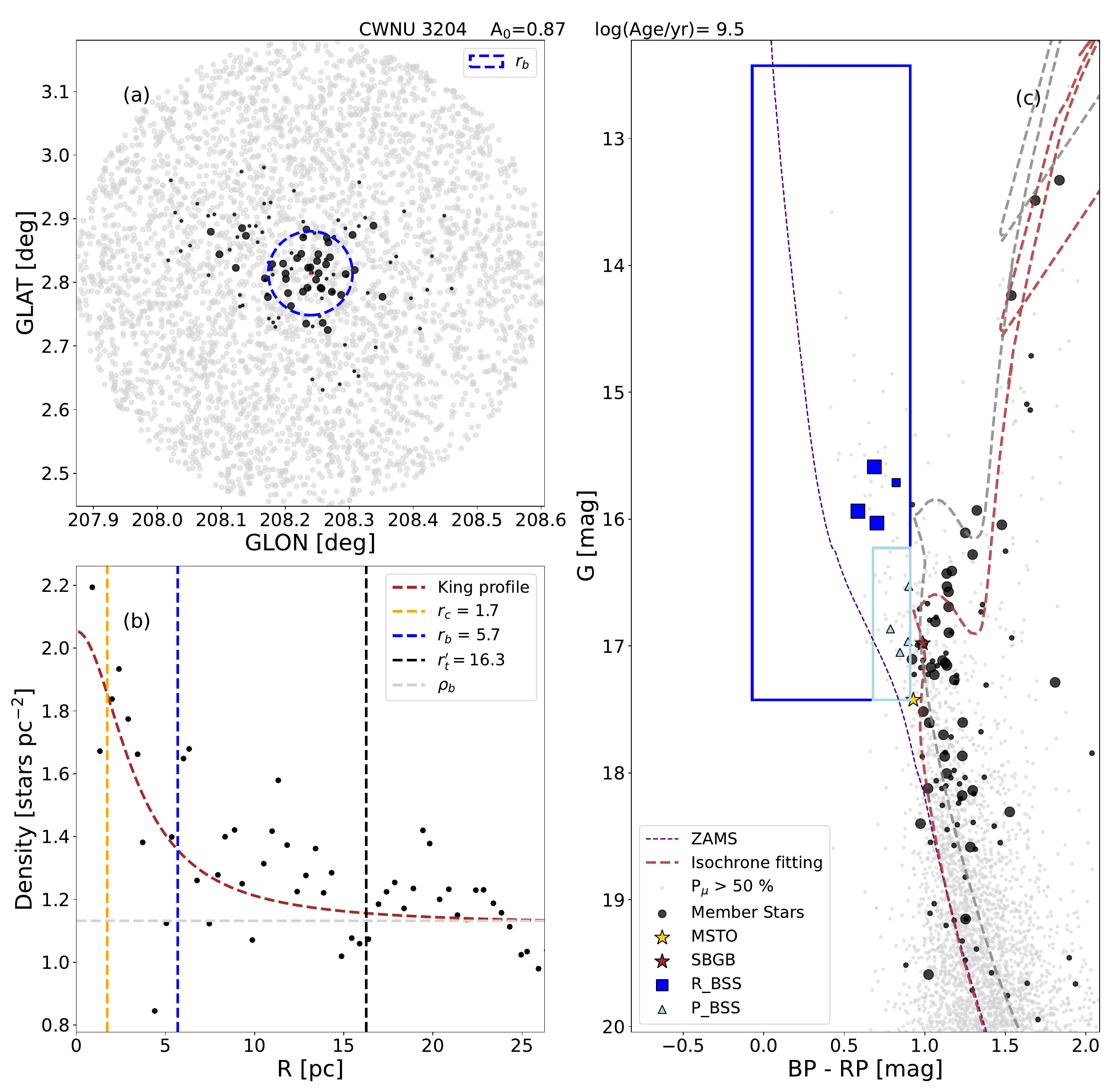}
\includegraphics[width=0.235\linewidth]{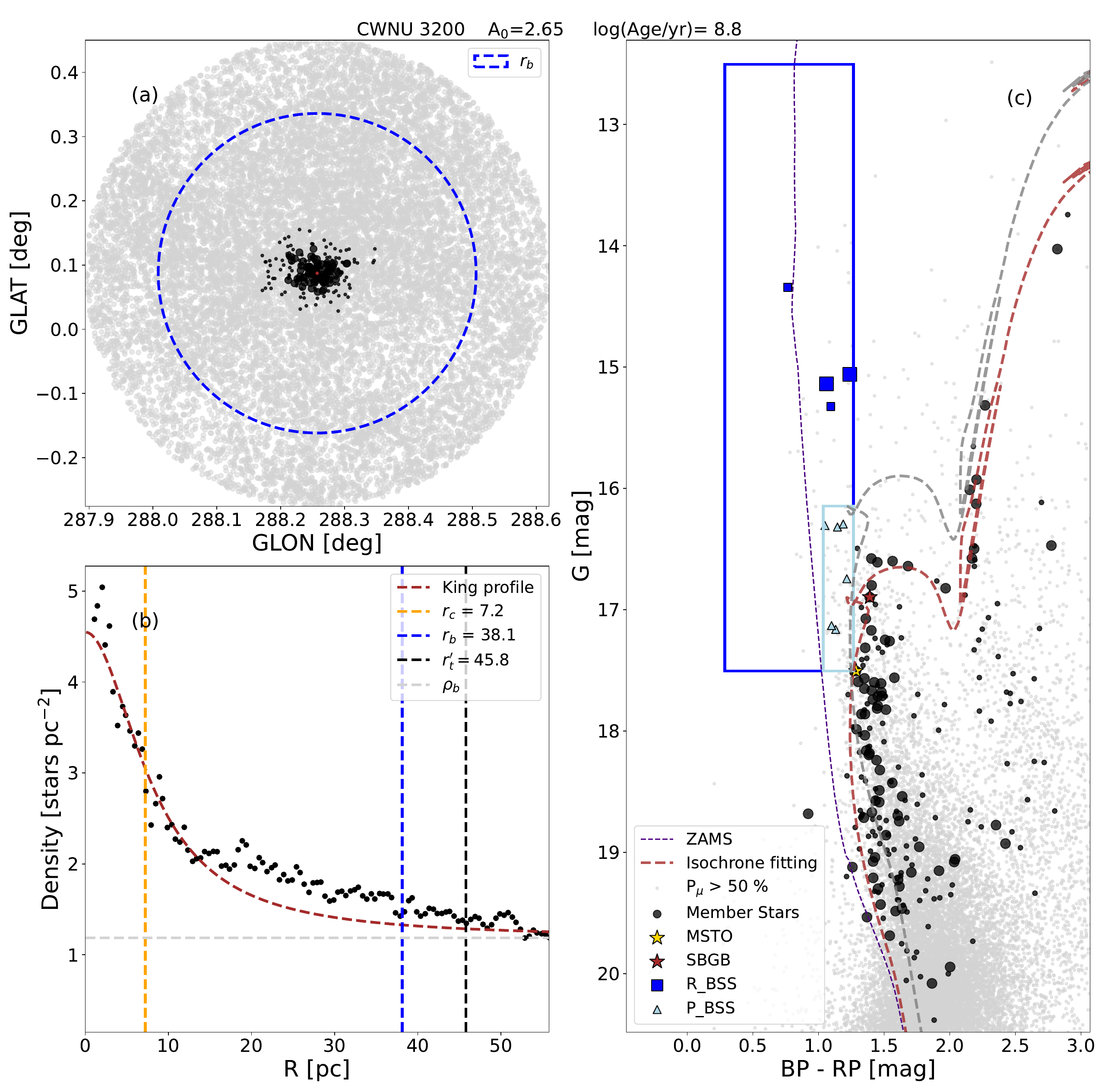}
\includegraphics[width=0.235\linewidth]{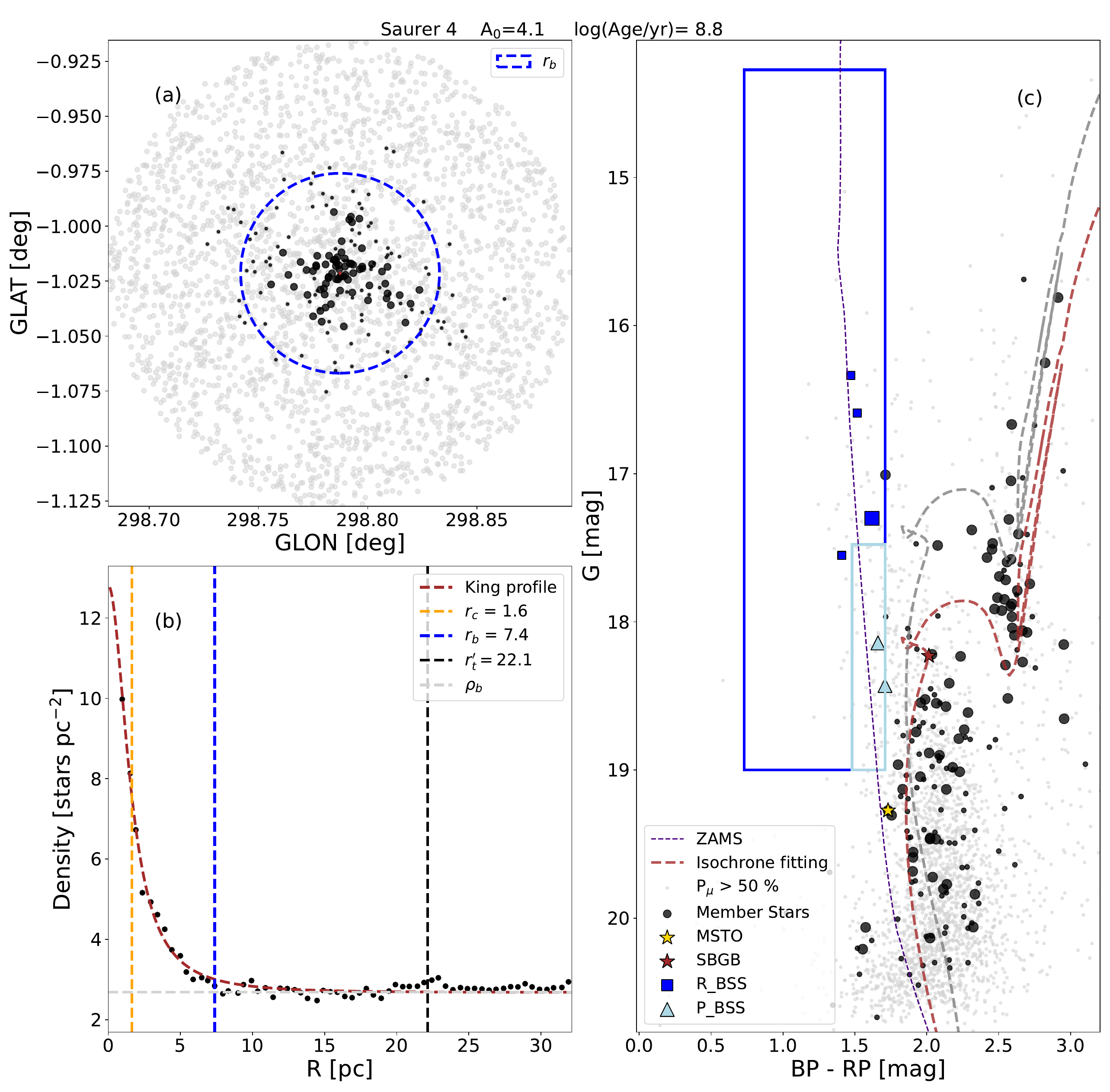}
\includegraphics[width=0.235\linewidth]{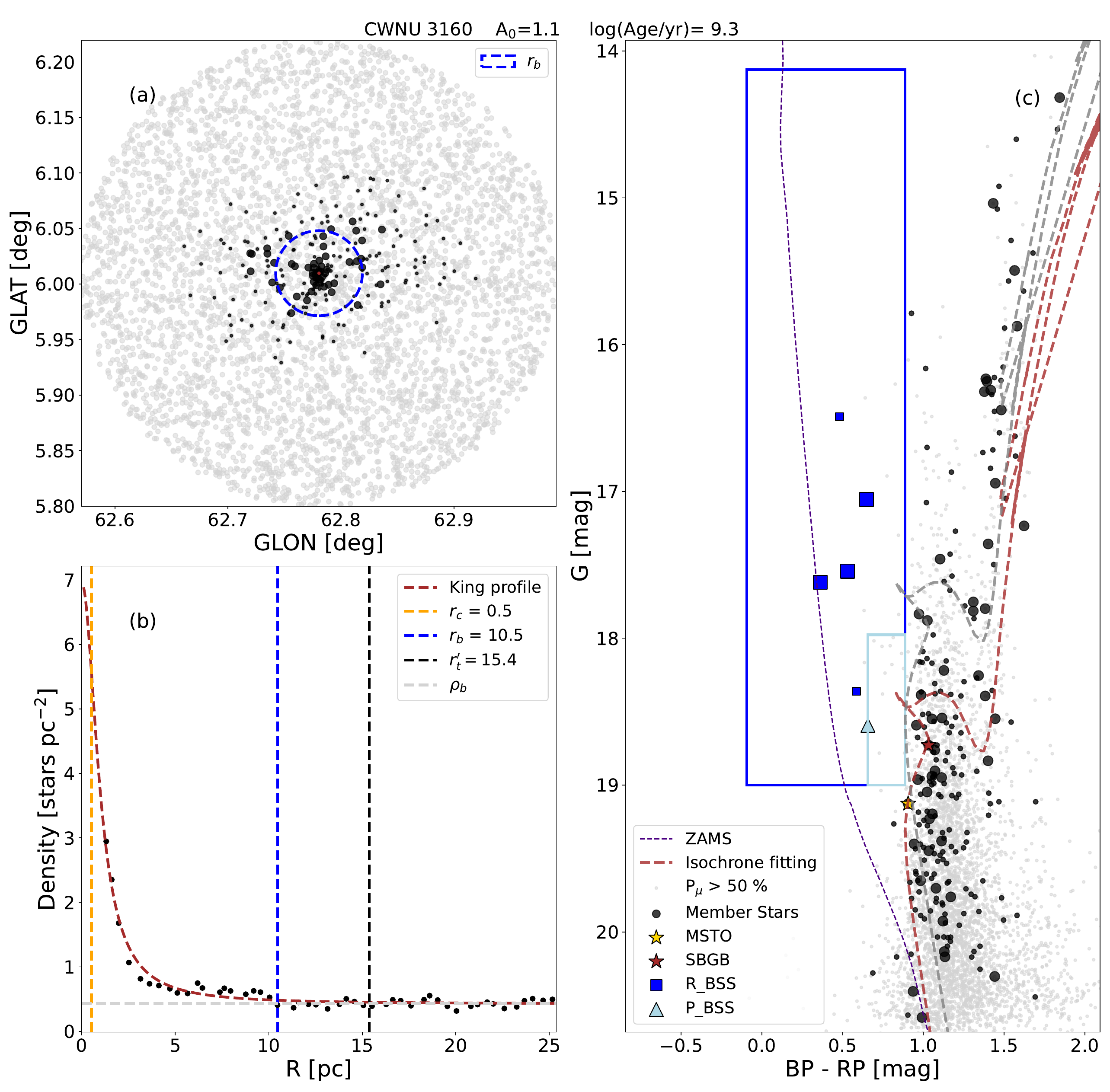}
 \includegraphics[width=0.235\linewidth]{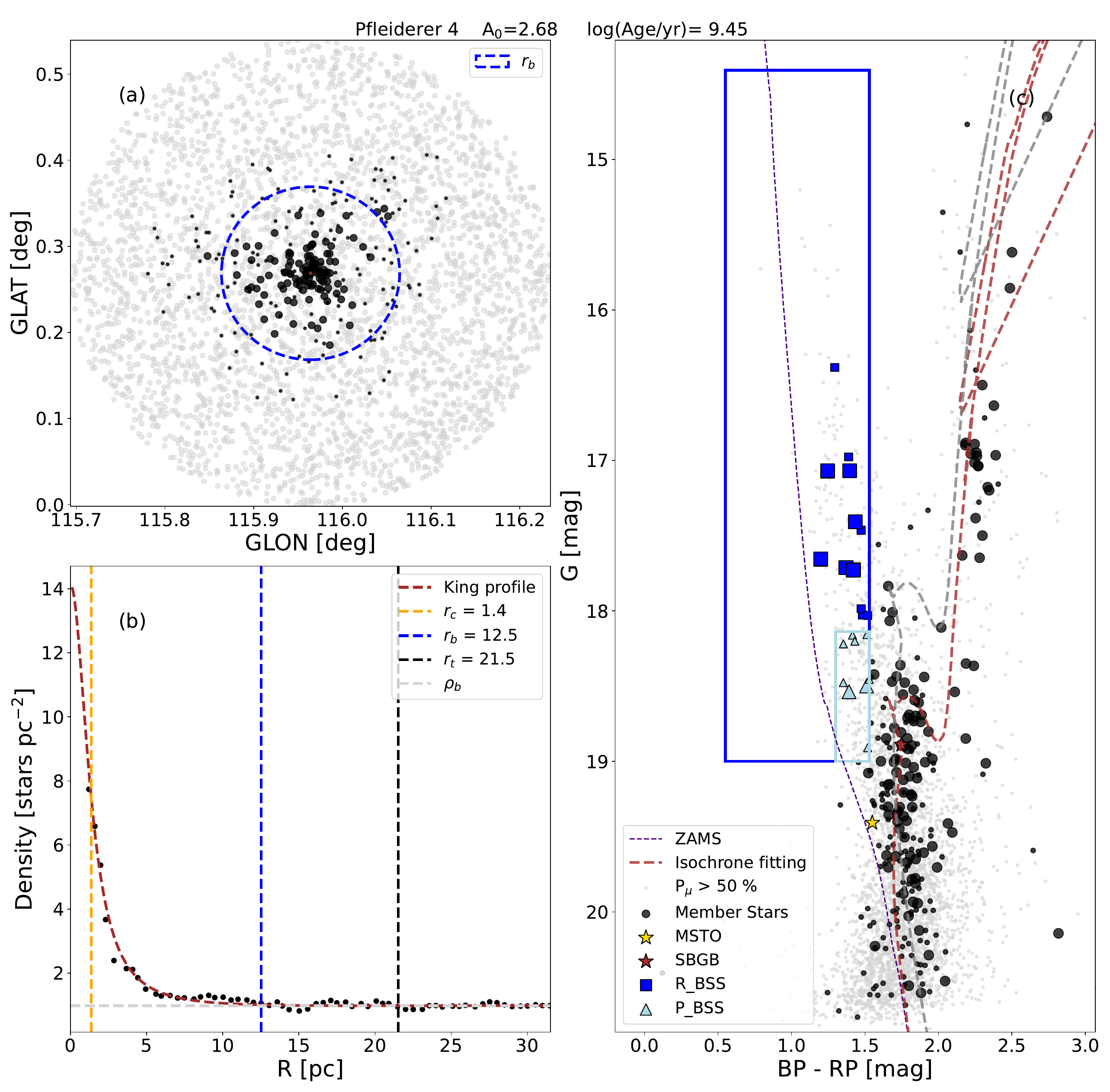}
\includegraphics[width=0.235\linewidth]{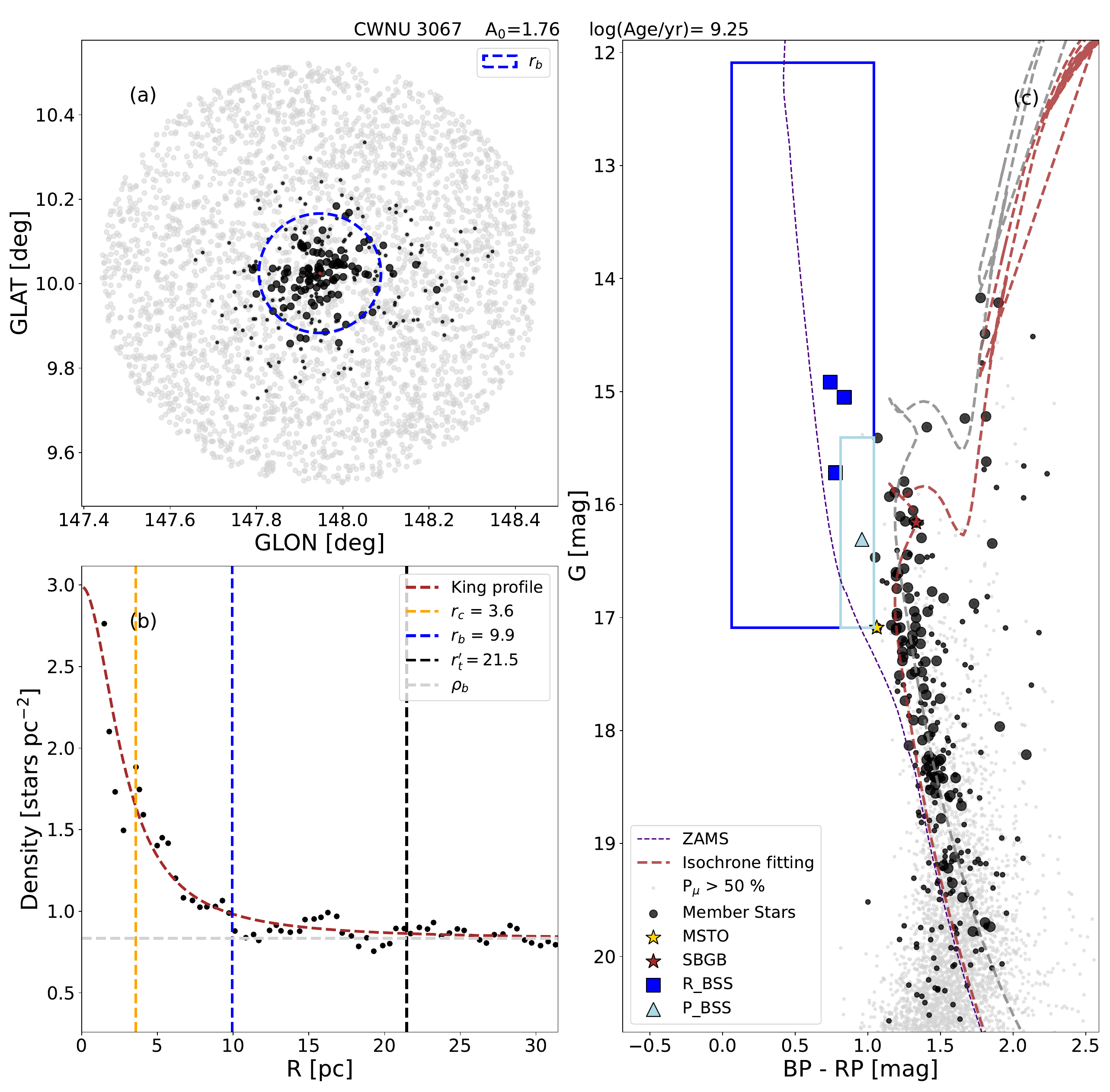}
\includegraphics[width=0.235\linewidth]{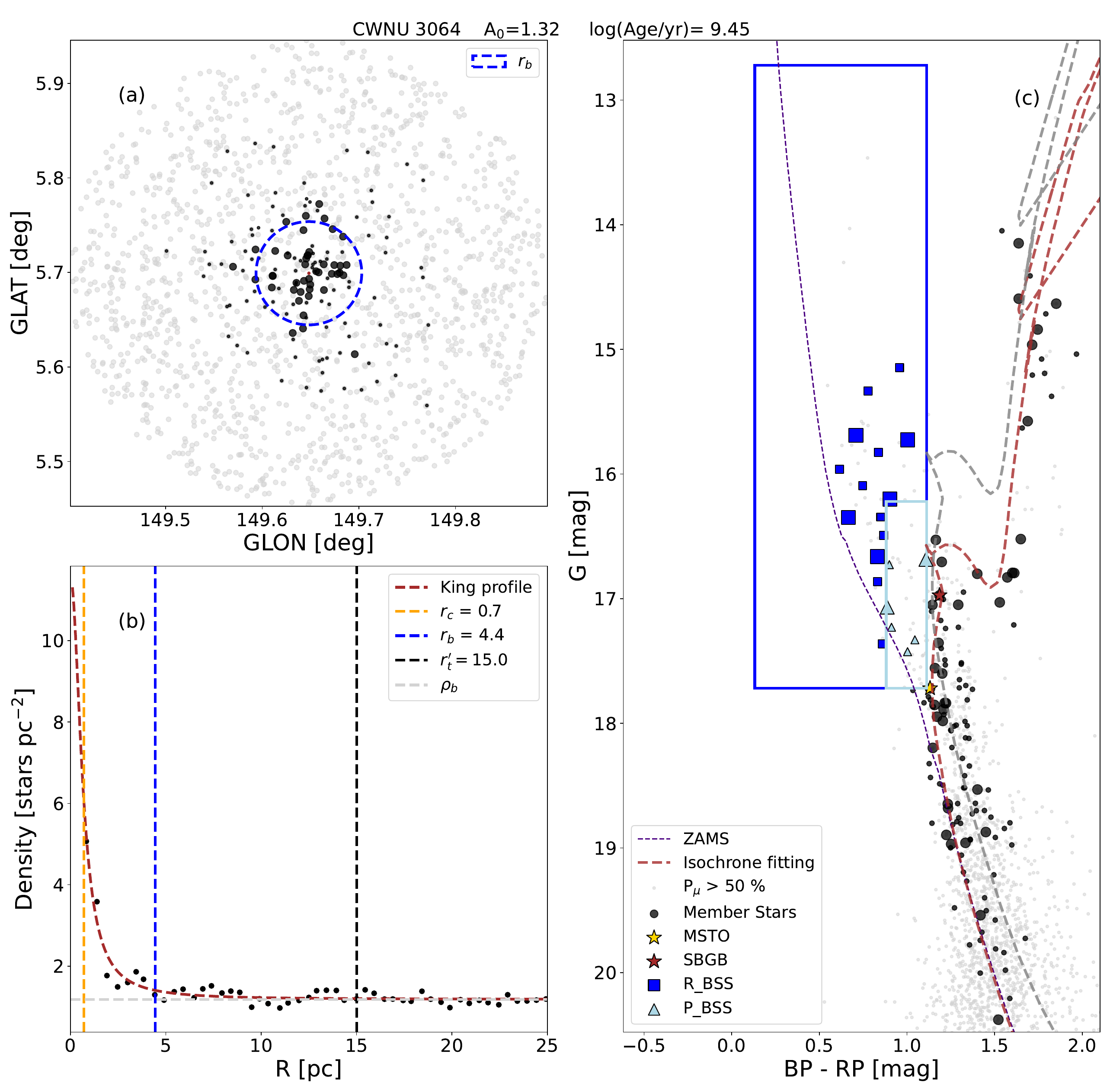}
\caption{Continued to Figure~\ref{figa1}.
}
\label{figa2}
\end{center}
\end{figure*}

\begin{figure*}
\begin{center}
\includegraphics[width=0.235\linewidth]{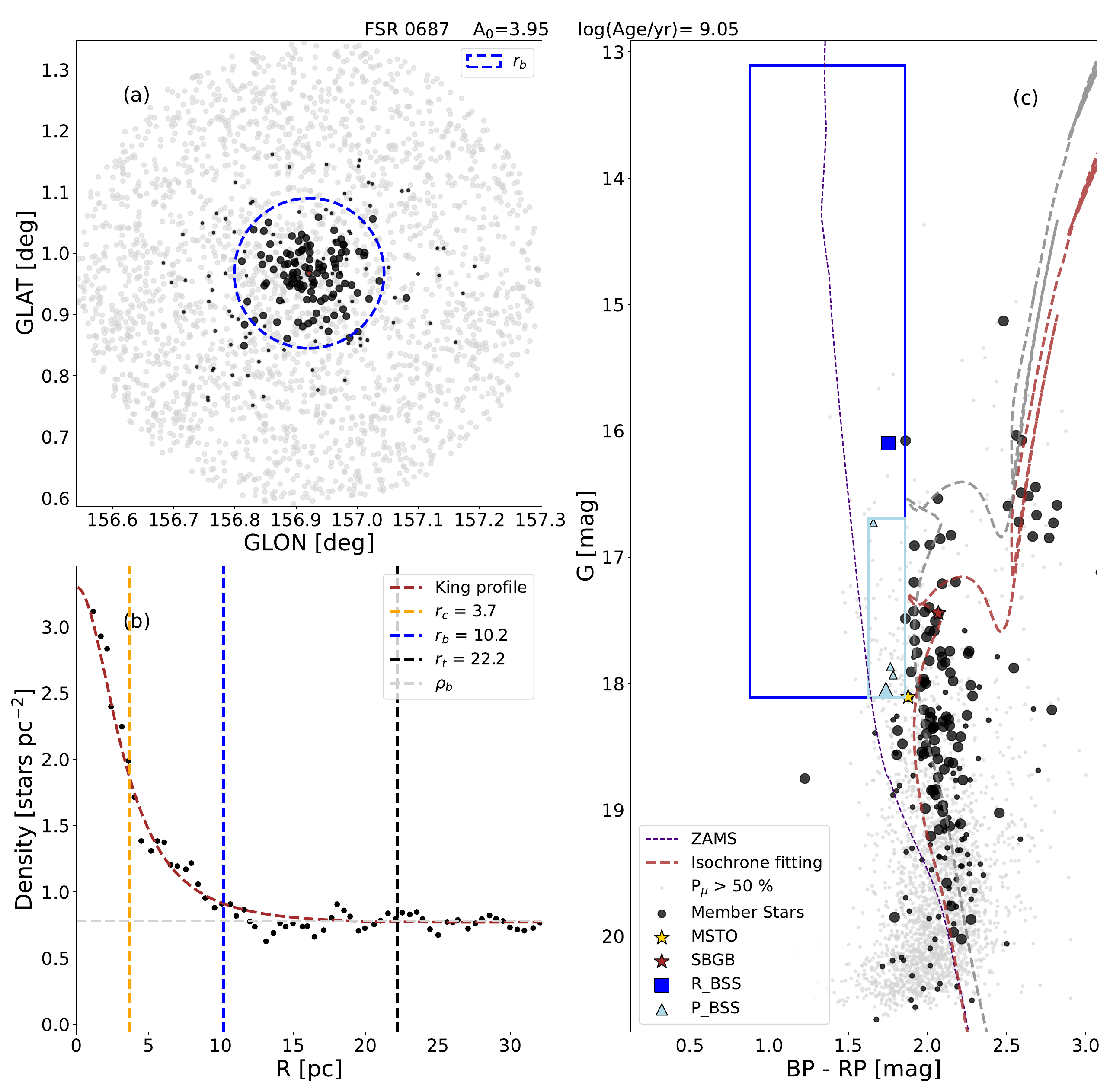}
\includegraphics[width=0.235\linewidth]{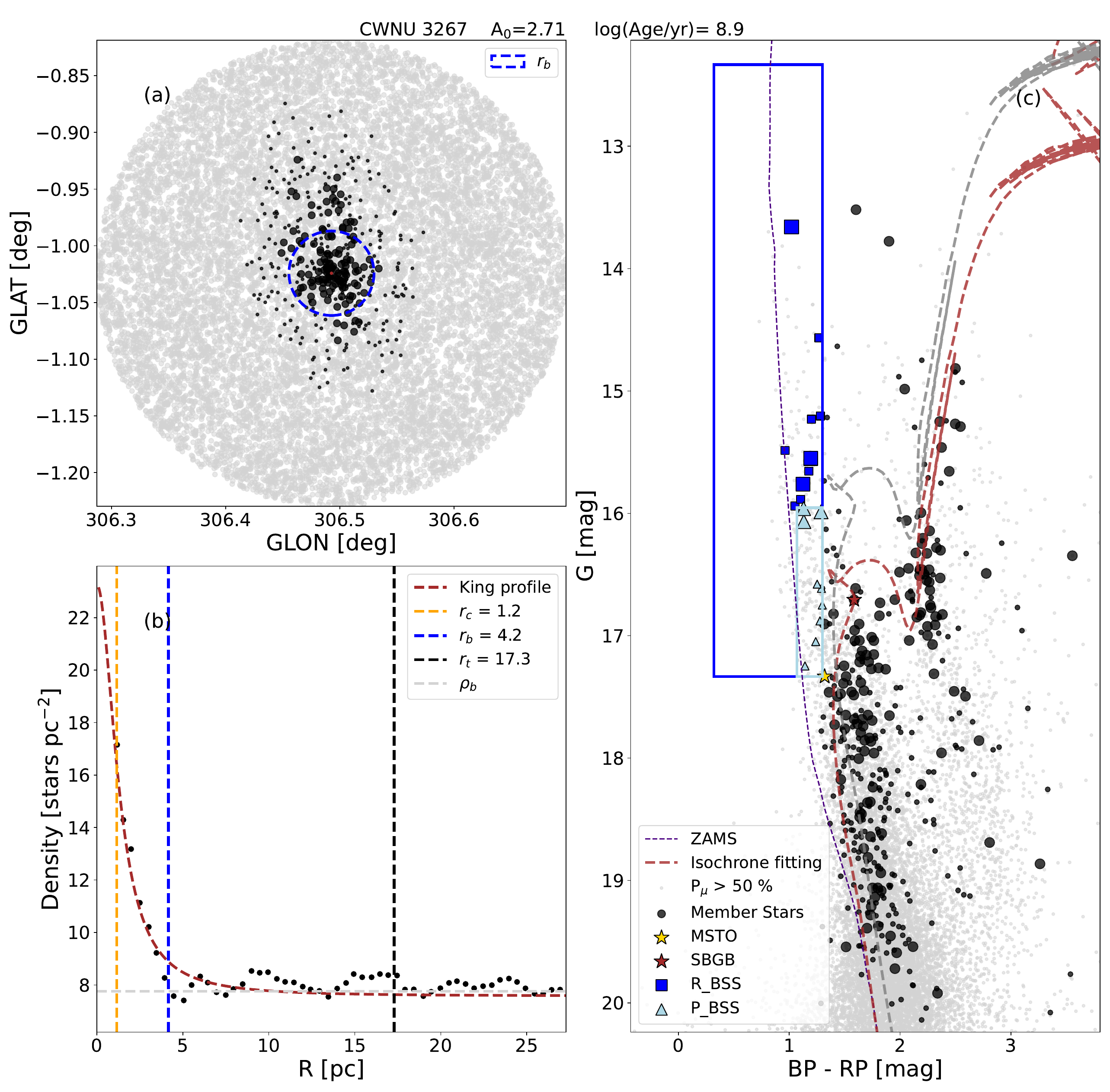}
\includegraphics[width=0.235\linewidth]{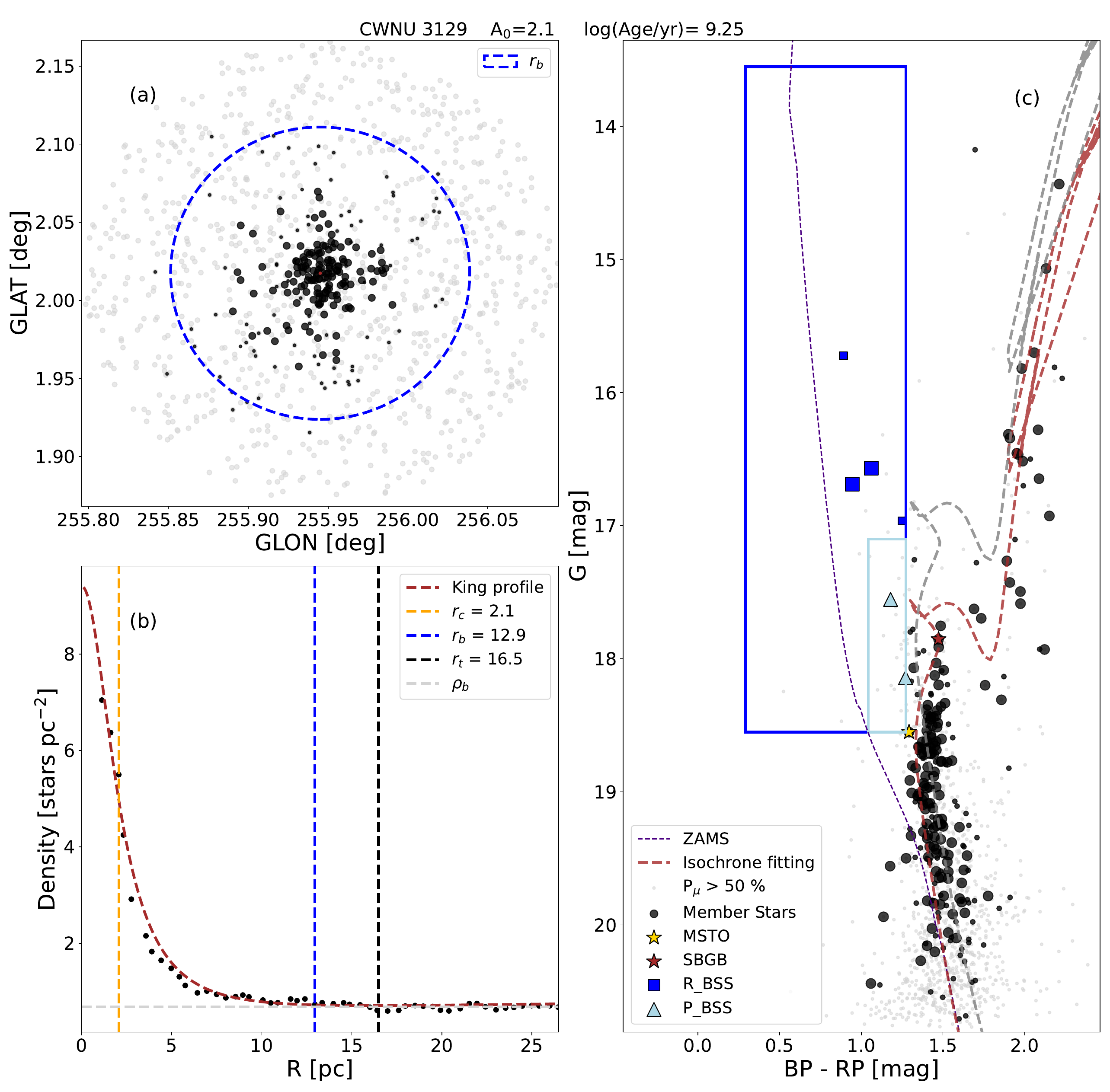}
\includegraphics[width=0.235\linewidth]{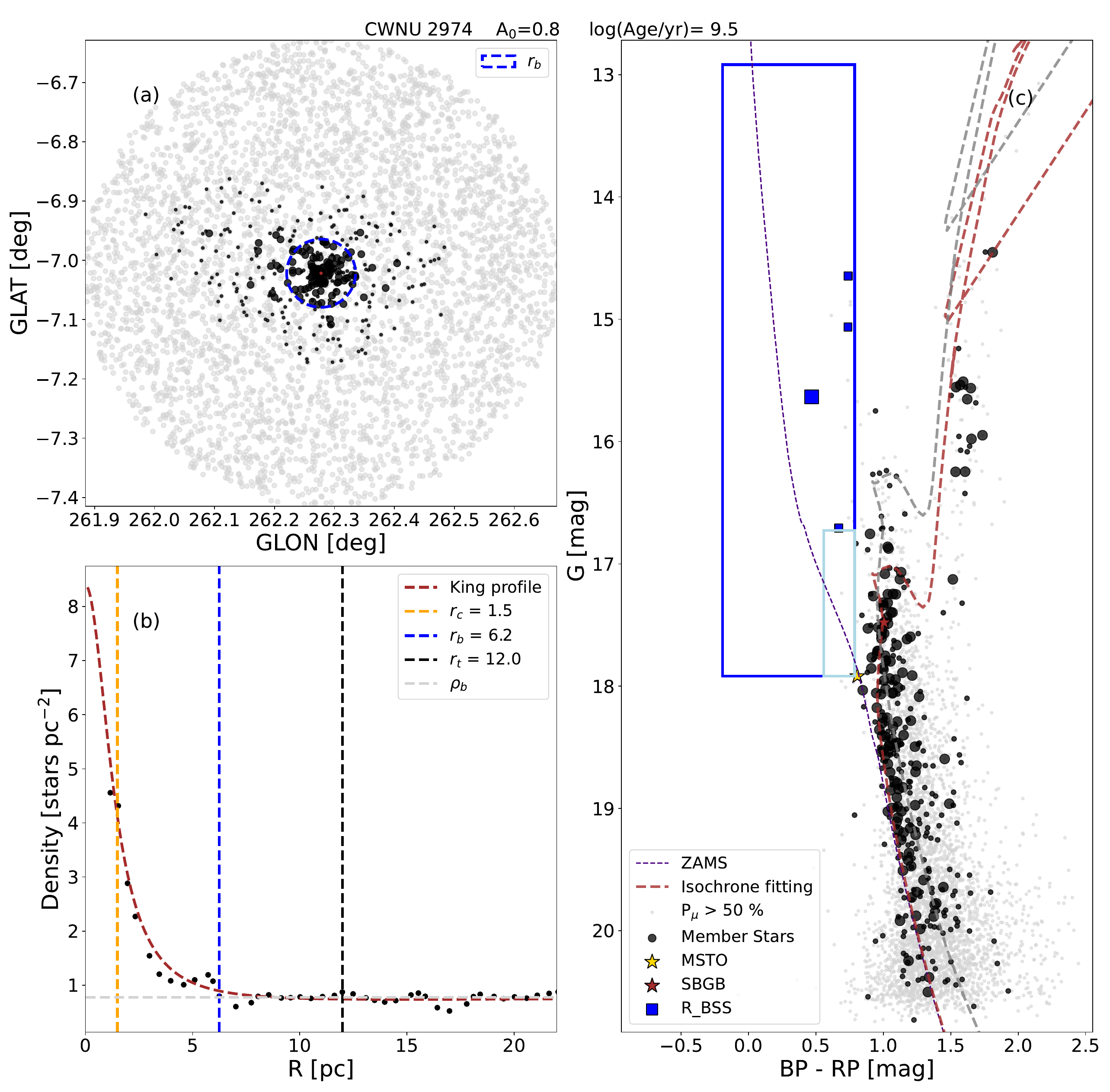}
\includegraphics[width=0.235\linewidth]{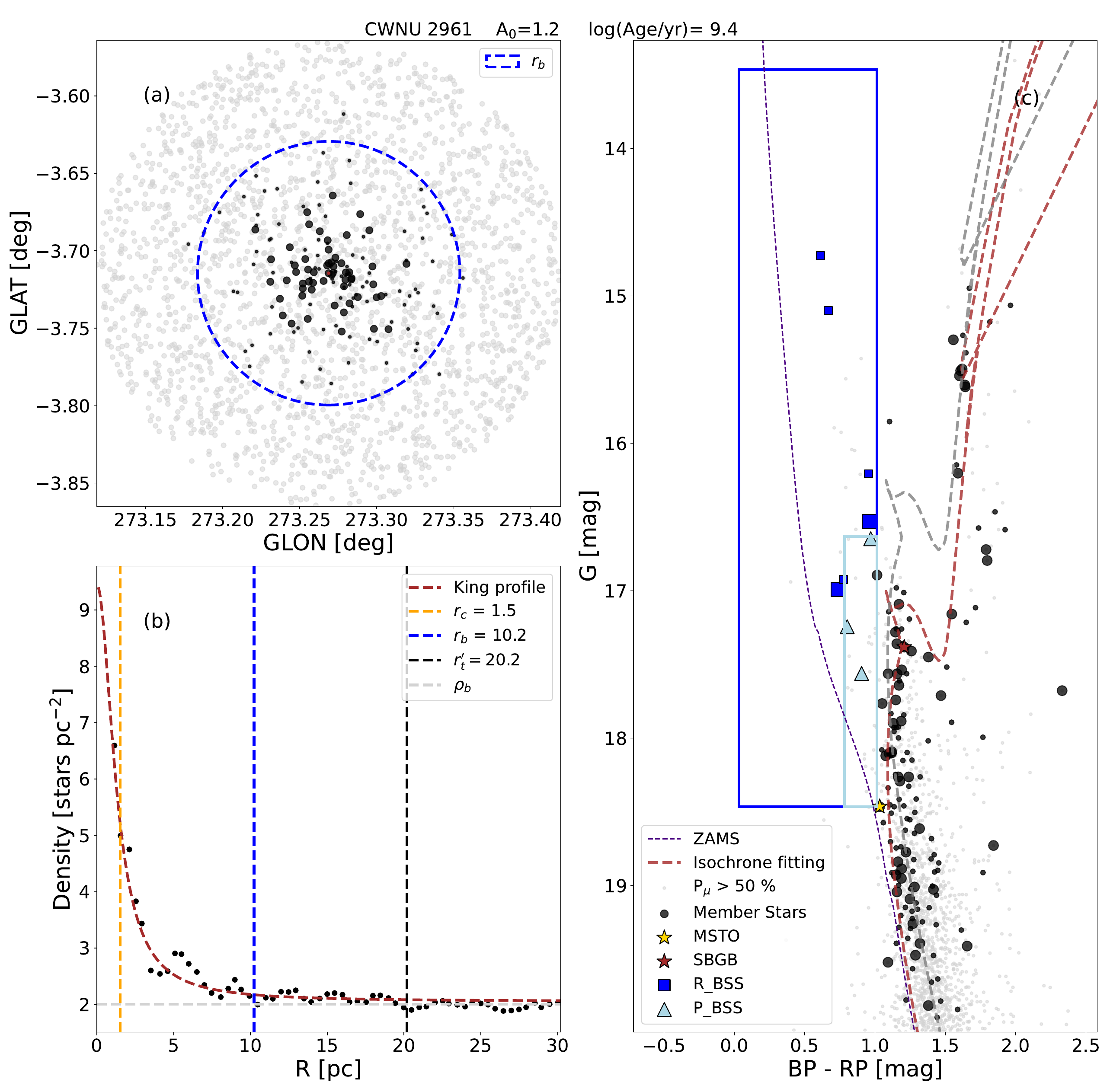}
\includegraphics[width=0.235\linewidth]{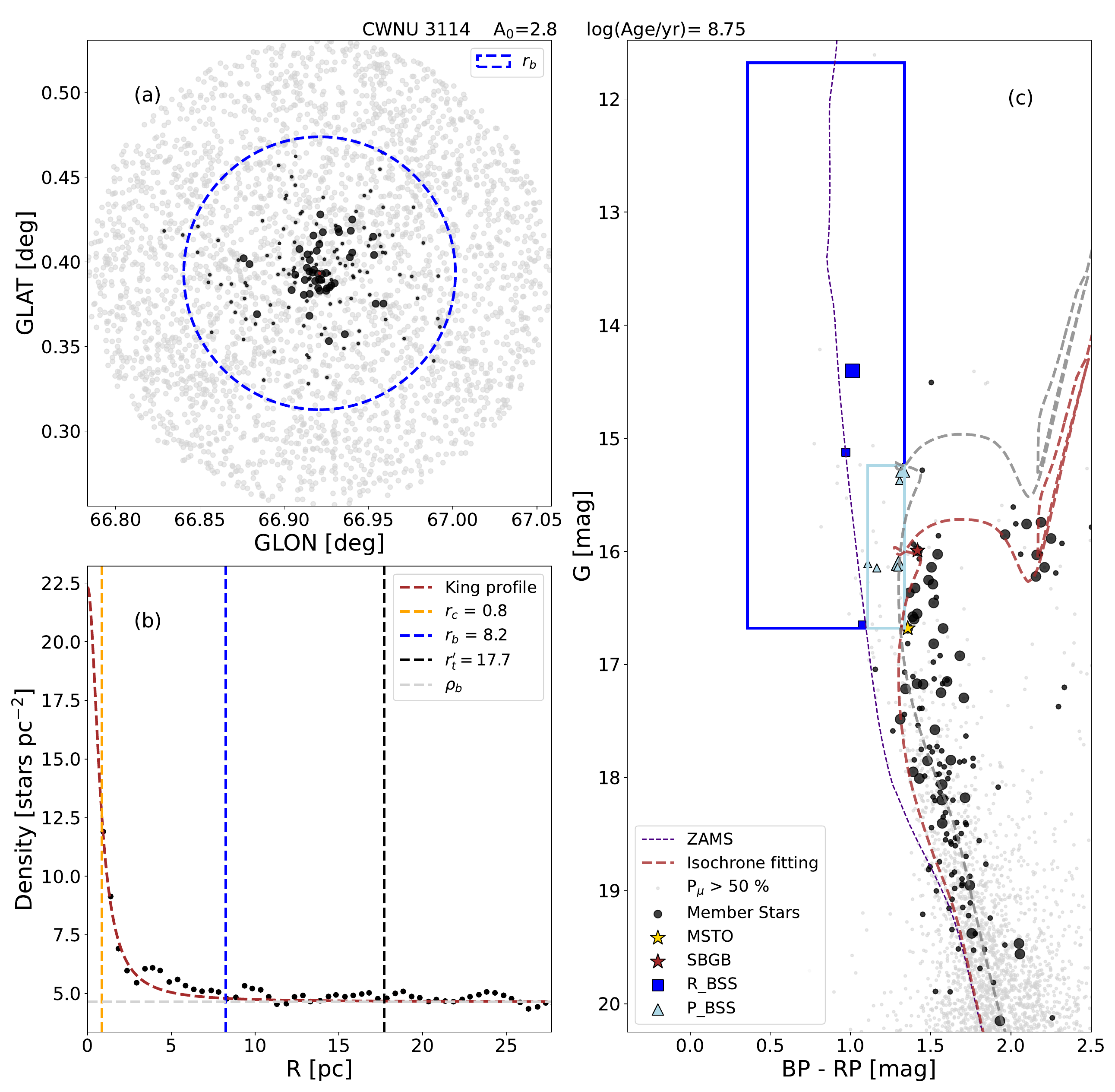}
\includegraphics[width=0.235\linewidth]{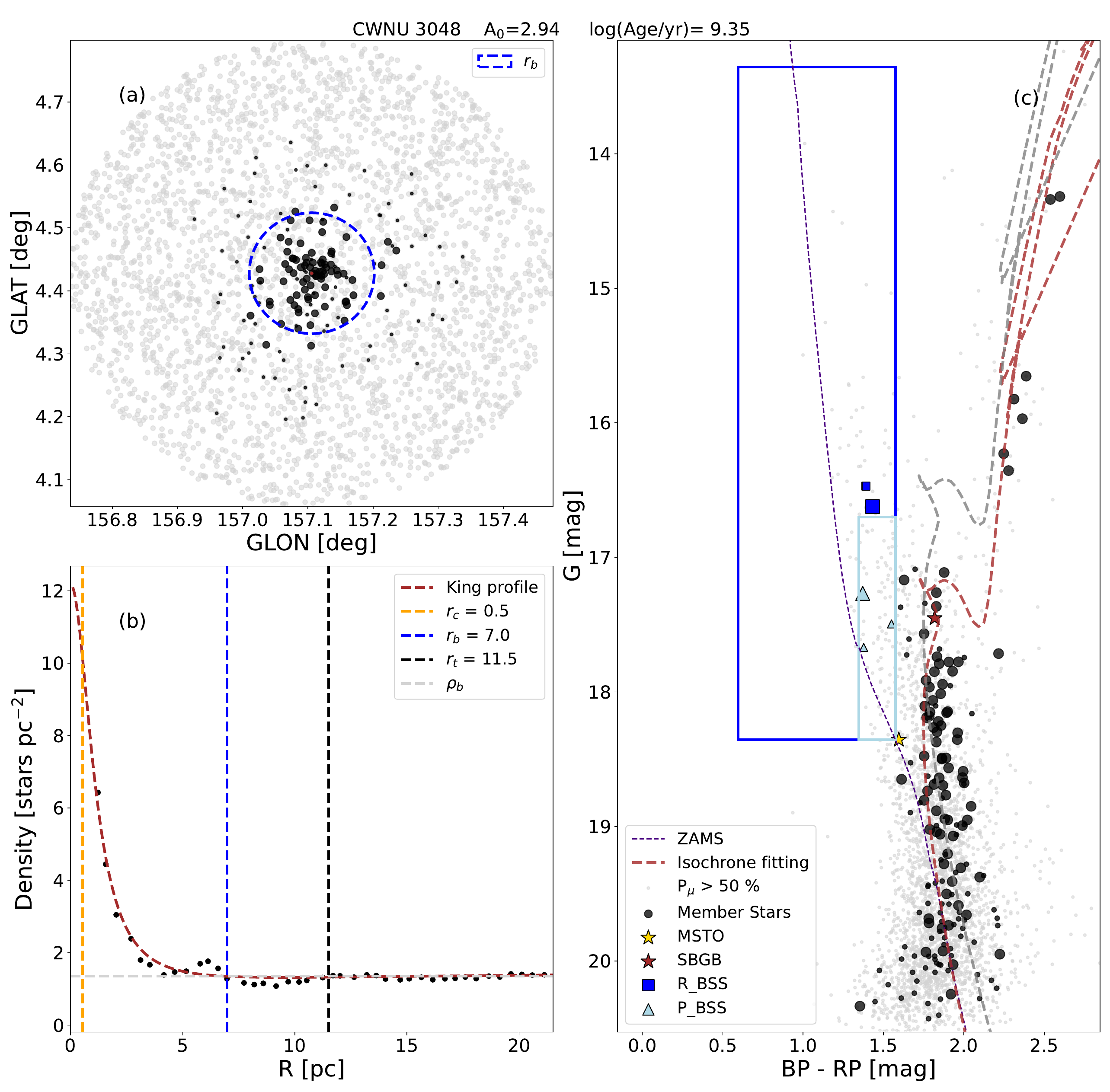}
\includegraphics[width=0.235\linewidth]{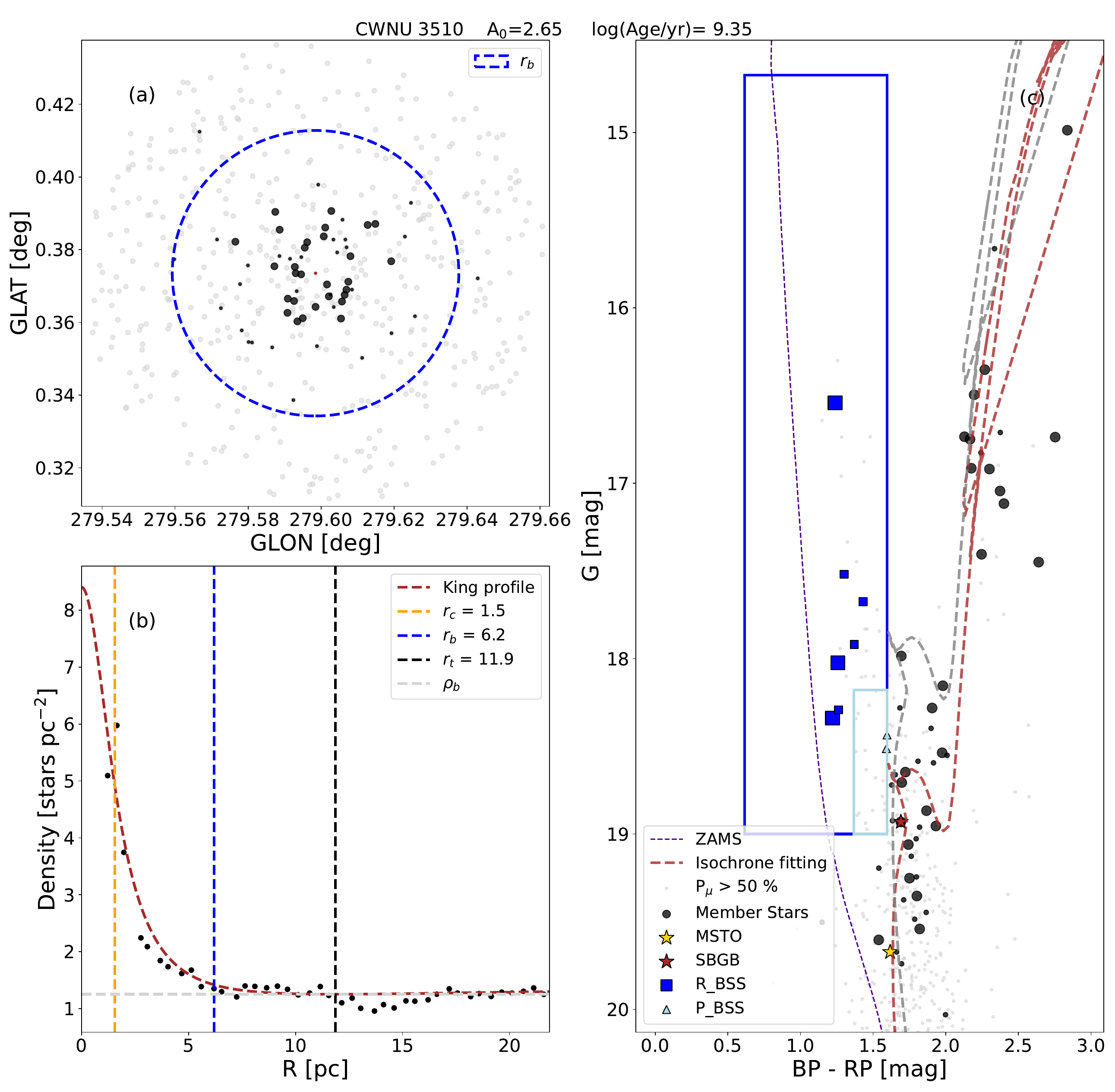}
\includegraphics[width=0.235\linewidth]{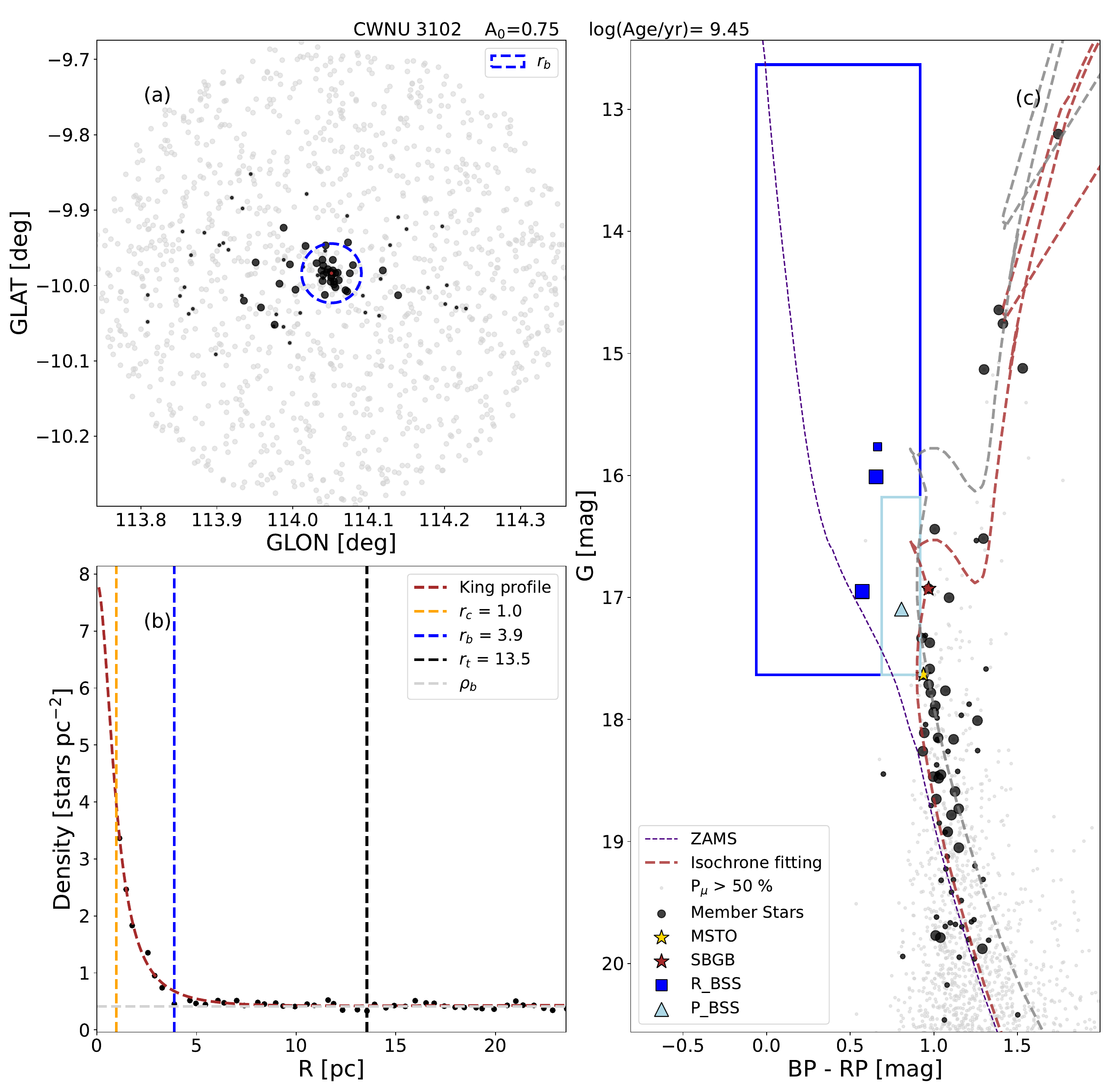}
\includegraphics[width=0.235\linewidth]{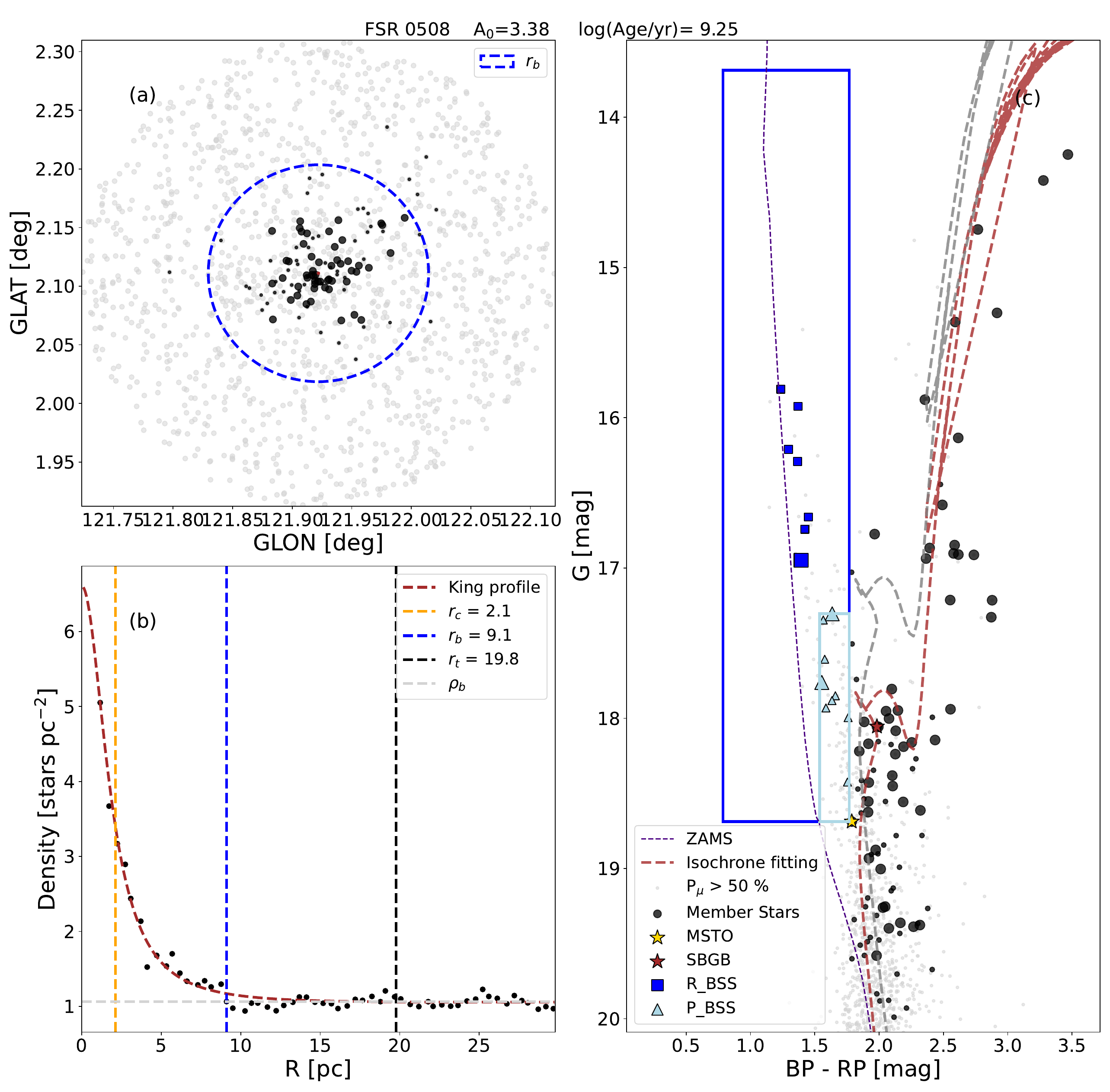}
\includegraphics[width=0.235\linewidth]{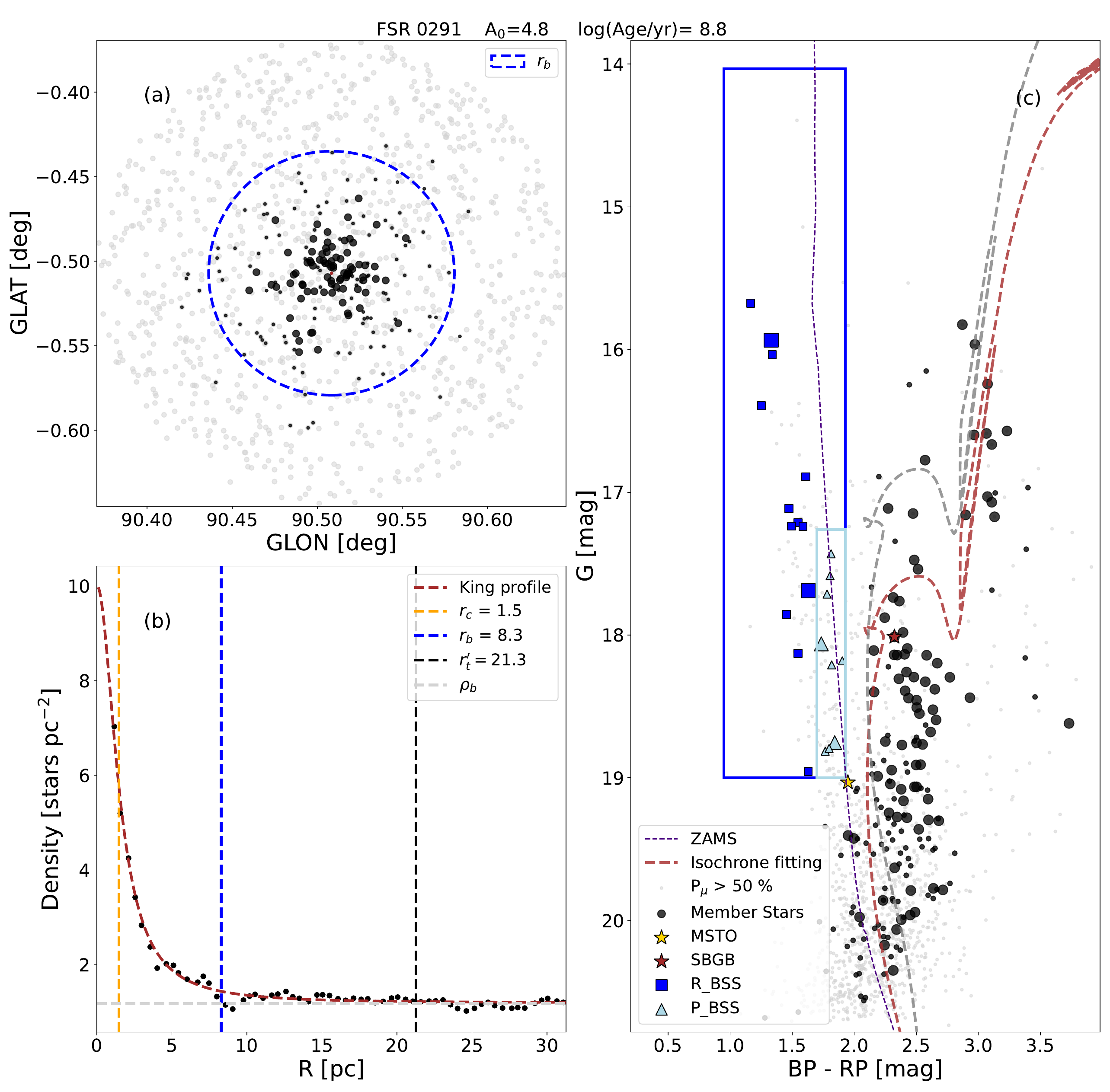}
\caption{Continued to Figure~\ref{figa1}.
}
\label{figa3}
\end{center}
\end{figure*}


\section{The Extinction Coefficient}\label{appendixb}
Table~\ref{extinction coefficient} presents the values $c_1$ to $c_{10}$ used in Equation~\ref{extinction_coefficient}. These values were derived from the publicly available auxiliary data provided by ESA/Gaia/DPAC/CU5, which was prepared by Carine Babusiaux.

\begin{table}[ht!]
    \centering
    \caption{The extinction coefficient in different bands}
    \resizebox{\textwidth}{!}{
        \begin{tabular}{ccccccccccc}
            \hline \hline
            $c_1$ & $c_2$ & $c_3$ & $c_4$ & $c_5$ & $c_6$ & $c_7$ & $c_8$ & $c_9$ & $c_{10}$ & Band \\ 
            \hline
            0.66320788 & -0.01798472 & 0.00049377 & -0.00267994 & -0.00651422 & 0.00003302 & 0.00000158 & -0.00007980 & 0.00025568 & 0.00001105 & RP \\ 
            1.15363197 & -0.08140130 & -0.03601302 & 0.01921436 & -0.02239755 & 0.00084056 & -0.00001310 & 0.00660124 & -0.00088225 & -0.00011122 & BP \\ 
            0.99596972 & -0.15972646 & 0.01223807 & 0.00090727 & -0.03771603 & 0.00151347 & -0.00002524 & 0.01145227 & -0.00093691 & -0.00026030 & G \\ 
            \hline  \hline
        \end{tabular}
    }
    \label{extinction coefficient} 
\end{table}

\section{The outliers of maximum $M_{e}$}\label{appendixc}
\begin{figure*} 
    \centering
    \includegraphics[width=0.4615\linewidth]{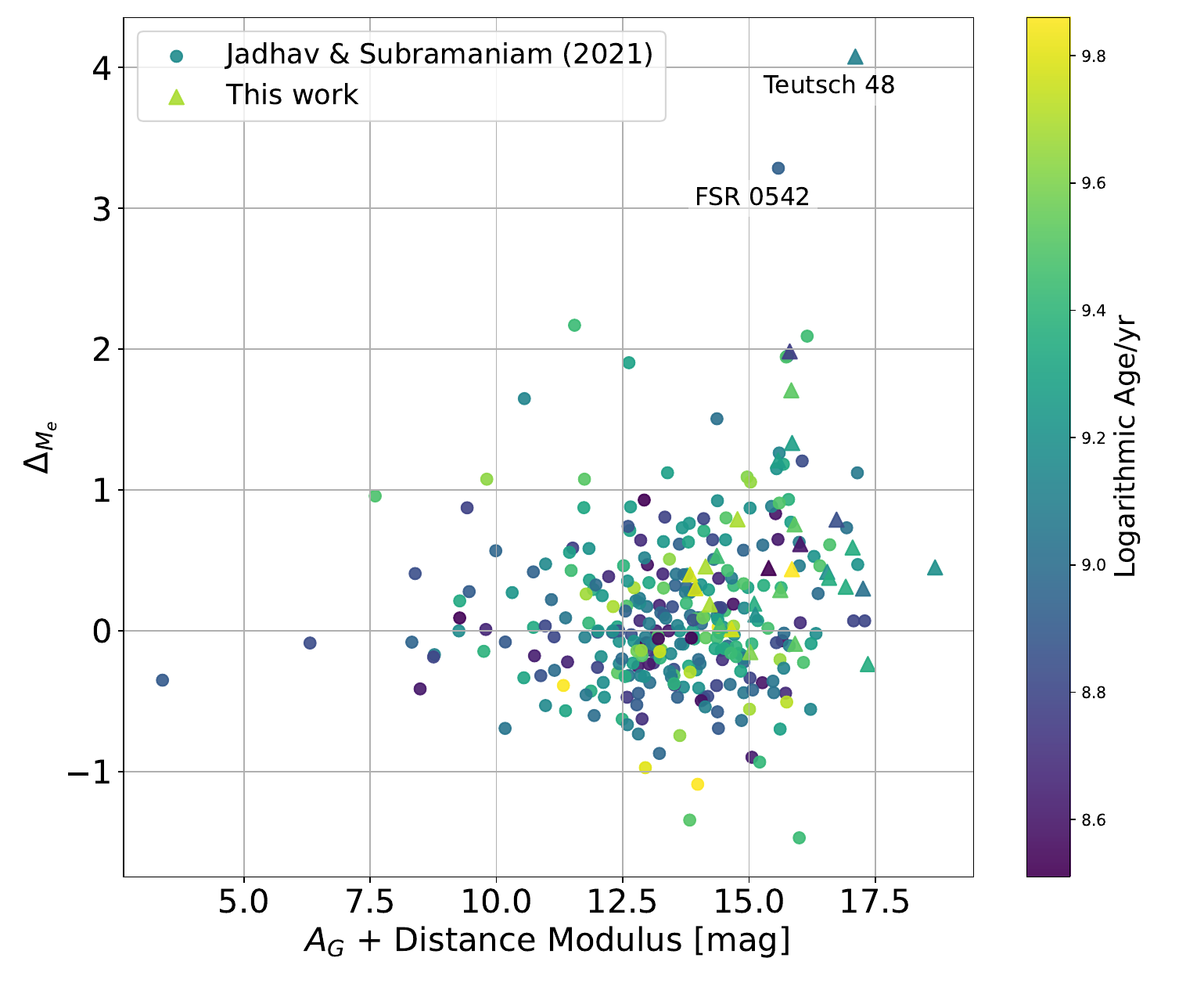}
\caption{The outliers deviating from the median $M_{e}$ values in Figure~\ref{fig_me_plot2}. The horizontal axis represents the sum of the extinction and distance modulus of the clusters, while the vertical axis indicates the deviation between the maximum $M_{e}$ of OC BSS and the median $M_{e}$ values in Figure~\ref{fig_me_plot2}. The color represents the logarithmic age of the clusters.}

    \label{fig_delta_me}
\end{figure*}

We observed that a few outliers in Figure~\ref{fig_me_plot2} deviate significantly from the maximum $M_{e}$ median line. As illustrated in Figure~\ref{fig_delta_me}, these notable outliers (such as FSR~0542 and Teutsch~48) exhibit greater distances and extinctions, and/or some of them possess considerable ages.

\newpage
\bibliography{ref_art}{}
\bibliographystyle{aasjournal}


\end{CJK*}
\end{document}